\documentclass[fleqn,usenatbib]{mnras}
\usepackage{amsmath}
\usepackage{txfonts}                % For good fonts
\usepackage{graphicx}               % Including figure files

\usepackage{amssymb}	% Extra maths symbols
\usepackage{epsfig}
\usepackage{epstopdf}
\usepackage{verbatim}
\usepackage{cancel}
\usepackage{lscape}
\usepackage{longtable}
\usepackage{deluxetable}
\usepackage{rotating}
\usepackage{lscape}
\usepackage{color}
\definecolor{dkcyan}{rgb}{0.000,0.600,0.600}
\definecolor{navy}{RGB}{0,0,102}
\definecolor{royal}{RGB}{0,0,204}
\definecolor{applegreen}{rgb}{0.55, 0.71, 0.0}
\definecolor{green}{rgb}{0.0, 0.5, 0.0}
\newcommand{\degree}{$^{\circ}$}

\hypersetup{draft}
\title[A quartet of black holes and a missing duo]{A quartet of black holes and a missing duo: probing the low-end of the M$_{\rm BH} - \sigma$ relation with the adaptive optics assisted integral-field spectroscopy }

\author[D. Krajnovi\'c et al.]{Davor Krajnovi\'c $^{1}$\thanks{E-mail:dkrajnovic@aip.de},  
Michele Cappellari$^{2}$,
Richard M. McDermid$^{3,4}$,\newauthor
Sabine Thater$^{1}$,
Kristina Nyland$^{5}$,
P. Tim de Zeeuw$^{6,7}$,
Jes\'us Falc\'on-Barroso$^{8,9}$,\newauthor
Sadegh Khochfar$^{10}$,
Harald Kuntschner$^{6}$,
Marc Sarzi$^{11}$, and
Lisa M. Young$^{12}$\\
$^{1}$Leibniz-Institut f\"ur Astrophysik Potsdam (AIP), An der Sternwarte 16, D-14482 Potsdam, Germany\\
$^{2}$Sub-Department of Astrophysics, Department of Physics, University of Oxford, Denys Wilkinson Building, Keble Road, Oxford OX1 3RH, UK\\
$^{3}$Department of Physics and Astronomy, Macquarie University, Sydney, NSW 2109, Australia\\
$^{4}$Australian Gemini Office, Australian Astronomical Observatory, PO Box 915, Sydney, NSW 1670, Australia\\
$^{5}$National Radio Astronomy Observatory, Charlottesville, VA 22903, USA\\
$^{6}$Max Planck Institut f\"ur extraterrestrische Physik, Giessenbachstrasse 1, D-85748 Garching, Germany\\
$^{7}$Leiden Observatory, Leiden University, Niels Bohrweg 2, 2333 CA Leiden, The Netherlands\\
$^{8}$Instituto de Astrof\'isica de Canarias, V\'ia L\'actea s/n, E-38200 La Laguna, Tenerife, Spain\\
$^{9}$Departamento de Astrof\'isica, Universidad de La Laguna, E-38206 La Laguna, Tenerife, Spain\\
$^{10}$Institute for Astronomy, University of Edinburgh, Royal Observatory, Edinburgh EH9 3HJ, UK\\
$^{11}$Centre for Astrophysics Research, University of Hertfordshire, Hatfield AL10 9AB, UK\\
$^{12}$Physics Department, New Mexico Institute of Mining and Technology, Socorro, NM 87801, USA\\
}

\date{Accepted 2018 March 20. Received 2018 February 23; in original form 2017 May 16}

\pubyear{2016}

\begin{document}
\label{firstpage}
\pagerange{\pageref{firstpage}--\pageref{lastpage}}
\maketitle

\begin{abstract}
We present mass estimates of supermassive black holes in six nearby fast rotating early-type galaxies (NGC\,4339, NGC\,4434, NGC\,4474, NGC\,4551, NGC\,4578 and NGC\,4762) with effective stellar velocity dispersion around 100 km/s. We use near-infrared laser-guide adaptive optics observations with the GEMINI/NIFS to derive stellar kinematics in the galactic nuclei, and SAURON observations from the ATLAS$^{\rm3D}$ Survey for large-scale kinematics. We build axisymmetric Jeans Anisotropic Models and axisymmetric Schwarzschild dynamical models. Both modelling approaches recover consistent orbital anisotropies and black hole masses within $1-2\sigma$ confidence level, except for one galaxy for which the difference is just above the $3\sigma$ level. Two black holes (NGC\,4339 and NGC\,4434) are amongst the largest outliers from the current black hole mass - velocity dispersion relation, with masses of $(4.3^{+4.8}_{-2.3})\times10^7$ and $(7.0^{+2.0}_{-2.8})\times10^7$ M$_\odot$, respectively ($3\sigma$ confidence level). The black holes in NGC\,4578 and NGC\,4762 lie on the scaling relation with masses of $(1.9^{+0.6}_{-1.4})\times10^7$ and $(2.3^{+0.9}_{-0.6})\times10^7$ M$_\odot$, respectively ($3\sigma$ confidence level). For two galaxies (NGC\,4474 and NGC\,4551) we are able to place upper limits on their black holes masses ($<7\times10^6$ and $<5\times10^6$ M$_\odot$, respectively, $3\sigma$ confidence level). The kinematics for these galaxies clearly indicate central velocity dispersion drops within a radius of 35 pc and 80 pc, respectively. These drops cannot be associated with cold stellar structures and our data do not have the resolution to exclude black holes with masses an order of magnitude smaller than the predictions. Parametrizing the orbital distribution in spherical coordinates, the vicinity of the black holes is characterized by isotropic or mildly tangential anisotropy.

\end{abstract}


%%%%%%%%%%%%%%%%%%%%%%%%%%%%%%%%%%%%%%%%%%%%%%%%%%

%%%%%%%%%%%%%%%%% BODY OF PAPER %%%%%%%%%%%%%%%%%%

%%%%%%%%%%%%%%%%%%%%%%%%%%%%%%%%%%%%%%%%%%%%%%%%%%%%%%%%%%%
%
% SECTION 1 SECTION 1 SECTION 1 SECTION 1 SECTION 1 SECTION 1
%
%%%%%%%%%%%%%%%%%%%%%%%%%%%%%%%%%%%%%%%%%%%%%%%%%%%%%%%%%%%

\section{Introduction}
\label{s:intro}

Determining the masses of black holes in the centres of galaxies is marred with difficulties. Galaxies are systems with $>10^{10}$ stars of different ages and metallicities, grouped in a number of structural components such as bulges, discs, rings, bars and spherical haloes. They also contain gas in various phases, and regions forming stars. Central black holes can be considered part of the invisible dark matter content of galaxies responsible for the changes to the total gravitational potential. All these components need to be taken into account at various levels of sophistication when constructing dynamical models and estimating masses of central black holes. 

There are several established methods for measuring the black hole masses, which have one thing in common: the use of luminous tracers of the gravitational potential, such as stars or clouds of ionized or molecular gas. The observed motions of these tracers are used to constrain dynamical models, which can separate the contribution to the potential of the black hole and the rest of the galaxy, and, therefore, provide the black hole mass, M$_{\rm BH}$. The various approaches of using different types of tracers are reviewed in \citet{2013ARA&A..51..511K}. For all types of dynamical models, and specifically for the stellar dynamical models used in the work presented here, it is essential that the assumed gravitational potential is a realistic representation of the galaxy, and that the data describing the motions of the tracer are adequate to distinguish the presence of a black hole\citep[see for reviews][]{2005SSRv..116..523F, 2013ARA&A..51..511K}. 

The mass of a black hole is added as a component in the total gravitational potential defined by the distribution of stars and, possibly, dark matter (the mass of the gas in the type of systems considered here is usually small and typically ignored). It dominates a region defined as the sphere of influence (SoI), with a characteristic radius $r_{\rm SoI} = GM_{\rm BH}/\sigma_e^2$, where $G$ is the gravitational constant and $\sigma_e$ is the characteristic velocity dispersion of the galaxy within an effective radius \citep{2001bhbg.conf...78D, 2005SSRv..116..523F}. Within SoI, the black hole is the dominant source of the gravitational potential, but its influence can impact the galaxy at larger radii. The reason for this is that the gravitational potential falls off relatively slowly with $1/r$, and that stars on non-circular orbits may pass within or close to the SoI, and therefore can be influenced by the black hole even though they spend most of their time well outside the SoI. This is why, to constrain a dynamical model, one needs to have good quality data covering the region close to the black hole, but also mapping the bulk of the galaxy body \citep[e.g.][]{2002MNRAS.335..517V}. 

The spectroscopic capabilities of the Hubble Space Telescope (HST) offered a way to probe the regions close to black holes at the high surface brightness centres of some galaxies \citep[see fig.~1 of][]{2013ARA&A..51..511K}, but it is the large collecting areas of the ground-based 8-10m telescopes assisted with natural guide star or laser guide star (LGS) adaptive optics (AO) that paved a way forward to extending  the types of galaxies with measured black hole masses \cite[e.g.][]{2006MNRAS.367....2H, 2006ApJ...643..226H, 2007MNRAS.379..909N, 2007ApJ...671.1329N,2009MNRAS.394..660C, 2009MNRAS.399.1839K, 2011Natur.480..215M, 2012ApJ...753...79W}. To this one should add the opening of the wavelength space with the observations of the nuclear masers \citep[e.g.][]{2010ApJ...721...26G,2011ApJ...727...20K,2016ApJ...826L..32G} and the sub-millimetre interferometric observations of circumnuclear gas discs \citep[e.g.][]{2013Natur.494..328D,2015ApJ...806...39O,2016ApJ...822L..28B}. Regarding dynamical models based on stellar kinematics, the advent of mapping galaxy properties with integral-field unit (IFUs) spectrographs has greatly improved the observational constraints on stellar orbital structure \citep{2005MNRAS.357.1113K}, resulting in more secure determinations of M$_{\rm BH}$ \citep{2010AIPC.1240..211C}.

The price that has to be paid to determine a single M$_{\rm BH}$ using stellar kinematics is, however, large: one typically needs observations coming from 3 to 4 different instruments usually mounted on the same number of different telescopes. These are the small-scale high-resolution IFU data [typically in the near-infrared (NIR) to increase the spatial resolution when observed from the ground], and the large scale IFU data of moderate (or poor) spatial resolution. These spectroscopic data are needed to determine the motions of the tracer, but imaging data are also needed to determine the spatial distribution of stars and infer the three-dimensional distribution of mass. Preferably, the imaging data should be of better spatial resolution than the spectroscopy, and currently the only source of such data is the HST. Large-scale (ground-based) imaging is also needed to map the stellar distribution (and ascertain the gravitational potential) to radii several times larger than the extent of the spectroscopic data. 

Obtaining such data sets, and ensuring they are of uniform and sufficient quality, is not a small challenge. The hard-earned data sets accumulated by the community over the past several decades, revealed and confirmed with increasing confidence the striking correlations between the M$_{\rm BH}$ and various properties of host galaxies. We refer the reader to in-depth descriptions and discussions on all black hole scaling relations in recent reviews by \citet{2005SSRv..116..523F}, \citet{2013ARA&A..51..511K} and \citet{2016ASSL..418..263G} and mention here only the M$_{\rm BH} - \sigma$ relation \citep{2000ApJ...539L...9F, 2000ApJ...539L..13G}. This relation, typically found to have the least scatter \citep[e.g.][]{2016ApJ...818...47S,2016ApJ...831..134V} reveals a close connection between two objects of very different sizes, implying a linked evolution of growth for both the host galaxy and its resident black hole. 

Recent compilations provide lists of more than 80 dynamically determined M$_{\rm BH}$ \citep[e.g.][]{2016ApJ...818...47S}, while the total number of black hole masses used in determining this relation approaches 200 \citep{2016ApJ...831..134V}. Increases in sample size have shown that, contrary to initial expectation, the M$_{\rm BH} - \sigma$ relation shows evidence of intrinsic scatter \citep{2009ApJ...698..198G}, and in particular that low- and high-$\sigma$ regions have increased scatter \citep{2008MNRAS.386.2242H, 2009ApJ...698..812G,2013ApJ...764..184M}. Low- and high-mass galaxies have different formation histories \citep[e.g.][]{2011MNRAS.417..845K} and it is not surprising that their black holes might have different masses \citep{2013ApJ...764..151G,2013ApJ...768...76S}, but as more black hole masses are gathered, there also seems to be a difference between galaxies of similar masses. On the high-mass side, brightest galaxies in cluster or groups seem to have more massive black holes than predicted by the relation \citep{2011Natur.480..215M,2012ApJ...756..179M,2016Natur.532..340T}, while among the low-mass systems, active galaxies, galaxies with bars or non-classical bulges show a large spread of M$_{\rm BH}$ \citep{2010ApJ...721...26G,2011Natur.469..374K,2011MNRAS.412.2211G,2016ApJ...826L..32G}. 

Determining the shape, scatter and extent of the M$_{\rm BH} - \sigma$ relation is important for our understanding of the galaxy evolution \citep[see for a review][]{2013ARA&A..51..511K}, growth of black holes, and their mutual connection. It is also crucial for calibrating the numerical simulations building virtual universes \citep[e.g.][]{2014MNRAS.443..648A, 2014MNRAS.444.1518V,2015MNRAS.446..521S}, and every addition to the relation is still very valuable, especially when one considers that the M$_{\rm BH}$ scaling relations are not representative of the general galaxy population \citep{2007ApJ...662..808L, 2007ApJ...660..267B, 2016MNRAS.460.3119S}. 

In this paper we investigate six early-type galaxies (ETGs) belonging to the low velocity dispersion part of the M$_{\rm BH} - \sigma$ relation, and as a result add four more measurements to the scaling relations. For two additional galaxies we are not able to detect black holes, making them rather curious, but exciting exceptions to the expectation that all ETGs have central black holes. In Section~\ref{s:obs} we define the sample, describe the observations, their quality and data reduction methods. In Section~\ref{s:kin} we present the stellar kinematics used to constrain the dynamical models which are described in Section~\ref{s:dyn}. Section~\ref{s:disc} is devoted to the discussion on the results, in particular, the possible caveats in the construction of dynamical models, the location of our M$_{\rm BH}$ with respect to other galaxies on the M$_{\rm BH} - \sigma$ relations, the internal orbital structure and the conjecture that two galaxies in our sample do not harbour black holes. We summarize our conclusions in Section~\ref{s:con}. Appendices~\ref{app:sauron}, \ref{app:PSF}, \ref{app:mge} and~\ref{app:comparison} present various material supporting the construction and validation of the dynamical models.

%%%%%%%%%%%%%%%%%%%%%%%%%%%%%%%%%%%%%%%%%%%%%%%%%%%%%%%%%%%
%
% SECTION 2 SECTION 2 SECTION 2 SECTION 2 SECTION 2 SECTION 2
%
%%%%%%%%%%%%%%%%%%%%%%%%%%%%%%%%%%%%%%%%%%%%%%%%%%%%%%%%%%%

\section{Sample selection, observations and data reduction}
\label{s:obs}

The core data set of this work was obtained using the Near-Infrared Integral Field Spectrograph (NIFS) on Gemini North Observatory on Hawaii and the SAURON\footnote{SAURON was decommissioned in 2016 and it was transferred to the Mus\`ee des Confluences in Lyon where it is on display.} IFU on the William Herschel Telescope of the Isaac Newton Group on La Palma. In addition to these spectroscopic observations we used images obtained with the Hubble Space Telescope (HST) and the Sloan Digital Sky Survey (SDSS).

\subsection{Sample selection}
\label{ss:sample}

Our goal was to obtain high spatial resolution kinematics of the nuclei of galaxies that have approximately $\sigma_e<130$ km/s, probing the low mass end of the black hole scaling relations. The ATLAS$^{\rm3D}$ sample \citep{2011MNRAS.413..813C} was a unique data base from which to select these galaxies, as it provided the large-scale IFU observations and accurate $\sigma_e$ estimates for a volume-limited sample of ETGs. However, not all low $\sigma_e$ ATLAS$^{\rm3D}$ galaxies could be included in the sample. The highest possible spatial resolution from the ground is achievable with an LGS AO system, which allows high spatial resolution observations of extended targets by relaxing the constraint for having a bright guide star close to the scientific target. However, a natural guide source is still required in order to correct for the so-called tip/tilt (bulk motion of the image) low-order distortion term, and to track the differential focus of the laser source (located in the atmospheric sodium layer at 90 km altitude, whose distance from the telescope changes with zenith distance) and the science target. In order to maximize the correction, the natural guide star should be relatively close to the target, and in the case of the GEMINI AO system Altair \citep{2006SPIE.6272E.114B} it can be as far as 25\arcsec\, from the target and down to 18.5 mag in {\it R} band (in low sky background conditions). The ATLAS$^{\rm3D}$ galaxies were not able to match even such relaxed restrictions. However, as the tip/tilt and focus corrections are also possible using the galaxy nucleus as the natural guide source provided there is a 1.5 mag drop within the central 1\arcsec, our targets were primarily selected to fulfil this requirement. 

Further restrictions were imposed by considering the possible results of the dynamical models. A rule of thumb says that to provide constraints on the mass of the black hole in stellar dynamical models, one should resolve the black hole SoI. In \citet{2009MNRAS.399.1839K} we showed that it is possible to constrain the lower limit for the mass of a black hole even if the SoI is 2 -- 3 times smaller than the nominal spatial resolution of the observations, provided one uses both the large-scale and high spatial resolution IFU data \citep[see also][]{2010AIPC.1240..211C}. This relaxation, nevertheless limits the number of possible galaxies as the SoI also decreases inversely with the distance of the galaxy. Selection based on SoI is only approximately robust as it relies on the choice of M$_{\rm BH} - \sigma$ scaling relation parameters, which are particularly uncertain in the low-mass regime. Our choice was to assume the \citet{2002ApJ...574..740T} relation, limited to galaxies with $\sigma_e<140$ km/s and $r_{\rm SoI}>0.04$\arcsec, yielding 44 galaxies\footnote{Note that the calculation was done on preliminary SAURON data and a number of $\sigma_e$ values were different from the final published in \citet{2013MNRAS.432.1709C} and used in the rest of the paper.}.

Additional restrictions were imposed to select only galaxies with archival HST imaging (in order to generate high resolution stellar mass models) and which showed no evidence of a bar (as bars introduce additional free parameters and degeneracies in the dynamical modelling). The combination of these considerations yielded a sample of 14 galaxies. Through several observing campaigns in 2009 and 2010 we obtained data for six galaxies: NGC\,4339, NGC\,4434, NGC\,4474, NGC\,4578 and NGC\,4762. Table~\ref{t:sample} presents the main properties of target galaxies and observational details. 

%%%%% Table 1. %%%%%%%%%%%%%%%%%%%%%%%%%%%%%%%%%%%%%%%%%%%%%%%%%%%%%
\begin{table}
   \caption{General properties of sample galaxies}
   \label{t:sample}
$$
  \begin{array}{llllllll}
    \hline
    \noalign{\smallskip}

%    \multicolumn{4}{c|}{$NIFS$ } &  \multicolumn{3}{c}{$HST$}\\ 
%   \hline
%    \noalign{\smallskip}
    $Galaxy$ & $R$_{e}  & \sigma_{e} & $D$ & $M$_K & \log ($M$_{bulge})& $i$ & Virgo\\
               & " & (km/s)          & (Mpc)        & $K-band$ & (\log(M_\odot))& (deg) &\\
    (1) & (2) & (3) & (4) & (5) & (6) & (7) & (8)\\
    \noalign{\smallskip} \hline \hline \noalign{\smallskip}
    $NGC$\,4339 & 29.9 & 95 & 16.0& -22.49 & 10.03&30  & 1\\ 
    $NGC$\,4434 & 14.6 & 98 & 22.4 & -22.55 & 9.86&45  & 0\\ 
    $NGC$\,4474 & 21.4 & 85 & 15.6 & -22.28 & 9.49&89 & 1\\ 
    $NGC$\,4551 & 20.4 & 94 & 16.1 & -22.18 & 10.00&65 & 1 \\ 
    $NGC$\,4578 & 39.4 &107& 16.3 & -22.66 & 9.92&50 & 1\\ 
    $NGC$\,4762 & 104.2 & 134 & 22.6 & -24.48 & 10.11&89 & 0\\ 

       \noalign{\smallskip}
    \hline
  \end{array}
$$ 
{Notes:  Column 1: galaxy name; Column 2: effective (half-light) radius in arcsec; Column 3: velocity dispersion within the effective radius; Column 4: distance to the galaxy; Column 5: 2MASS {\it K}-band magnitude from \citet{2000AJ....119.2498J}; Column 6: Bulge mass, obtained by multiplying the total dynamical mass from \citet{2013MNRAS.432.1862C} and bulge-to-total ratio from \citet{2013MNRAS.432.1768K}; Column 7: assumed inclination. Column 8: Virgo membership. Distances are taken from \citet{2011MNRAS.413..813C}, Virgo membership from \citet{2011MNRAS.416.1680C}, while all other properties are from \citet{2013MNRAS.432.1709C}. 
}
\end{table}
%%%%%%%%%%%%%%%%%%%%%%%%%%%%%%%%%%%%%%%%%%%%%%%%%%%%%%%%%%%%%%%%%%

\subsection{Photometric data}
\label{ss:phot}

Dynamical models depend on detailed parametrization of the stellar light distributions. Specifically, it is important to have high-resolution imaging of the central regions around the supermassive black hole, as well as deeper observations of the large radii. The former are important to describe the stellar potential close to the black hole, while the latter is critical for tracing the total stellar mass. The extent of the large-scale imaging should be such that it traces the vast majority of the stellar mass and we used the SDSS DR7 $r$-band images \citep{2009ApJS..182..543A}, which were already assembled during the ATLAS$^{\rm3D}$ Survey and are presented in \citep{2013MNRAS.432.1894S}. The highest spatial resolution imaging was obtained using the HST archival data, using both the Wide-Field Planetary Camera \citep[WFPC2,][]{1995PASP..107.1065H} and Advanced Camera for Survey \citep[ACS,][]{1998SPIE.3356..234F} cameras in filters most similar to the SDSS r-band (more details can be found in Table~\ref{t:obs}). Both WFPC2 and ACS calibrated data were requested through the ESA/HST Data Archive. The individual WFPC2 CR-SPLIT images were aligned and combined removing the cosmic rays. The ACS data were re-processed through the pyIRAF task {\tt multidrizzle}.

\subsection{Wide-field SAURON spectroscopy}
\label{ss:sauron}

The SAURON data were observed as part of the ATLAS$^{\rm3D}$ Survey presented in \citet{2011MNRAS.413..813C}, where the data reduction and the extraction of kinematics are also described. The velocity maps of our galaxies were already presented in \cite{2011MNRAS.414.2923K}, while the higher order moments of the line-of-sight velocity distribution (LOSVD) are presented here in Appendix~\ref{app:sauron}. Observations of the ATLAS$^{\rm3D}$ Survey were designed such that the SAURON field of view (FoV) covers at least one effective radius. For five of our galaxies a nominal $30\arcsec \times40\arcsec$ SAURON FoV was sufficient, but in one case (NGC4762) the galaxy was covered with an adjacent mosaic of two SAURON footprints, resulting in approximately $75\arcsec \times 30\arcsec$ FoV. We use the SAURON kinematics available online\footnote{The data are available on the public ATLAS$^{\rm 3D}$ site: \href{http://purl.org/atlas3d}{http://purl.org/atlas3d}} directly, with no modifications. More details on the instrument can be found in \citet{2001MNRAS.326...23B}.

%%%%% Table 1. %%%%%%%%%%%%%%%%%%%%%%%%%%%%%%%%%%%%%%%%%%%%%%%%%%%%%
\begin{table*}
   \caption{Summary of observations}
   \label{t:obs}
$$
  \begin{array}{ccccc|ccc}
    \hline
    \noalign{\smallskip}

    \multicolumn{5}{c|}{$NIFS$ } &  \multicolumn{3}{c}{$HST$}\\ 
   \hline
    \noalign{\smallskip}
    $Galaxy$ & PID&$Numb.$  & $Exp.$ & $t$_{exp} & $PID$ & $Instrument$ & $Filter$ \\
                   & &$of exp.$ & $comb.$ & (h)  &  & &  \\
    (1) & (2) & (3) & (4) & (5) & (6) & (7) & (8)\\
    \noalign{\smallskip} \hline \hline \noalign{\smallskip}
    $NGC$\,4339 & $GN-2010A-Q-19$&26/42 & 24 & 7    & 5446 & $WFPC2$ & F606W \\ 
    $NGC$\,4434 & $GN-2009A-Q-54$&25/38 & 23 & 6.3 & 9401 & $ACS$ & F475W \\ 
    $NGC$\,4474 & $GN-2009A-Q-54$&20/33 & 17 & 5.5 & 6357 & $WFPC2$ & F702W \\ 
    $NGC$\,4551 & $GN-2010A-Q-19$&21/33 & 12 & 5.5 & 9401 & $ACS$ & F475W \\ 
    $NGC$\,4578 & $GN-2010A-Q-19$&26/43 & 25 & 7.2 & 5446 & $WFPC2$ & F606W \\ 
    $NGC$\,4762 & $GN-2010A-Q-19$&16/24 & 12 & 4    & 9401 & $ACS$ & F475W \\ 

       \noalign{\smallskip}
    \hline
  \end{array}
$$ 
{Notes: Column 1: galaxy name; Column 2: Gemini Proposal ID number; Column 3: total number of object exposures / total number of science exposures (object and sky); Column 4: total number of exposures merged into the final data cube; Column 5: total exposure time including both object and sky exposures (in hours); Column 6: HST proposal ID number; Column 7: HST Camera; and Column 8: filter.}\looseness=-1
\end{table*}
%%%%%%%%%%%%%%%%%%%%%%%%%%%%%%%%%%%%%%%%%%%%%%%%%%%%%%%%%%%%%%%%%%

\subsection{NIFS LGS AO Spectroscopy}
\label{ss:red}

The small-scale, high spatial resolution IFU data were obtained using the NIFS \citep{2003SPIE.4841.1581M}. All galaxies were observed in the {\it K} band with a {\it H}+{\it K} filter and a spectral resolution of $R \sim 5000$. The observations were done in a O-S-O-O-S-O sequence, where O is an observation of the object (galaxy) and S is an observation of a sky field. NIFS pixels are rectangular ($0.04\arcsec \times 0.103\arcsec$) and we dithered the individual object frames by a non-integer number of pixels in both directions, to provide redundancy against bad detector pixels, and to oversample the point spread function (PSF). Each observation was for 600 s. In addition to galaxies, a set of two telluric stars were observed before and after the science observations, covering A0 V and G2 V types. Their observation followed the same strategy of object and sky interchange as for the science targets. A summary of the NIFS observations are given in Table~\ref{t:obs}. 

The reduction of the NIFS data was identical to that described in \citet{2009MNRAS.399.1839K}, with one exception pertaining to the correction of the heliocentric velocity. We used the templates of the IRAF scripts provided by the GEMINI Observatory\footnote{\href{https://www.gemini.edu/sciops/instruments/nifs/data-format-and-reduction}{https://www.gemini.edu/sciops/instruments/nifs/data-format-and-reduction}}. The initial reduction steps included flat fielding, bad pixel correction, cosmic ray cleaning, sky subtraction, preparation of the Ronchi mask used in the spatial rectification of the data and wavelength calibration using arc lamp exposures. 

As some of the galaxies were observed over a period of a few months, the heliocentric velocities of individual object frames are significantly different. For NGC\,4339 and NGC\,4578, the differences were of the order of 15 and 25 km/s, respectively, while for other galaxies the differences were less than a few km/s. We performed the correction at the stage of the wavelength calibration. To avoid resampling the data multiple times, we corrected for the heliocentric motion while computing the dispersion solution by modifying the list of arc line reference wavelengths using the relativistic Doppler shift formula $\lambda_{new} = \lambda_{\rm old} (1+\beta)$. Where $\beta=v/c$, $v=v_{\rm helio}$, $c$ is the speed of light and $v_{\rm helio}$ is the heliocentric velocity of the Earth at a given science frame.

The telluric features in the spectra were corrected using the observed telluric stars, which were reduced following the standard reduction. As we had two telluric stars bracketing each set of four science observations, we used the star that was closest in airmass to a given science frame. Sometimes this was a A0 V star and at other instances a G2 V star. To remove the intrinsic stellar features from the G2 V observations, we used a high-resolution solar template\footnote{\href{ftp://nsokp.nso.edu/pub/atlas/photatl/}{ftp://nsokp.nso.edu/pub/atlas/photatl/}} \citep{1991aass.book.....L}, while for A0 V stars we used a similarly high resolution model spectrum of Vega\footnote{\href{http://kurucz.harvard.edu/stars.html}{http://kurucz.harvard.edu/stars.html}} \citep{1991ppag.proc...27K}. In both cases, the template was fitted to the observed star using the penalized Pixel Fitting (pPXF) method\footnote{\label{ft:CapSoft}\href{http://purl.org/cappellari/software}{http://purl.org/cappellari/software}} \citep{2004PASP..116..138C,2017MNRAS.466..798C} to match the velocity shift and instrumental broadening, before taking the ratio of the observed and fitted spectra to derive the telluric correction spectrum. The final telluric correction of the object frames was performed within the GEMINI NIFS pipeline using the prepared correction curves. 

As the last step of the data reduction, we merged individual object frames into the final data cube. The merging procedure followed that of \citet{2009MNRAS.399.1839K} and consisted of recentring of all frames to a common centre and merging of all exposures on to a grid that covers the extent of all object frames. The recentring was performed on images reconstructed by summing the data cubes along the wavelength direction. The image which had the highest resolution and most regular surface brightness isophotes in the centre was assumed as a reference, while other images were shifted in $x-$ and $y-$directions until their outer isphotes matched those of the reference image. We rejected a few  data cubes from final merging if they showed elongated or non-regular isophotes, which were evidence that the guiding on the nucleus was not always successful during the 600 s exposures. Except for NGC\,4551 which had $\sim$40 per cent of its frames elongated, this typically resulted in removal of a few data cubes with the poorest seeing (see Table~\ref{t:obs}). The final cubes covered approximately $3\arcsec \times 3\arcsec$ mapped with squared pixels of $0.^{\prime\prime}05$ on the side. These pixels oversample the cross-slice direction and slightly reduce the sampling along the slice, but this is justified given our dither pattern and the final PSF. 

%%%%% Table 1. %%%%%%%%%%%%%%%%%%%%%%%%%%%%%%%%%%%%%%%%%%%%%%%%%%%%%
\begin{table}
   \caption{PSF of NIFS observations}
   \label{t:psf}
$$
  \begin{array}{lllll}
    \hline
    \hline
    \noalign{\smallskip}

        $Galaxy$ & $FWHM$_N & $FWHM$_B & $Int$_N & Strehl \\
                   & (arcsec)        & (arcsec)       & &\\
    (1) & (2) & (3) & (4) & (5) \\
    \noalign{\smallskip} \hline \hline \noalign{\smallskip}
    $NGC$\,4339 & 0.22 \pm0.01 & 0.81 \pm 0.02& 0.42 \pm 0.02& 0.10\\ 
    $NGC$\,4434 & 0.17 \pm0.01& 0.75 \pm 0.02& 0.52 \pm 0.01& 0.17\\ 
    $NGC$\,4474 & 0.15 \pm0.02& 0.90   \pm 0.1& 0.45 \pm 0.01&  0.22\\ 
    $NGC$\,4551 & 0.18 \pm0.01& 0.74 \pm 0.02& 0.51 \pm 0.01&  0.15 \\ 
    $NGC$\,4578 & 0.15 \pm0.03& 0.88 \pm 0.02& 0.55 \pm 0.02&  0.22\\ 
    $NGC$\,4762 & 0.17 \pm0.01& 0.55 \pm 0.03& 0.70 \pm 0.01& 0.17\\ 

       \noalign{\smallskip}
    \hline
  \end{array}
$$ 
{Notes -- Column 1: galaxy name; Column 2: FWHM of the narrow Gaussian component ; Column 3: FWHM of the broad Gaussian component; Column 4: intensity of the broad Gaussian, where the intensity of the broad Gaussian is equal to $1 - $Int$_N$ ; Column 5: an estimate of the Strehl ratio, calculated as the ratio of peak intensity in the narrow Gaussian of the PSF and the peak intensity of the ideal diffraction limited PSF of NIFS using FWHM=$0.07\arcsec$ as the diffraction limit of the Gemini 8m telescope at $2.2\mu$m. Uncertainties are derived as a standard deviation of the results of fits with different initial parameters or set ups (see the text for details). }\looseness=-1
\end{table}
%%%%%%%%%%%%%%%%%%%%%%%%%%%%%%%%%%%%%%%%%%%%%%%%%%%%%%%%%%%%%%%%%%

\subsection{Determination of the point spread function}
\label{ss:psf}

We made use of the HST imaging to determine the point-spread function (PSF) of the NIFS observations. The HST imaging is typically of higher (or comparable) resolution as the LGS AO observations, and has a well known and stable PSF. We used the Multi-Gaussian Expansion (MGE) method \citep{1992A&A...253..366M, 1994A&A...285..723E} to parameterize the HST images and deconvolve the MGE models. We prepared images of the HST PSFs for both the WFPC2 and ACS cameras using the {\tt TinyTim} software \citep{2011SPIE.8127E..16K}, taking into account the position of the centre of the galaxy on the camera chip, the imaging filter and using a K giant spectrum as input. To obtain the PSF of an ACS image, due to the camera off centre position, we followed a more complicated procedure \citep[see also e.g.][]{2013AJ....146...45R}. We first constructed a distorted PSF using the standard setup of {\tt TinyTim}. We substituted the central part ($6\arcsec \times 6\arcsec$) of the ACS image of a target that still needs to be corrected by the {\tt multidrizzle} task, with the distorted PSF image. Then we run {\tt multidrizzle} with the same set up as when preparing the images. This achieved the same distortion correction on the PSF image as it is at the location of the galaxy centre of the ACS image. We then cut out the PSF image and prepared it for the final processing, which involved a parametrization of the PSF image with concentric and circular Gaussians using the MGE software$^{\ref{ft:CapSoft}}$ of \citet{2002MNRAS.333..400C}. The MGE parametrizations of the PSFs are given in Table~\ref{tapp:psf}.%

Deconvolved MGE models of the HST images were compared with the reconstructed NIFS images. The method is the same as in \citet{2006MNRAS.370..559S} and \citet{2009MNRAS.399.1839K} and it consists of convolving the MGE model with a test PSF made of a concentric and circular double Gaussian. The double Gaussian is parameterized with the dispersions of the two components (a narrow and a broad one) and a relative weight. The convolved image is rebinned to the same size as the NIFS image and the parameters of the test PSF are varied until the best-fitting double Gaussian is found. As the fit is strongly degenerate we approached it in different ways: by keeping the centre of the test PSF free or fixed, changing the initial values of the parameters of the test PSF, as well as changing the size of the NIFS map used in the comparison. The difference between the obtained results provide an estimate of the uncertainty of the process.

Comparison between the NIFS light profiles and the convolved MGE models (of the HST images) is shown in Fig.~\ref{fapp:psf}. As the MGE models were oriented as the NIFS images (north up, east left), the profiles are shown along a column and a row cut passing through the centre (not necessarily along the major or minor axes). The agreement is generally good, suggesting that this degenerate process of fitting two Gaussians worked reasonably well. In some cases (e.g. NGC\,4762) there is evidence that the PSF might not be circular at about 5 - 10 per cent level. Assuming a PSF different to that order from our best estimate, would change the black hole by about 20 - 30 per cent (based on a dynamical model such as described in Section~\ref{ss:jam}.), and is fully consistent with typical uncertainties on black hole masses. The final PSF parameters of our merged data cubes are given in Table~\ref{t:psf}. Generally speaking, the narrow component Gaussian are typically below 0.2\arcsec (full width at half-maximum, FWHM), while the broad component Gaussians are between 0.75 and 0.9\arcsec  (FWHM). Strehl ratios, approximated as the ratio between the peak intensity in the normalized narrow-Gaussian component and the expected, diffraction limited Gaussian PSF of NIFS (with FWHM of 0.07\arcsec), are between 10 and 20 per cent. These results confirm the expected improvement in the spatial resolution using the LGS AO and guiding on the galactic nuclei. 

%%%%%%%%%%%%%%%%%%%%%%%%%%%%%%%%%%%%%%%%%%%%%%%%%%%%%%%%%%%
\begin{figure*}
%Fig made by make_single_kinematics_fits_LGS.pro
\includegraphics[width=\textwidth]{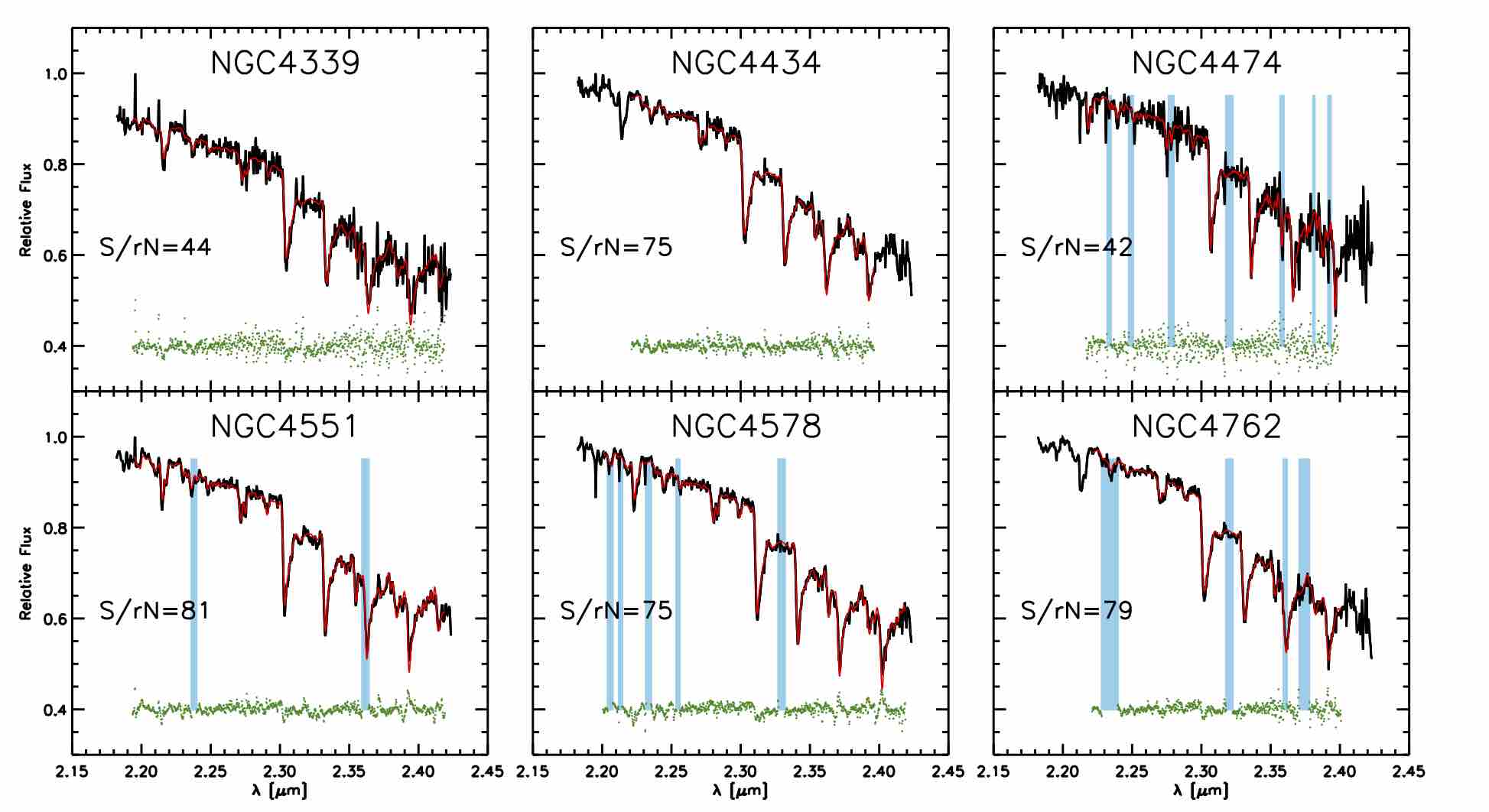}
\caption{Example NIFS spectra of target galaxies obtained by summing the data cubes within a circular aperture of 1\arcsec\, radius. Red line is the pPXF best-fitting model, while the green dots are the residuals between the data and the fit. Shaded light blue area indicate regions that were excluded from the fit. These fits are representative in terms of the wavelength range and the fit parameters of pPXF of fits for each Voronoi-binned spectrum providing the kinematics for each galaxy (Fig.~\ref{f:maps}). S/rN ratios for these spectra are given on each panel. }
\label{f:spec}
\end{figure*}
%%%%%%%%%%%%%%%%%%%%%%%%%%%%%%%%%%%%%%%%%%%%%%%%%%%%%%%%%%%

%%%%%%%%%%%%%%%%%%%%%%%%%%%%%%%%%%%%%%%%%%%%%%%%%%%%%%%%%%%
%
% SECTION 3 SECTION 3 SECTION 3 SECTION 3 SECTION 3 SECTION 3
%
%%%%%%%%%%%%%%%%%%%%%%%%%%%%%%%%%%%%%%%%%%%%%%%%%%%%%%%%%%%

\section{Extraction of stellar kinematics}
\label{s:kin}

\subsection{Stellar kinematics in the near-infrared}
\label{ss:kinNIFS}

Before we determined the stellar kinematics, the NIFS data cubes were spatially binned using the adaptive Voronoi-binning method$^{\ref{ft:CapSoft}}$ of \citet{2003MNRAS.342..345C}. The goal was to ensure that all spectra have a uniform distribution of signal-to-noise ratios (S/N) across the field. The error spectra were not propagated during the reduction, therefore we used an estimate of the noise (eN), obtained as the standard deviation of the difference between the spectrum and its median smoothed version (smoothed over 30 pixels). As this noise determination is only approximate, the targeted S/N level, which is passed to the Voronoi-binning code, should be taken as an approximation of the actual S/N. A measure of the real S/N was estimated a posteriori after the extraction of kinematics, and the binning iteratively improved by changing the target S/N. The choice of the target S/N is driven by the wish to keep the spatial bins as small as possible, especially in the very centre of the NIFS FoV, which directly probes the black hole surrounding, and increasing the quality of the spectra for extraction of kinematics. We finally converged to the typical bin size (in the centre) of $\lesssim 0.1$\arcsec, while at the distance of 1\arcsec\, bins are 0.2-0.3\arcsec\, in diameter.  This was achieved by setting a target S/N of 60 for NGC4339, NGC4551 and NGC4578, while for NGC4474 target S/N was set to 50, and for NGC4434 and NGC4762 to 80.

We extracted the stellar kinematics using the penalised Pixel Fitting (pPXF) method of \citet{2004PASP..116..138C}. The line-of-sight velocity distribution (LOSVD) of stars was parameterized by a Gauss--Hermite polynomials \citep{1993MNRAS.265..213G,1993ApJ...407..525V}, quantifying the mean velocity, $V$, velocity dispersion, $\sigma$, and the asymmetric and symmetric deviations of the LOSVD from a Gaussian, specified with the $h_3$ and $h_4$ Gauss-Hermite moments, respectively. The pPXF software fits a galaxy spectrum by convolving a template spectrum with the corresponding LOSVD, where the template spectrum is derived as a linear combination of spectra from a library of stellar templates. In order to minimize the template mismatch one wishes to use as many as possible stars spanning the range of stellar populations expected in target galaxies. \citet{2009ApJS..185..186W} presented two near-infrared libraries of stars observed with GNIRS and NIFS instruments. We experimented with both, and while they gave consistent kinematics, using the GNIRS templates typically had an effect of reducing the template mismatch manifested in spatially asymmetric features on the maps of even moments ($h_4$) of the LOSVD. A certain level of template mismatch in some galaxies is still visible, as will be discussed below. 

For each galaxy we constructed an optimal template by running the pPXF fit on a global NIFS spectrum (obtained by summing the full cube). Typically 2--5 stars were given non-zero weight from the GNIRS library. This optimal template was then used for fitting the spectra of each individual bin. While running pPXF, we also add a fourth-order additive polynomial and, in some cases, mask regions of spectra contaminated by imperfect sky subtraction or telluric correction. 

In Fig.~\ref{f:spec} we show fits to the global NIFS spectra, summed within a circle of 1\arcsec~radius, as an illustration of the fitting process. The residuals to the fit (shown as green dots), calculated as the difference between the best-fitting pPXF model and the input spectrum, are used in two ways. First, their standard deviation defines a residual noise level (rN). We use this to define the signal-to-residual noise (S/rN), which measures both the quality of the data and the quality of the fit. For each of the global spectra shown in Fig~\ref{f:spec}, the S/rN is higher than the S/eN. This shows only partial reliability of the S/eN and a need to re-iterate the binning process until a right balance between the S/rN and the bin sizes is achieved. Therefore, when the achieved S/rN was too small (i.e $<25$) across a large fraction of the field, we increased the target S/N and rebinned the data until a sufficient S/rN was obtained across the field. 

%%%%% Table 3. %%%%%%%%%%%%%%%%%%%%%%%%%%%%%%%%%%%%%%%%%%%%%%%%%%%%%
\begin{table}
   \caption{Mean kinematic errors for NIFS and SAURON data}
   \label{t:errors}
$$
  \begin{array}{c cccc}
    $NIFS$&$ $& $$&$$&$$\\
   \hline
    $galaxy$ & \Delta V & \Delta \sigma & \Delta h_3 & \Delta h_4 \\
                   & km/s        & km/s       & &   \\
    (1) & (2) & (3) & (4) & (5) \\
    \noalign{\smallskip} \hline \hline \noalign{\smallskip}
    $NGC$\,4339 & 6 & 8 & 0.04 & 0.04\\ 
    $NGC$\,4434 & 4 & 6 & 0.03 & 0.03\\ 
    $NGC$\,4474 & 7 & 8 & 0.07 &  0.07\\ 
    $NGC$\,4551 & 5 & 7 & 0.04 &  0.05\\ 
    $NGC$\,4578 & 4 & 5 & 0.03 &  0.04\\ 
    $NGC$\,4762 & 4 & 4 & 0.03 & 0.04\\ 
		\hline
   \noalign{\smallskip}	
   \noalign{\smallskip}	
   \noalign{\smallskip}	
    $SAURON$&$ $& $$&$$&$$\\
   \hline
    $galaxy$ & \Delta V & \Delta \sigma & \Delta h_3 & \Delta h_4 \\
                   & km/s        & km/s       & &   \\
    (1) & (2) & (3) & (4) & (5) \\
    \noalign{\smallskip} \hline \hline \noalign{\smallskip}
    $NGC$\,4339 & 8 & 12 & 0.07 & 0.07\\ 
    $NGC$\,4434 & 9 & 11 & 0.08 & 0.08\\ 
    $NGC$\,4474 & 8 & 10 & 0.08 &  0.07\\ 
    $NGC$\,4551 & 7 & 9 & 0.07 &  0.06\\ 
    $NGC$\,4578 & 7 & 9 & 0.07 &  0.06\\ 
    $NGC$\,4762 & 7 & 9 & 0.07 & 0.06\\ 
       \noalign{\smallskip}
    \hline
  \end{array}
$$ 
{Notes -- Column 1: galaxy name; Column 2: the mean error in the velocity ; Column 3: the mean error in the velocity dispersion; Column 4: the mean error in Gauss-Hermite coefficient $h_3$; Column 5: the mean error in the Gauss-Hermite coefficient $h_4$.}\looseness=-1
\end{table}
%%%%%%%%%%%%%%%%%%%%%%%%%%%%%%%%%%%%%%%%%%%%%%%%%%%%%%%%%%%%%%%%%%

The second use of the residuals to the fit is to estimate the errors to kinematics parameters. This is done by means of Monte Carlo simulations where each spectrum has an added perturbation consistent with the random noise of amplitude set by the standard deviation of the residuals (rN).  Errors on $V$, $\sigma$, $h_3$ and $h_4$ were calculated as the standard deviation of 500 realization for each bin. Kinematic errors are similar between galaxies and spatially closely follow the S/rN distribution. The mean errors for each galaxy are given in Table~\ref{t:errors}.

%%%%%%%%%%%%%%%%%%%%%%%%%%%%%%%%%%%%%%%%%%%%%%%%%%%%%%%%%%%
\begin{figure*}
%Fig made by make_plots_all_kin_maps_LGS_NIFS
\includegraphics[width=\textwidth]{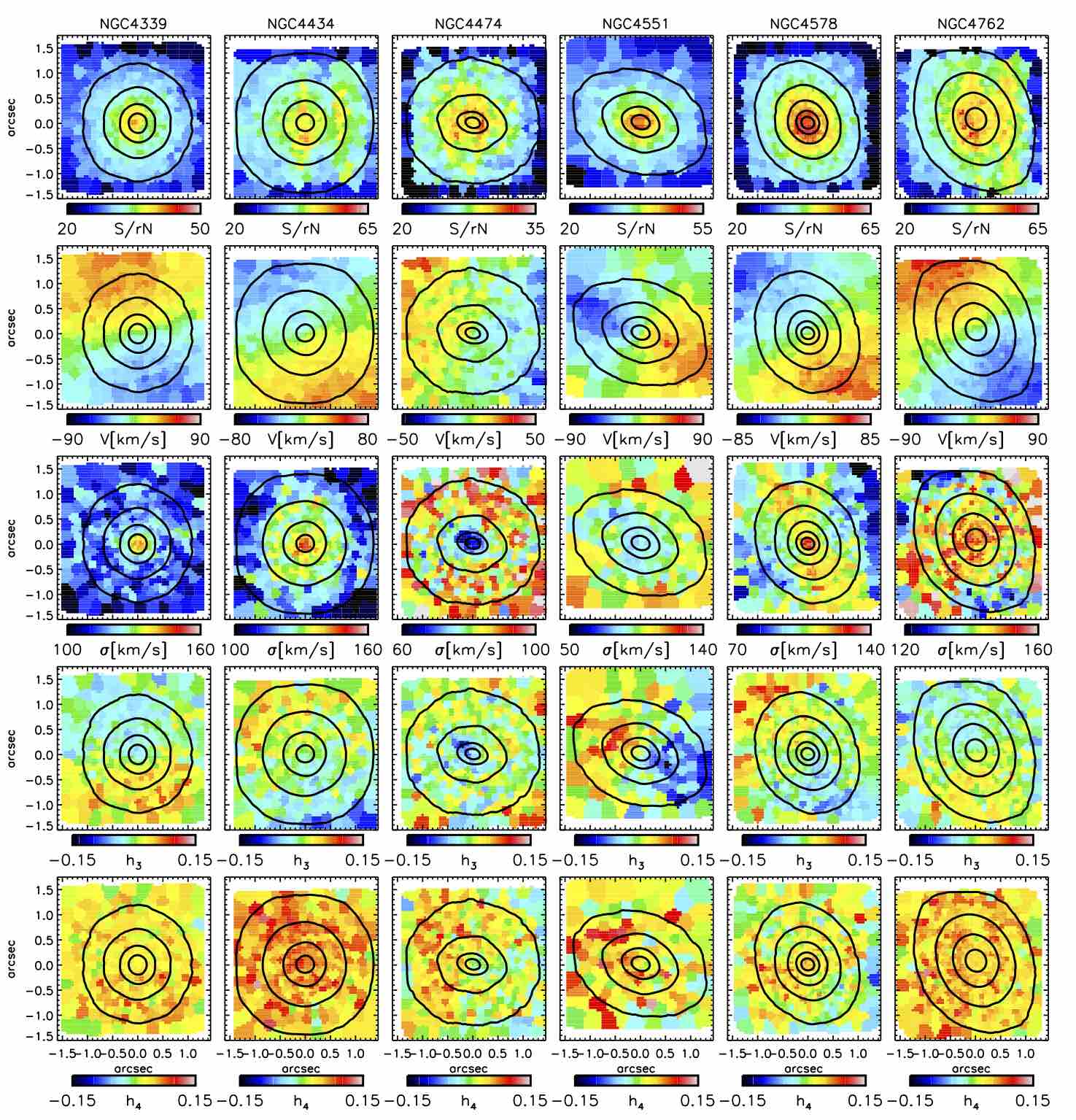}
\caption{NIFS kinematics for sample galaxies (from left to right) NGC\,4339, NGC\,4434, NGC\,4474, NGC\,4551, NGC\,4578 and NGC4762. From top to bottom each panel shows maps of: S/rN, the mean velocity (with the systemic velocity subtracted), the velocity dispersion, and the Gauss--Hermite coefficients $h_3$ and $h_4$. A median value of $-0.054$ was subtracted from the $h_3$ for NGC\,4474. Colour bars indicate the range of scales shown on maps. Black contours are isophotes, shown in steps of half a magnitude. North is up and east to the left.  }
\label{f:maps}
\end{figure*}
%%%%%%%%%%%%%%%%%%%%%%%%%%%%%%%%%%%%%%%%%%%%%%%%%%%%%%%%%%%

Fig.~\ref{f:maps} presents the NIFS kinematics of our sample galaxies, as well as the achieved S/rN across the NIFS field. The lowest S/rN are obtained in NGC\,4474, NGC\,4339 and NGC\,4551. In NGC\,4474 the S/rN$>$25 is achieved for bins within the central 1\arcsec, while NGC\,4339 and NGC\,4551 have S/rN$>$30 within the same region, but steeply rising to 40 and 50, respectively. The kinematics follow the properties seen on the SAURON large-scale kinematic maps (see Figs.~\ref{fapp:sauron1} and~\ref{fapp:sauron2} and Section~\ref{ss:kinSAU}): galaxies show regular rotation, a velocity dispersion peak in the centre, anticorrelated $V$ and $h_3$ maps and typically flat and positive $h_4$ maps. The high-resolution data, however, present additional features for two galaxies: NGC\,4474 and NGC\,4551. In both cases, the velocity dispersion maps show a significant decrease in the centre ($\sim20$ km/s), where the spatial extent of the feature in NGC\,4474 is about half the size of the one in NGC\,4551 (the galaxies are at similar distances). The structures are within the region of highest S/rN on the maps. For NGC\,4474 the typical S/rN is, however, only 30. Nevertheless, at that S/rN, the velocity and velocity dispersion are robustly recovered. We confirmed this by extracting kinematics assuming only a Gaussian LOSVD, as well as extracting kinematics using larger spatial bins and increasing the S/rN. The kinematic components seen in NGC\,4474 and NGC\,4551 could be associated with dynamically cold structures (e.g. nuclear discs) or could indicate the lack of black holes. Regardless of the origin, they have a profound influence on the determination of the M$_{\rm BH}$ in these galaxies, as will be discussed in Section~\ref{ss:nondetect}. The velocity dispersion maps of NGC\,4578 and NGC\,4762 are also somewhat unusual, but consistent with the SAURON observations. NGC\,4578 shows an elongated structure along the major axis, while the NGC\,4762 velocity dispersion map is dominated by an extension along the minor axis. 

Aforementioned template mismatch-like features are traced in $h_3$ and $h_4$ maps of NGC4474 and, to a lesser degree, in the $h_4$ map of NGC4551. The $h_4$ maps are not symmetric, as they should be for an even moment of the LOSVD. Similarly, the $h_3$ map of NGC4474 does not show the expected anticorrelation with the velocity map. In order to improve on the high-order moments, we explored a range of pPXF parameters while fitting the spectra, as well as used various combinations of template libraries and extracted the kinematics to an even higher Gauss--Hermite order, but these tests did not improve the fits. In the case of NGC\,4474, the most likely reason for the unusual $h_3$ and $h_4$ maps is a combination of the low S/N of the spectra (only about 30), the low inclination, which is likely responsible for the low-level rotation, and therefore an expected low level of anticorrelation between $V$ and $h_3$, and a possible template mismatch. The later is supported also by the test where we forced a high target S/N while binning, which results in a uniform S/rN$\sim45$ across the field, and bin sizes of approximately 0.3--0.4\arcsec\, in diameter. The kinematics extracted from these spectra have the same features as the kinematics presented in Fig.~\ref{f:maps}: the dip in velocity dispersion, uniform $h_3$ and a non-symmetric $h_4$. We conclude that the higher order LOSVD moments of NGC\,4474 are likely not reliable, which should be kept in mind while interpreting the results, but we use the presented kinematics. 

\begin{figure}
%Fig made by make_plots_all_kin_maps_LGS_NIFS
\includegraphics[width=\columnwidth]{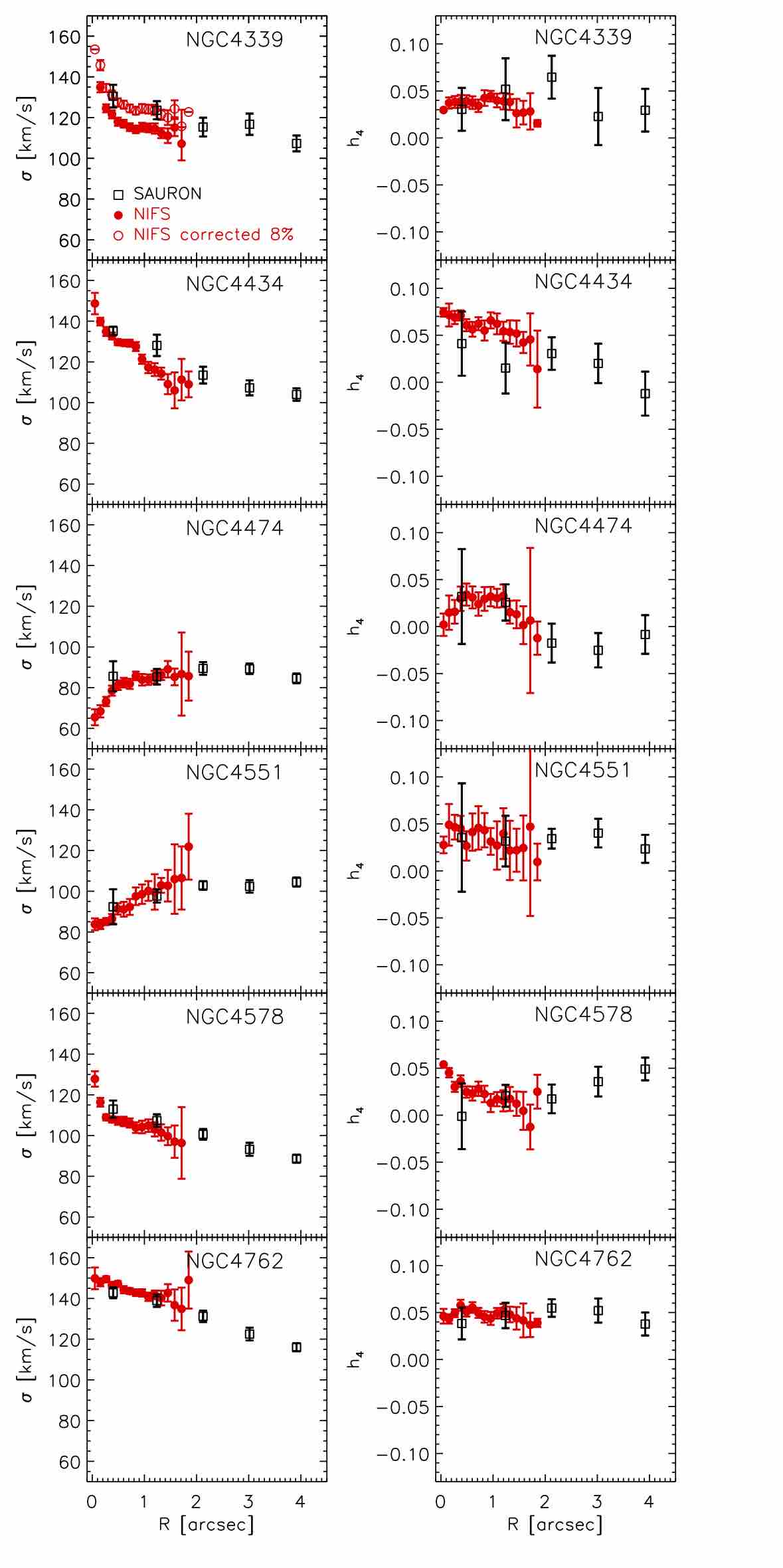}
\caption{Comparison of radial profiles of the velocity dispersion (left hand column) and $h_4$ (right hand column) between the NIFS (red circles) and SAURON (open squares) data. The radial profiles were obtained by averaging using the bisquare weight of the values within concentric circular rings. The error bars show the half-width of the 95 per cent confidence interval around the sample mean. For NGC\,4339 the NIFS velocity dispersion is lower by about 8 per cent, while for all other galaxies the agreement between the optical and NIR kinematics is remarkable. }
\label{f:SAU_NIFS}
\end{figure}

In Fig.~\ref{f:SAU_NIFS} we compare the radial profiles of the velocity dispersion and $h_4$ of SAURON and NIFS kinematics. As is evident, the two kinematic data sets are well matched, with some small deviations of the NIFS kinematics. These are noticeable only for the velocity dispersion profiles of NGC4339, which are about 8 per cent lower than those measured with SAURON. In cases of NGC\,4434 and NGC\,4478 there is a potential offset of less than 5 per cent, but this is within the dispersion of the data points and we do not consider it significant. The NGC\,4474 velocity dispersion and $h_4$ compare well with the SAURON data in the overlap region, ensuring at least that the data sets are consistent, if not fully reliable. The influence of the offset for NGC\,4339 on the determination of the M$_{\rm BH}$ will be discussed later in Section~\ref{s:dyn}, but our general conclusion is that the two sets of kinematics compare well and can be used as they are.  

%%%%%%%%%%%%%%%%%%%%%%%%%%%%%%%%%%%%%%%%%%%%%%%%%%%%%%%%%%%
%%%%%%%%%%%%%%%%%%%%%%%%%%%%%%%%%%%%%%%%%%%%%%%%%%%%%%%%%%%

\subsection{SAURON stellar kinematics}
\label{ss:kinSAU}

Observations, data reduction and the extraction of stellar kinematics for the ATLAS$^{\rm 3D}$ Survey is described in detail in \citet{2011MNRAS.413..813C}, and here we only briefly repeat the important steps of the extraction of stellar kinematics\footnote{Available from \href{http://purl.org/atlas3d}{http://purl.org/atlas3d}}. The SAURON data were spatially binned using the adaptive Voroni binning method of \citet{2003MNRAS.342..345C} using a target S/N of 40. The stellar kinematics were extracted using pPXF \citep{2004PASP..116..138C} employing as stellar templates the stars from the MILES library \citep{2006MNRAS.371..703S}. The SAURON kinematic maps are presented in Figs.~\ref{fapp:sauron1} and \ref{fapp:sauron2}. The errors were estimated using a Monte Carlo simulation, and the mean values are given in Table~\ref{t:errors} for comparison with the NIFS data.

As discussed in detail in \citet{2004PASP..116..138C}, once the galaxy velocity dispersion falls below the instrumental velocity dispersion ($\sigma_{\rm inst}$), the extraction of the full LOSVD becomes an unconstrained problem. For SAURON data, $\sigma_{\rm inst}=98$ km/s, and spectra with an intrinsic $\sigma <\sigma_{\rm inst}$ will essentially not have reliable measurements of the $h_3$ and $h_4$ moments. The pPXF penalizes them towards zero to keep the noise in $V$ and $\sigma$ under control. At larger radii covered by SAURON FoV, all our galaxies fall within this case, which is visible on the maps of $h_3$ and $h_4$ in Figs.~\ref{fapp:sauron1} and~\ref{fapp:sauron2}. Even within the central 3\arcsec$\times3$\arcsec\, this is true for NGC\,4474, NGC\,4551 and partially for NGC\,4578. This problem does not arise for NIFS data as the instrumental resolution is about 30km/s. This means that large-scale SAURON $h_3$ and $h_4$ values for our galaxies are at least partially unconstrained. The comparison of the radial profiles in Fig.~\ref{f:SAU_NIFS} suggests that the SAURON data, at least within the central regions, crucial for the recovery of the central black hole mass, are acceptable. Still, as the full LOSVD is necessary to constrain the construction of orbit-based dynamical models employed in this paper, the results of this modelling should be verified in an independent way. This can be achieved with dynamical models that use only the first two moments of LOSVD ($V$ and $\sigma$), specifically their combination $V_{\rm rms}=\sqrt{V^2 + \sigma^2}$. Therefore we also extracted the mean velocity and the velocity dispersion parameterizing the LOSVD in the pPXF with a Gaussian, for both NIFS and SAURON data. The $V$ and $\sigma$ extracted in such way are fully consistent with those presented in Figs.~\ref{f:maps},~\ref{fapp:sauron1} and~\ref{fapp:sauron2}. The uncertainties were calculated using the Monte Carlo simulation as before, but with penalization switched off. In this way, even if the LOSVDs are penalized their uncertainties carry the full information on the possible non-Gaussian shapes.

%%%%%%%%%%%%%%%%%%%%%%%%%%%%%%%%%%%%%%%%%%%%%%%%%%%%%%%%%%%
%
% SECTION 4 SECTION 4 SECTION 4 SECTION 4 SECTION 4 SECTION 4
%
%%%%%%%%%%%%%%%%%%%%%%%%%%%%%%%%%%%%%%%%%%%%%%%%%%%%%%%%%%%

\section{Dynamical models}
\label{s:dyn}

\subsection{Methods}
\label{ss:methods}

The current method of choice for determining M$_{\rm BH}$ is an extension of the \citet{1979ApJ...232..236S} method, which builds a galaxy by a superposition of representative orbits in a potential of a given symmetry. In axisymmetric models, the orbits are specified by three integrals of motion: energy $E$, the component of the angular momentum vector along the symmetry axis $L_z$, and the analytically unspecified third integral $I_3$. This method was further developed by a number of groups to be applied on axisymmetric galaxies when both photometric (the distribution of mass) and kinematics (the LOSVD) constraints are used \citep{1988ApJ...327...82R,1997ApJ...488..702R,1998ApJ...493..613V,1999ApJS..124..383C, 2003ApJ...583...92G, 2004ApJ...602...66V, 2004MNRAS.353..391T}, using IFU data \citep{2002MNRAS.335..517V,  2006MNRAS.366.1126C}, as well as extended to a more general triaxial geometry \citep{2008MNRAS.385..647V}. 

Both the strengths and the weaknesses of the Schwarzschild method lie in its generality. Earlier papers pointed out possible issues with black hole mass determinations \citep{2004ApJ...602...66V, 2004MNRAS.347L..31C}, but detailed stellar dynamical models of the two benchmark galaxies with the most reliable independent M$_{\rm BH}$ estimates NGC4258 \citep{2009ApJ...693..946S,2015MNRAS.450..128D} and the Milky Way \citep{2014A&A...570A...2F,2017MNRAS.466.4040F}, using both anisotropic Jeans \citep{2008MNRAS.390...71C} and Schwarzschild's models, demonstrated that, in practice, both methods can recover consistent and reliable masses. The main source of error are systematics in the determination of the stellar mass distribution within the black hole SoI, which is generally not included in the error budget.

The extent of the kinematic data used to constrain Schwarzschild models is also of high importance \citep{2005MNRAS.357.1113K}. Outside the regions covered by, for example, a few long slits, the Schwarzschild method, due to its generality, is a poor predictor of stellar kinematics \citep{2005CQGra..22S.347C}. The IFUs have helped decrease this problem, but to robustly recover M$_{\rm BH}$ one still needs to cover at least the area within a half-light radius of the galaxy \citep{2005MNRAS.357.1113K}, but also map the stellar LOSVDs in the vicinity of the black hole \citep{2009MNRAS.399.1839K}. It is also important to allow for sufficient freedom in the models, for the shape of the total mass density to properly describe the true one, within the region where kinematics is fitted. This implies that, if one includes in the models kinematics at large radii (i.e. $>2$R$_e$), where dark matter is expected to significantly affect the mass profile, one should explicitly model its contribution, to avoid possible biases in the black hole masses \citep{2009ApJ...700.1690G,2011ApJ...729...21S,2013AJ....146...45R}. Finally, the recovery of the intrinsic shape of the galaxy is only possible for specific cases \citep{2009MNRAS.398.1117V}, as the Schwarzschild method, even when constrained by large-scale IFU data, suffers from the degeneracy in recovery of the inclination \citep{2005MNRAS.357.1113K}. As shown by \citet{2009MNRAS.398.1117V}, while it is possible to determine whether the potential has an axial or triaxial symmetry, only the lower limit to the inclination of an axisymmetric potential imposed by photometry is constrained \citep[one should also keep in mind the mathematical non-uniqueness of the photometric deprojection, e.g.][]{1987IAUS..127..397R}. Similarly, the viewing angles of a triaxial system can be determined only if there are strong features in the kinematic maps such as kinematically distinct cores. 

%%%%%%%%%%%%%%%%%%%%%%%%%%%%%%%%%%%%%%%%%%%%%%%%%%%%%%%%%%%
\begin{figure*}
%Fig made by mge_overplot_hst.pro
\includegraphics[width=\textwidth]{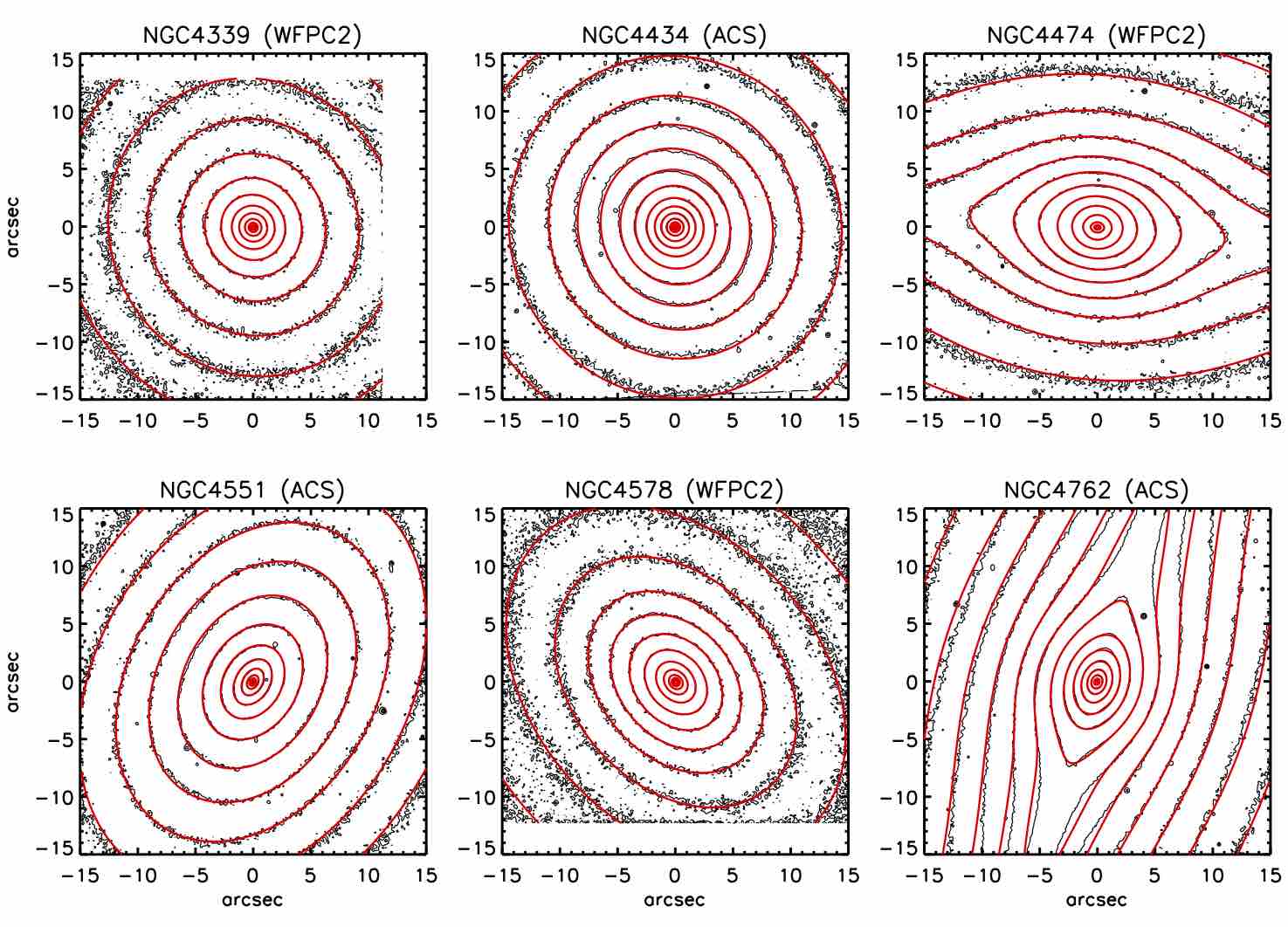}
\caption{MGE models (red smooth contours) are plotted over the HST ACS and WFPC2 imaging of our galaxies. The WFPC2 and ACS images are binned ($3\times3$ and $2\times2$ pixels, respectively) to reduce the noise for the comparison purposes only. HST imaging was used only within the central 10 \arcsec, while the SDSS images (not shown) were used to constrain the MGE models at large radii. Flux levels are normalized to the central brightnesses of the HST images and the contours are spaced by 0.5 magnitudes.}
\label{f:mge}
\end{figure*}
%%%%%%%%%%%%%%%%%%%%%%%%%%%%%%%%%%%%%%%%%%%%%%%%%%%%%%%%%%%

An alternative, less general but consequently less degenerate, is to solve the Jeans equations. The standard approach consists of assuming a distribution function which depends only on the two classic integrals of motion ($E$, $L_z$) \citep{1922MNRAS..82..132J}. In this case the velocity ellipsoid is semi-isotropic: $\sigma_z^2 = \sigma_R^2$ and $\overline{v_Rv_z} = 0$, where $\overline{v_z^2}=\sigma_z^2$ and $\overline{v_R^2}=\sigma_R^2$ are the velocity dispersions along the cylindrical coordinates $R$ and $z$ \citep[e.g.][]{1998AJ....115.2285M}. Allowing for the anisotropy of the velocity ellipsoid introduces two additional unknowns: the orientation and the shape of the velocity ellipsoid. One approach to introduce the anisotropy is based on an empirical finding that the velocity ellipsoid is flattened in the $z$-direction (the symmetry axis) and to first order oriented along the cylindrical coordinates \citep{2007MNRAS.379..418C}\footnote{\citet{2007MNRAS.379..418C} work was based on Schwarzschild models of 24 galaxies. JAM models were in the mean time successfully applied on the 260 galaxies of the ATLAS$^{3D}$ Survey confirming that the assumptions built in the JAM models are adequate for ETGs.}. Jeans anisotropic modelling \citep[JAM;][]{2008MNRAS.390...71C} follows an approach where the velocity anisotropy is introduced as $\beta=1-\sigma_z^2/\sigma_R^2$, defining the shape of the velocity ellipsoid, oriented along the cylindrical coordinates. Characterizing the surface brightness in detail leaves four unknowns that have to be constrained by the IFU kinematics: mass-to-light ratio ($M/L$), inclination $i$, anisotropy $\beta$, and, if the data support it, mass of the black hole, M$_{\rm BH}$. This assumption on the velocity ellipsoid, while not exactly valid away from the equatorial plane or far from the minor axis, seems to work remarkably well on real galaxies \citep{2013MNRAS.432.1709C}, even allowing for a determination of the inclination \citep{2008MNRAS.390...71C}, at least for fast rotators \citep[see for a review section 3.4 of][]{2016ARA&A..54..597C}, as well as oblate galaxies in numerical simulations \citep{2012MNRAS.424.1495L,2016MNRAS.455.3680L}. Recently, a major comparison between Schwarzschild and JAM modelling \citep{doi:10.1093/mnras/sty288}, for a sample of 54 S0--Sd galaxies with integral-field kinematics from the EDGE-CALIFA survey \citep{2017ApJ...846..159B}, found that the two methods recover fully consistent mass density profiles.

A further difference between the orbit and Jeans equation-based modelling is that the latter is constructed such that it is constrained by the second velocity moment only, without the need for the higher parametrization of the LOSVD. The second velocity moment can be approximated by the combination of the observed mean velocity and the velocity dispersion, $V_{\rm rms} =  \sqrt{V^2 + \sigma^2}$ \citep{2008MNRAS.390...71C}, simplifying the requirements on the data quality. For these reasons, we will use both modelling approaches in determining M$_{\rm BH}$ of our targets. We will fit the NIFS data only with JAM models and then both the NIFS and SAURON data with Schwarzschild models. 

We note that in a number of other studies where the two methods were compared in detail \citep{2010AIPC.1240..211C,2014Natur.513..398S, 2015MNRAS.450..128D, 2017MNRAS.466.4040F, 2017A&A...597A..18T}, black hole masses from JAM and Schwarzschild modelling were found to agree well. NGC4258 and the Milky Way deserve a special attention as their M$_{\rm BH}$ are the most secure and based on methods different from those discussed here. In the case of NGC\,4258, the \citet{2009ApJ...693..946S} result is within 15 per cent of the maser M$_{\rm BH}$, while the \citet{2015MNRAS.450..128D} result is within about 25 per cent. The difference between Siopis et al. and Drehmer et al. black hole masses are consistent at $3\sigma$ level. In the case of the Milky Way, \citet{2017MNRAS.466.4040F} modelled the black hole with both Schwarzschild and JAM methods and presented results that are consistent within $1\sigma$ level.

These results are fully consistent with tests between Schwarzschild methods based on the same data. Such studies are regrettably rare, but the most recent were done for two galaxies: M32 \citep{2002MNRAS.335..517V,2010MNRAS.401.1770V} and NGC\,3379 \citep{2006MNRAS.370..559S,2010MNRAS.401.1770V}. In the case of M32 the results are consistent at $1\sigma$ confidence level, while M$_{\rm BH}$ estimates for NGC\,3379 are within $3\sigma$ confidence level, but differ for more than a factor of 2.

In this work, we also add NGC\,1277, for which we show in Appendix~\ref{app:n1277} that JAM can provide results consistent with the Schwazschild models. This last example demonstrates the usefulness of applying independent approaches to the same data, as we do here, to increase the confidence in our results. Both methods can potentially produce results of limited fidelity. In case of the more general Schwarzschild models the numerical noise, as well as the issues discussed above, can limit the quality of the data, as much as the lack of generality and possible degeneracies \citep[i.e. mass -- anisotropy, but see][]{1997MNRAS.288..618G} are limiting JAM models. The agreement between these different methods provides a certain level of security in the robustness of the results. A disagreement in the modelling results, however, would be inconclusive as to which solution is more trustworthy beyond the statement that JAM models lack generality.

Note that we do not include dark matter in any of our models, and we postpone the discussion on possible consequences to Sections~\ref{ss:jam} and~\ref{ss:sys}.

%%%%%%%%%%%%%%%%%%%%%%%%%%%%%%%%%%%%%%%%%%%%%%%%%%%%%%%%%%%
\begin{figure*}
%Fig made by make_jam_grids_nifs.pro
\includegraphics[width=\textwidth]{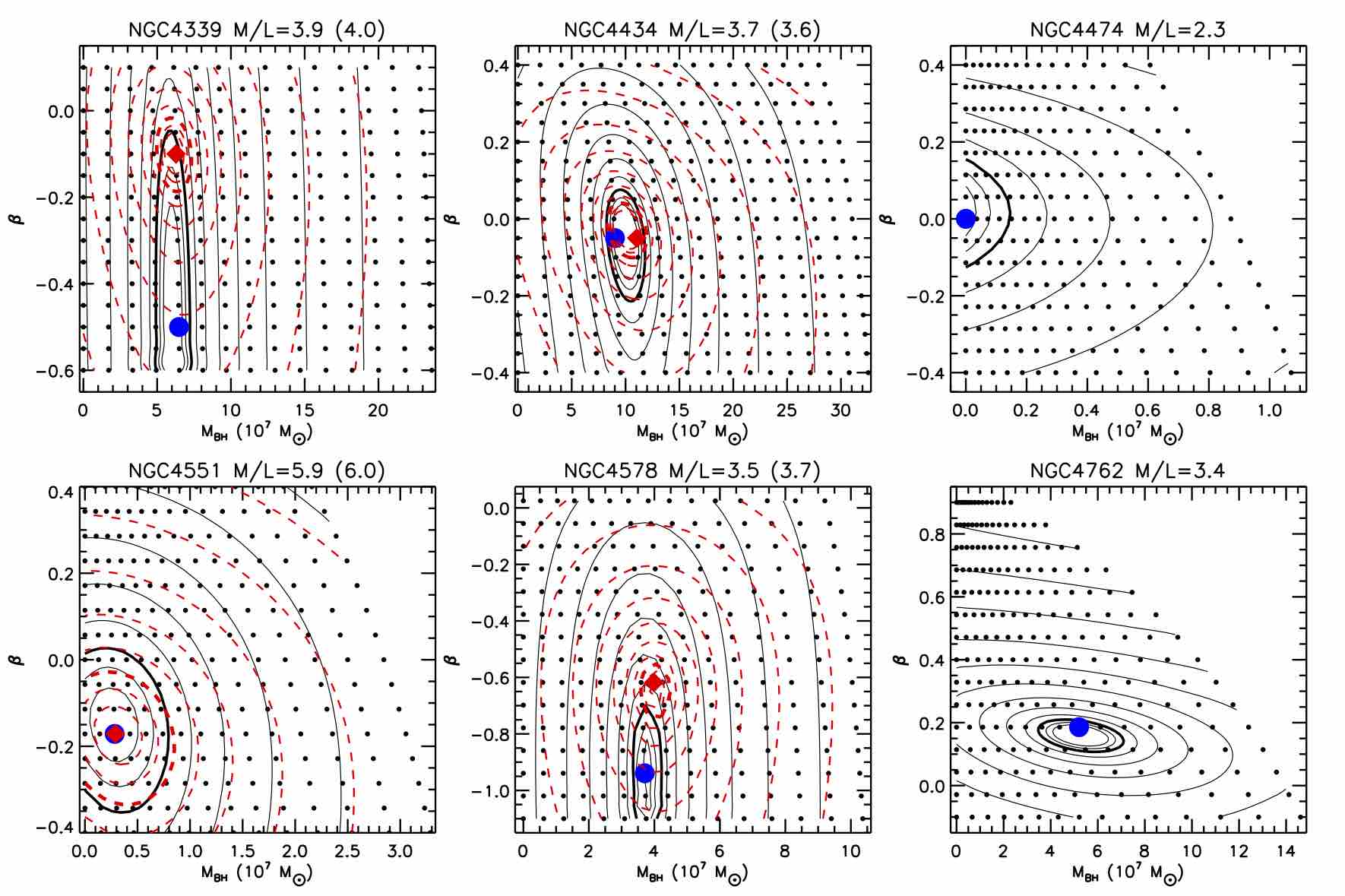}
\caption{Grids of JAM models (small round symbols) for different black hole mass and velocity anisotropies. Contours show the distribution of $\Delta \chi^2 = \chi^2- \chi^2_{min}$, where the continuous (black) contours are for models at the best-fitting inclination given in Table~\ref{t:sample}, while the dashed contours (red) are for edge-on models at the inclination of 89\degree\, (the best-fitting inclinations for NGC\,4474 and NGC\,4762 are already at 89\degree). The distribution of $\Delta \chi^2$ is fitted with a minimum curvature surface for presentation purposes. The thick contours indicate the $3\sigma$ confidence level for two degrees of freedom. The best-fitting model is shown with a large blue circle and the best-fitting model at $i=89$\degree is shown with a large red diamond. $M/L$ ratios of these best-fitting models are given on top of each panel, with values for the edge-on cases given in parenthesis.}
\label{f:jams}
\end{figure*}
%%%%%%%%%%%%%%%%%%%%%%%%%%%%%%%%%%%%%%%%%%%%%%%%%%%%%%%%%%%

\subsection{Mass models}
\label{ss:mass}

The first step in the construction of dynamical models is a detailed parametrization of the surface brightness distribution. Our approach is to use the MGE method \citep{1994A&A...285..723E} and the fitting method and software of \citet[][see footnote \ref{ft:CapSoft} for the software]{2002MNRAS.333..400C}. We used both the HST imaging and SDSS data, as they were presented in \citet{2013MNRAS.432.1894S}. The SDSS images were in the $r$ band while the HST images were obtained with two instruments (WFPC2 and ACS), and we selected those filters that provided the closest match to the SDSS images (see Table~\ref{t:obs}). We fitted the MGE to both images simultaneously, fixing the centres, ellipticities and the position angles of the Gaussian components, and scaling the outer SDSS light profiles to the inner HST profiles by ensuring that the outer parts of the HST profiles smoothly join with the SDSS data. In this way the HST images provide the reference for the photometric calibration. 

When moving to physical units, we followed the WFPC2 Photometry Cookbook\footnote{\href{http://www.stsci.edu/hst/wfpc2/analysis/wfpc2_cookbook.html}{http://www.stsci.edu/hst/wfpc2/analysis/wfpc2$\_$cookbook.html}} and converted from the STMAG to Johnson R band (Vega mag), assuming M$_{\rm R}$=4.41 for the absolute magnitude of the Sun \citep{2007AJ....133..734B}, and a colour term of 0.69 mag (for a K0V stellar type). For ACS images we followed the standard conversion to AB magnitude system using the zero-points from \citet{2005PASP..117.1049S} and assuming a M$_{\rm F450W}=$5.22 mag for the absolute magnitude of the Sun\footnote{\href{http://www.ucolick.org/$\sim$cnaw/sun.html}{http://www.ucolick.org/$\sim$cnaw/sun.html}}. For all galaxies we accounted for the galactic extinction \citep{2011ApJ...737..103S}. We list the parameters of the MGE models in Table~\ref{tapp:mge} and in Fig.~\ref{f:mge} we show the comparison between the MGE models and the HST data. 

Fig.~\ref{f:mge} shows that MGE models reproduce well the central regions of our galaxies, except partially the disc in NGC\,4762, where the largest deviations are less than 10 per cent. Reproducing this transition between the bulge and the disc along the major axis would require negative Gaussians. As these are not accepted by our modelling techniques, we do not attempt to improve the fit in this way. As we show later in the dynamical models this does not have an impact on the results. Overall, the surface brightness distributions of our galaxies are consistent with axisymmetry, showing no evidence of changes in the photometric or kinematic position angles with radius).

\subsection{JAMs}
\label{ss:jam}

As our galaxies are part of the ATLAS$^{\rm 3D}$ sample, they were already modelled with JAM in \citet{2013MNRAS.432.1709C}. These models were constrained by the SAURON kinematics only and used SDSS images for parametrization of light. The models also included various parametrization of the dark matter haloes. Alternative JAM models of ATLAS$^{\rm 3D}$ galaxies, with no direct parametrization of the dark matter, but instead fitting for the total mass, were also presented in \citet{2017MNRAS.467.1397P}. These previous works explored global parameters of our galaxies, including the dark matter fraction and the inclination, assuming axisymmetry. We build slightly different JAM models, constrained with only the NIFS kinematics, and MGE models fitted to the combined HST and SDSS imaging data (see Section~\ref{ss:mass}). We assume the inclination given by models from \citet{2013MNRAS.432.1709C}, listed in Table~\ref{t:obs}. To constrain the JAM models we use the second velocity moment, as described in Section~\ref{ss:kinSAU}. Unlike for the Schwarzschild models, in the case of the JAM models, there is no need for large scale kinematics to constrain the fraction of stars on radial orbits. This is because the kinematics of the whole model is already uniquely defined by the adopted model parameters. For this reason, the best estimates of black hole masses are obtained when fitting the kinematics over the smallest field that is sufficient to uniquely constrain the anisotropy, M$_{\rm BH}$ and $M/L$ \citep[e.g.][]{2015MNRAS.450..128D}. In this way one minimizes the possible biases in the JAM models caused by spatial variations in anisotropy or $M/L$ in the galaxy, without the need to actually allow for these parameters to vary in the models.

%%%%%%%%%%%%%%%%%%%%%%%%%%%%%%%%%%%%%%%%%%%%%%%%%%%%%%%%%%%
\begin{figure}
%Fig made by make_all_jam_model_plots
\includegraphics[width=0.86\columnwidth]{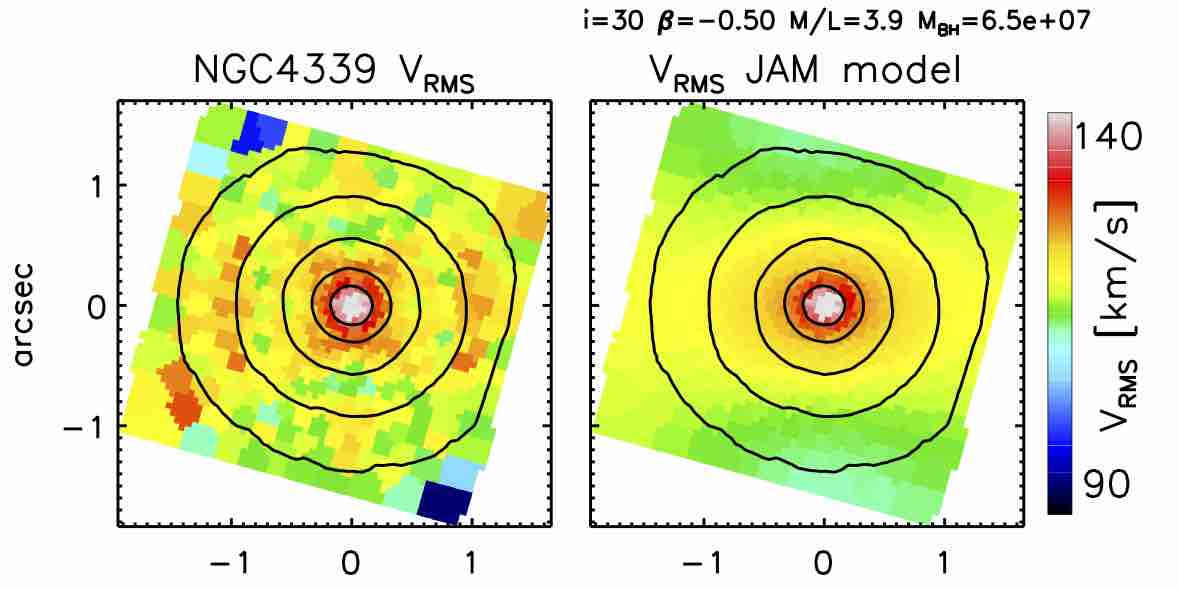}
\includegraphics[width=0.86\columnwidth]{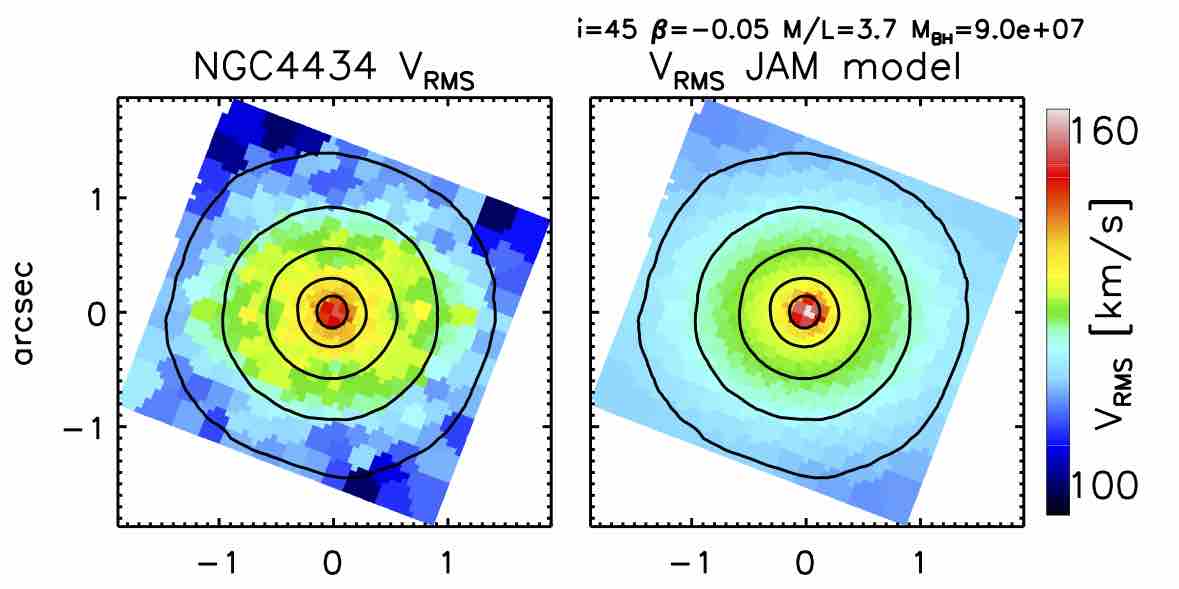}
\includegraphics[width=0.86\columnwidth]{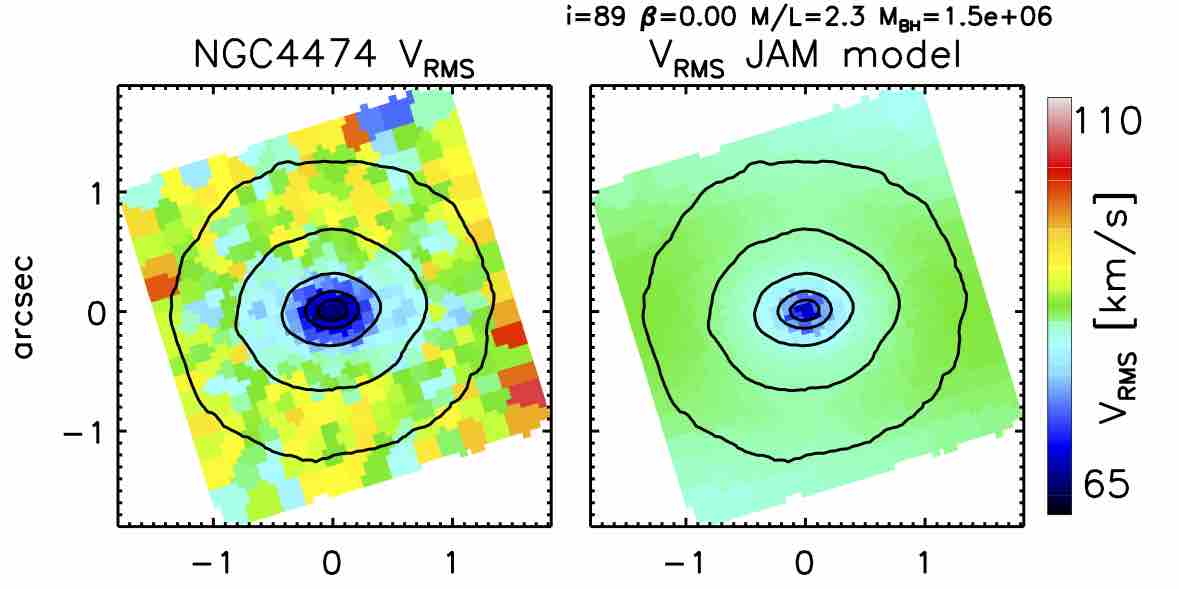}
\includegraphics[width=0.86\columnwidth]{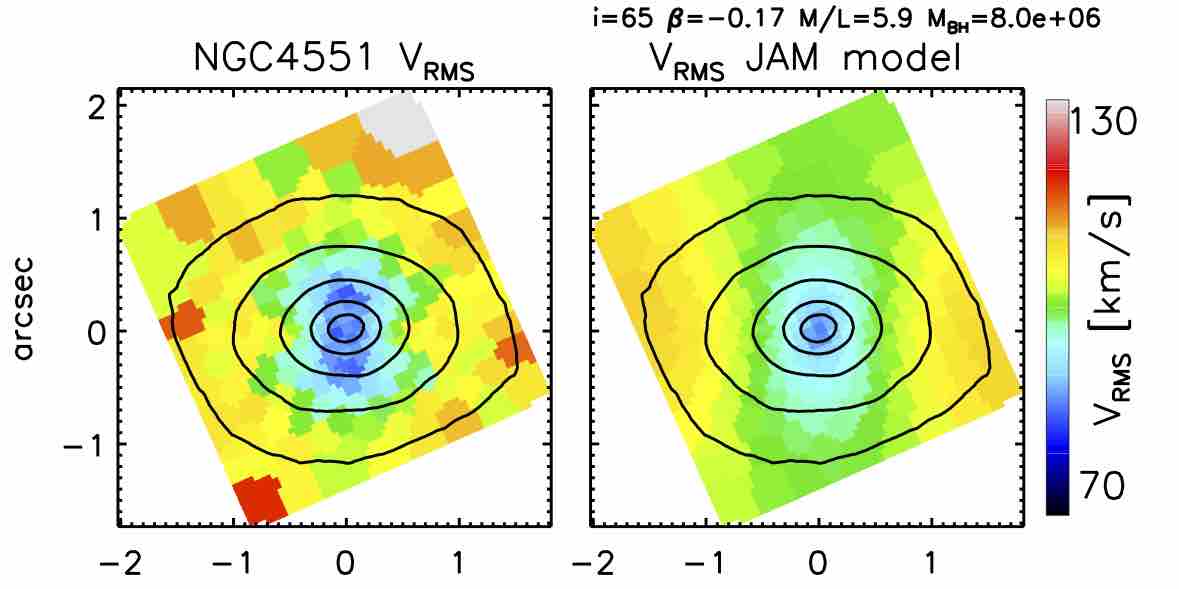}
\includegraphics[width=0.86\columnwidth]{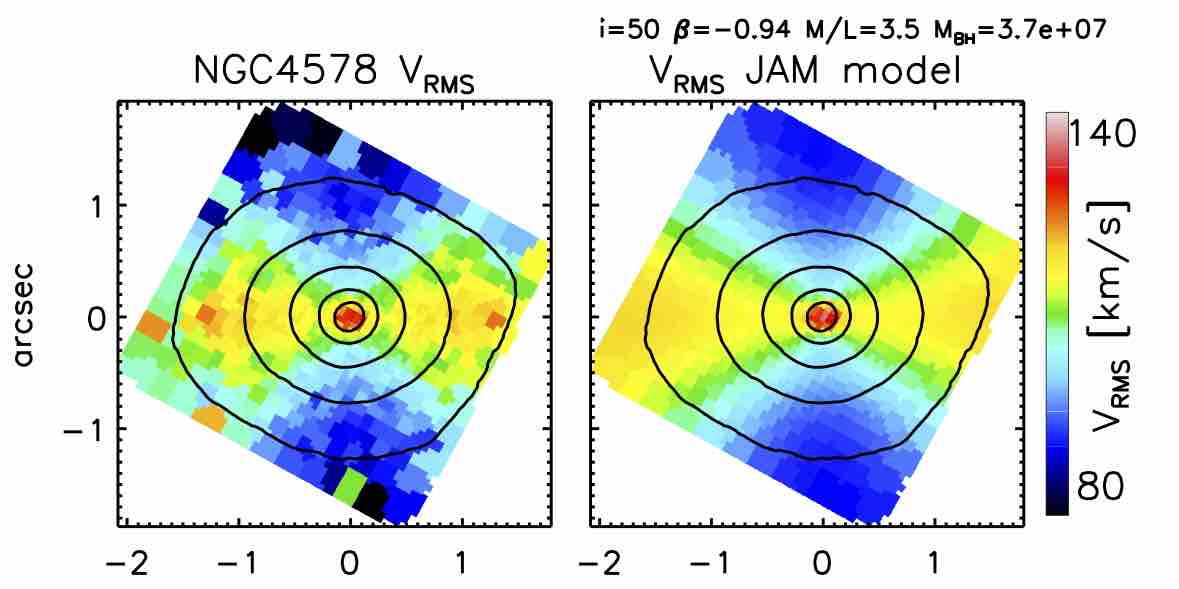}
\includegraphics[width=0.86\columnwidth]{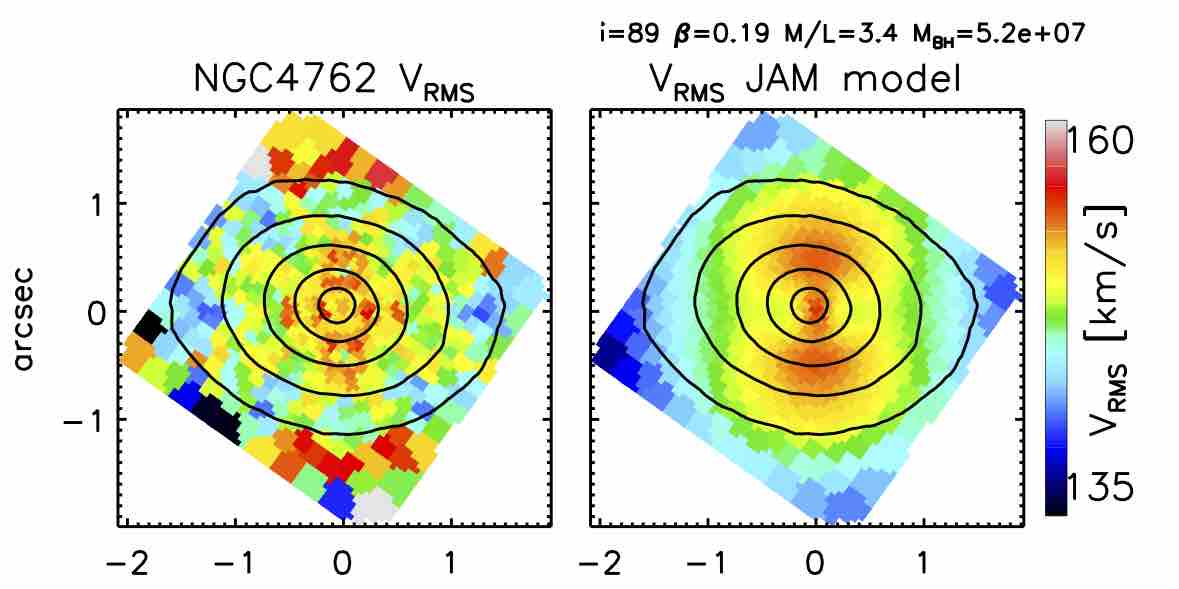}
\caption{Comparison between the second velocity moment as parametrized by V$_{\rm rms}$ from the NIFS data and the JAM models, for the best-fitting parameters (for NGC\,4474 and NGC\,4551 we use the upper limit for the black hole mass), which are shown on top of the model V$_{\rm rms}$ maps. While the models were constrained using the original kinematics, V$_{\rm rms}$ maps shown here are symmetrized as described in Section~\ref{ss:schw} for comparison purposes with bi-symmetric maps produced by the JAM models. }
\label{f:jam_comp}
\end{figure}
%%%%%%%%%%%%%%%%%%%%%%%%%%%%%%%%%%%%%%%%%%%%%%%%%%%%%%%%%%%

The JAM models based on SAURON data constrain the inclination of two galaxies to be edge-on (close to 90\degree), but most galaxies have a low inclination. As models on low inclinations are more degenerate, to explore the parameter space we also run models assuming an edge-on orientation (in practice, $i=89$\degree) for these galaxies. Therefore, each JAM model assumes an axisymmetric light distribution at a given inclination, and has additional three free parameters. Two of those are used to fully specify the distribution of the second velocity moment (at a given inclination): a black hole mass M$_{\rm BH}^{\rm JAM}$ and a constant velocity anisotropy parameter $\beta_{\rm JAM}$. The $M/L$ ratio is then used to linearly scale the predicted second velocity moment to the observed V$_{\rm rms}$. We build a grid of models varying M$_{\rm BH}^{\rm JAM}$ and $\beta_{\rm JAM}$. These are shown in Fig~\ref{f:jams} and the best-fitting parameters are presented in Table~\ref{t:results}.

NIFS data have a small angular coverage, but in the majority of cases they are able to constrain the black hole and the velocity anisotropy. The JAM models provide only an upper limit at the $3\sigma$ confidence level for black hole masses in NGC\,4474 and NGC\,4551, although in the latter case there is also a lower limit at $1\sigma$ confidence level, suggesting a low-mass black hole of $2.8\times10^6$ M$_\odot$. The velocity anisotropy is poorly constrained for NGC\,4339 and NGC\,4578, also being unusually low for ETGs. These values are however unreliable as the galaxies are oriented at low inclinations ($<50$\degree) and cannot be trusted \citep{2012MNRAS.424.1495L}. In cases where the anisotropy can be trusted, galaxies are either isotropic (NGC\,4474) or mildly anisotropic with negative (NGC\,4551) and positive (NGC\,4762) $\beta_{\rm JAM}$.
%%%%%%%%%%%%%%%%%%%%%%%%%%%%%%%%%%%%%%%%%%%%%%%%%%%%%%%%%%%
\begin{figure}
%Fig made by make_all_jam_model_plots_major_minor_axis
\includegraphics[width=0.88\columnwidth]{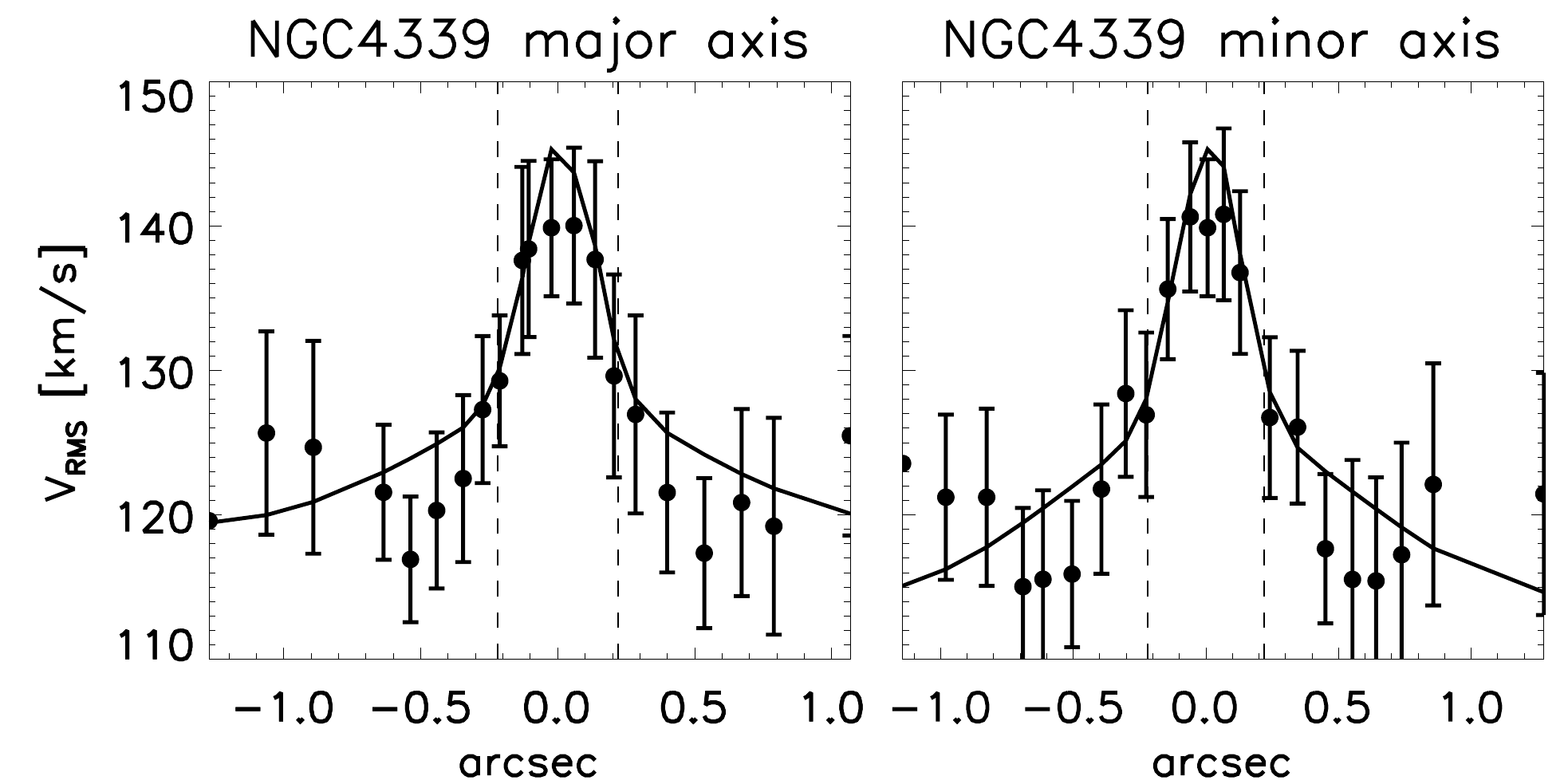}
\includegraphics[width=0.88\columnwidth]{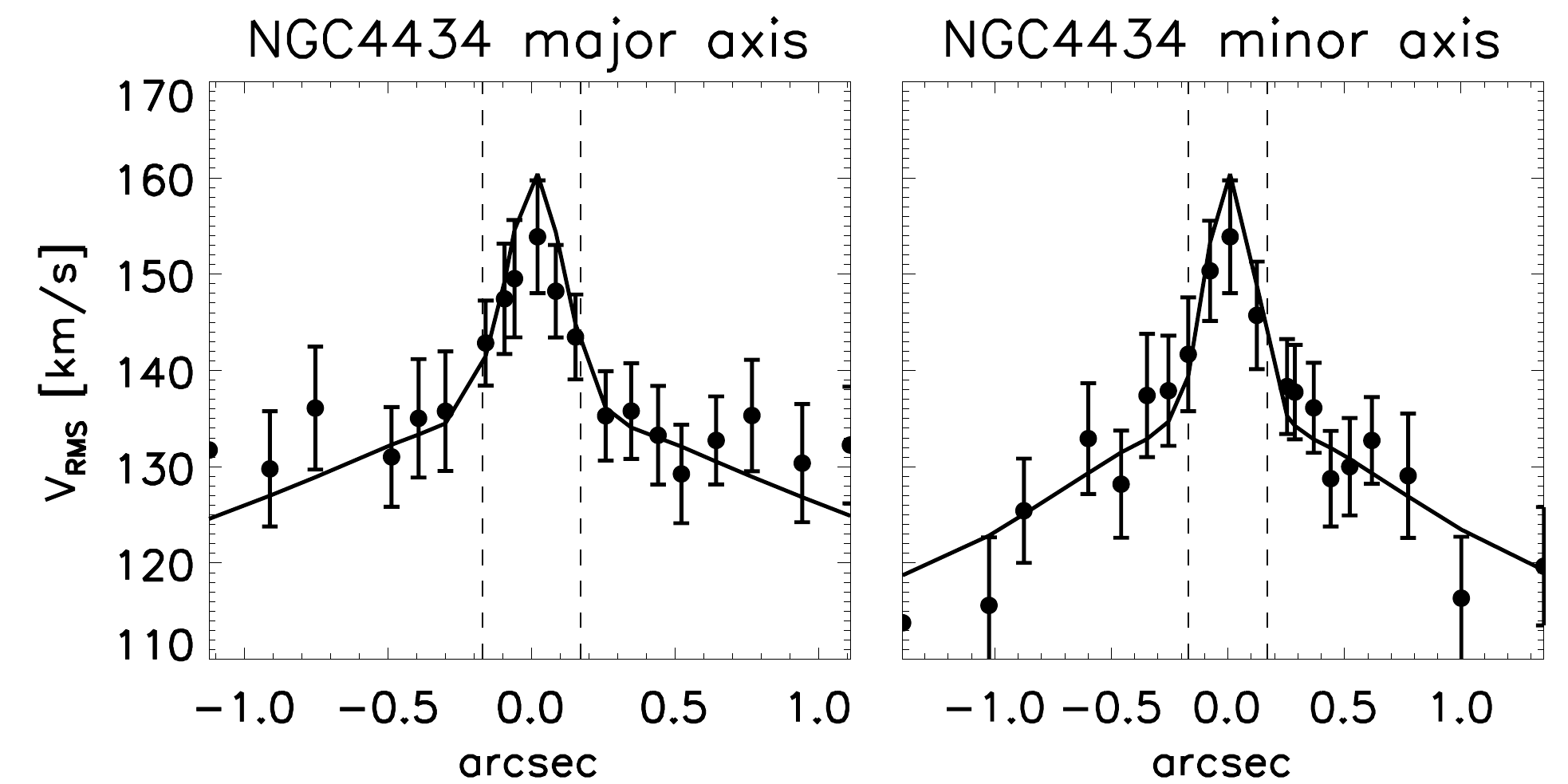}
\includegraphics[width=0.88\columnwidth]{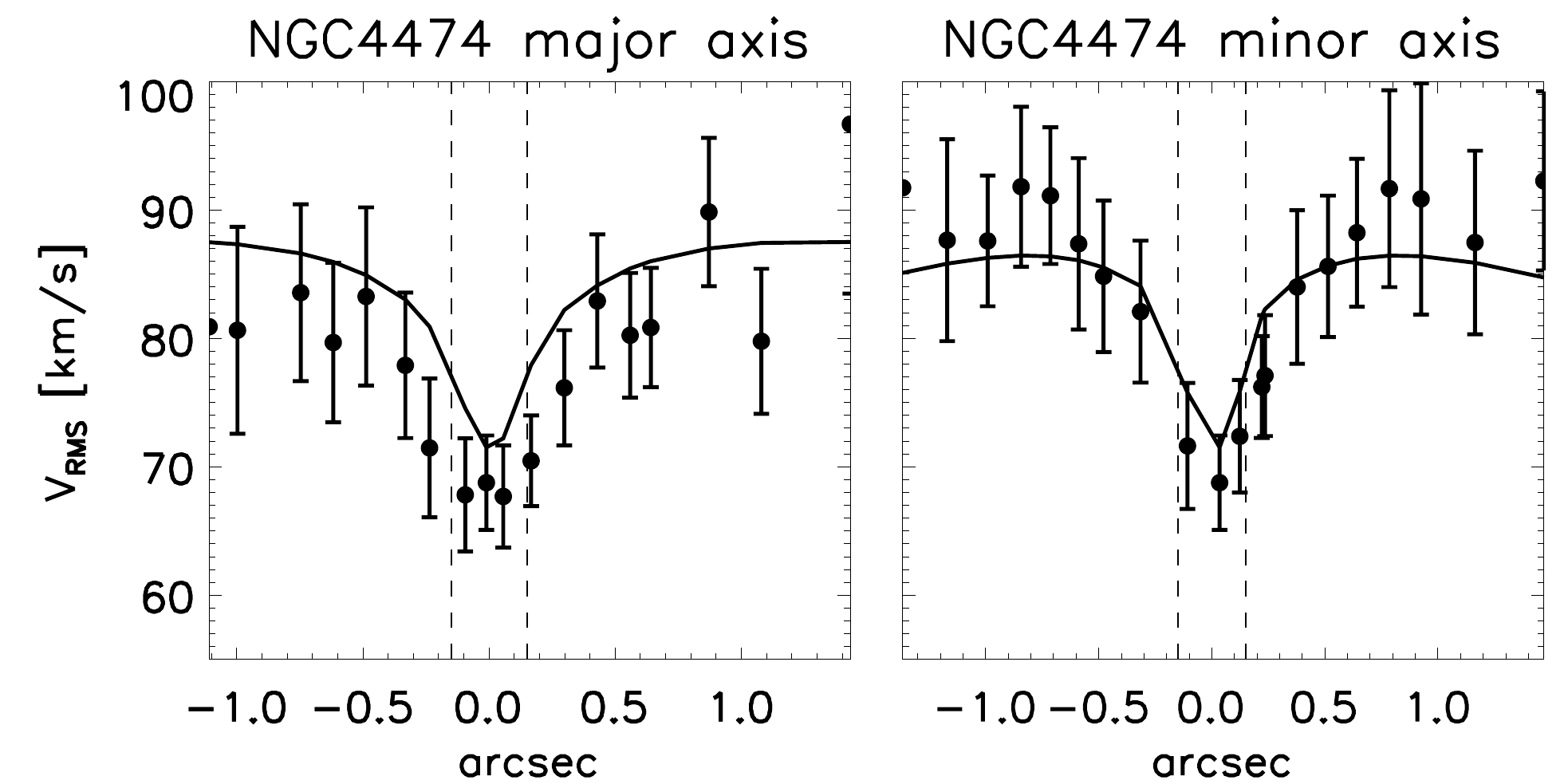}
\includegraphics[width=0.88\columnwidth]{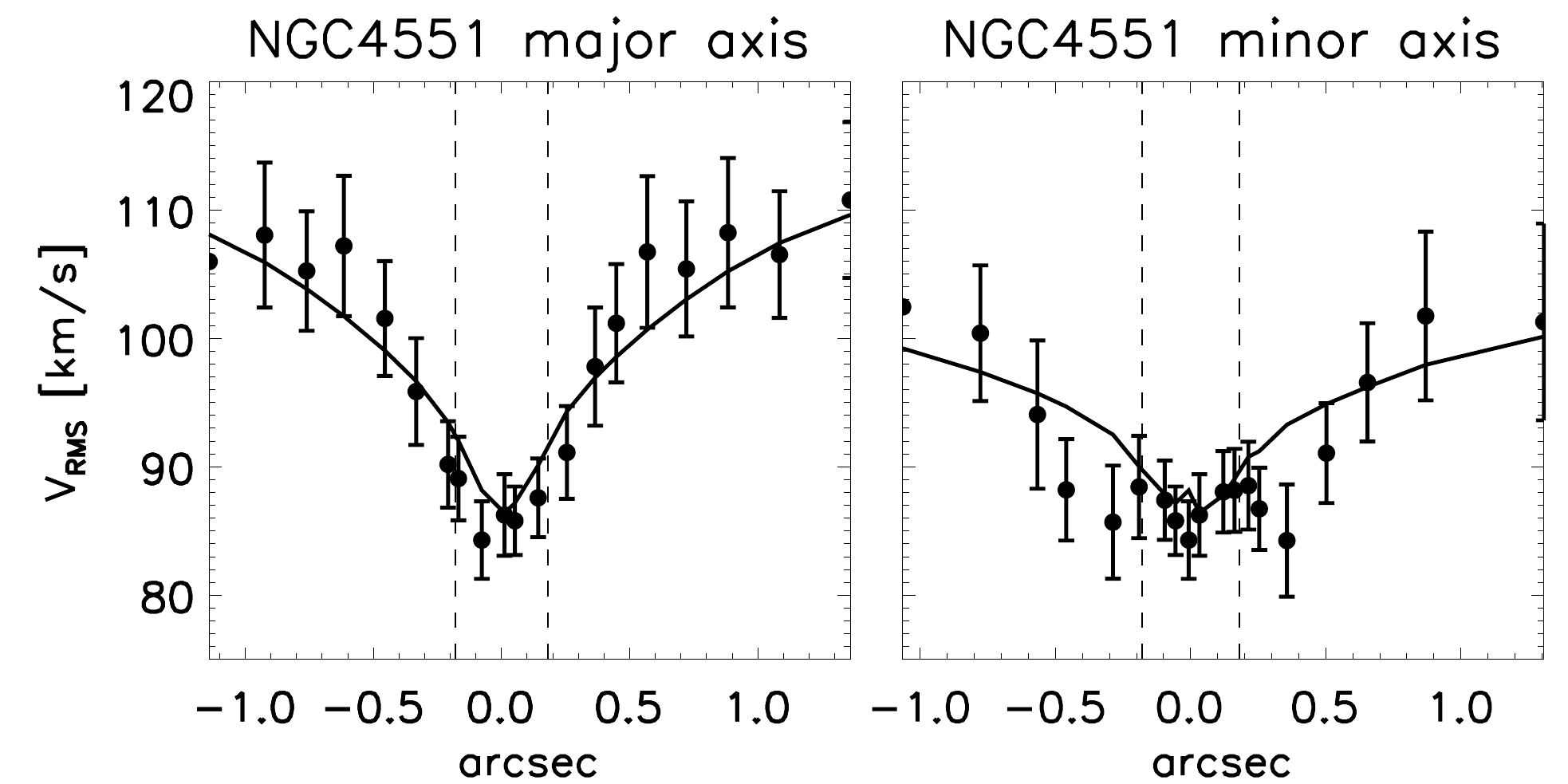}
\includegraphics[width=0.88\columnwidth]{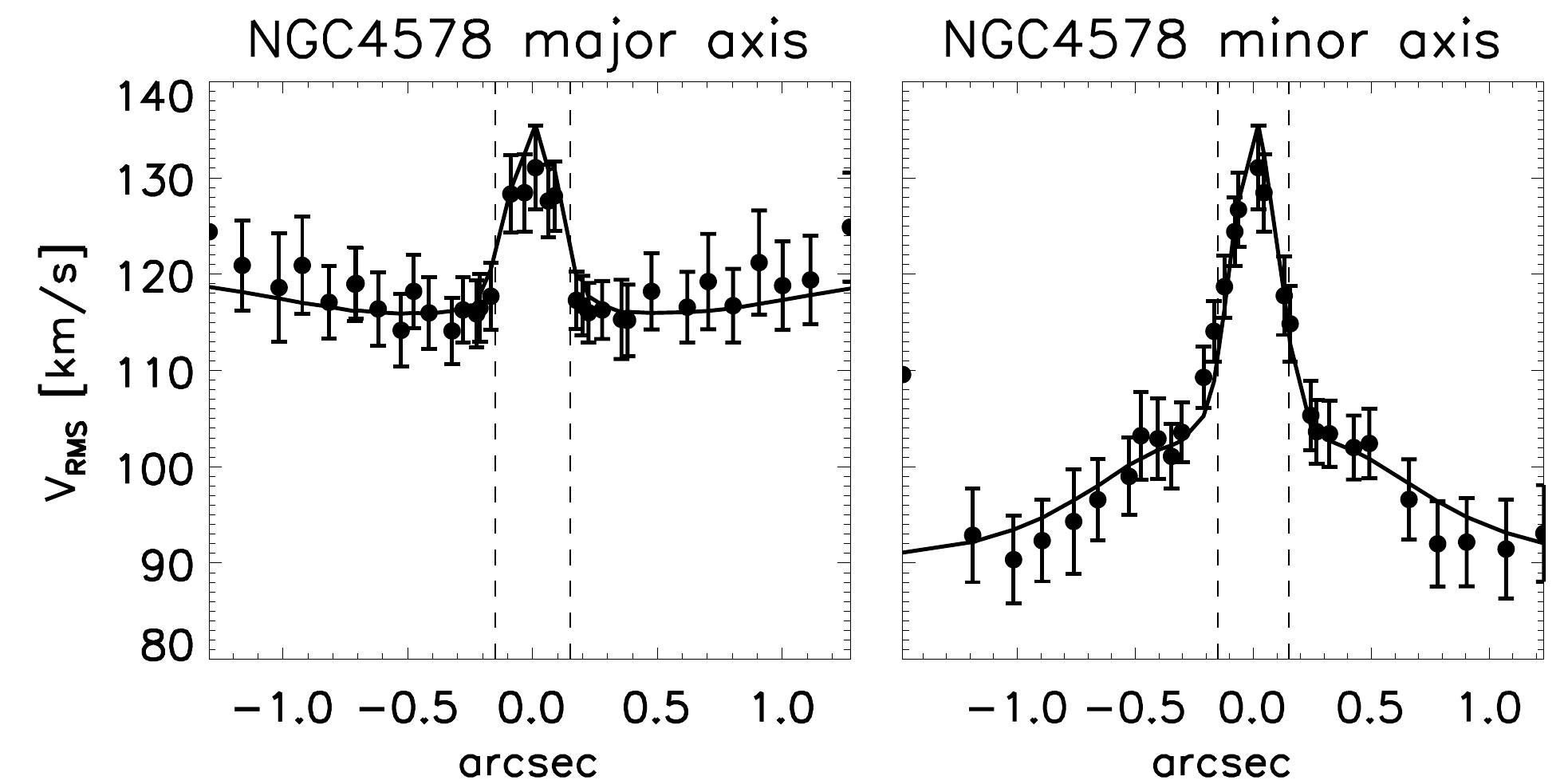}
\includegraphics[width=0.88\columnwidth]{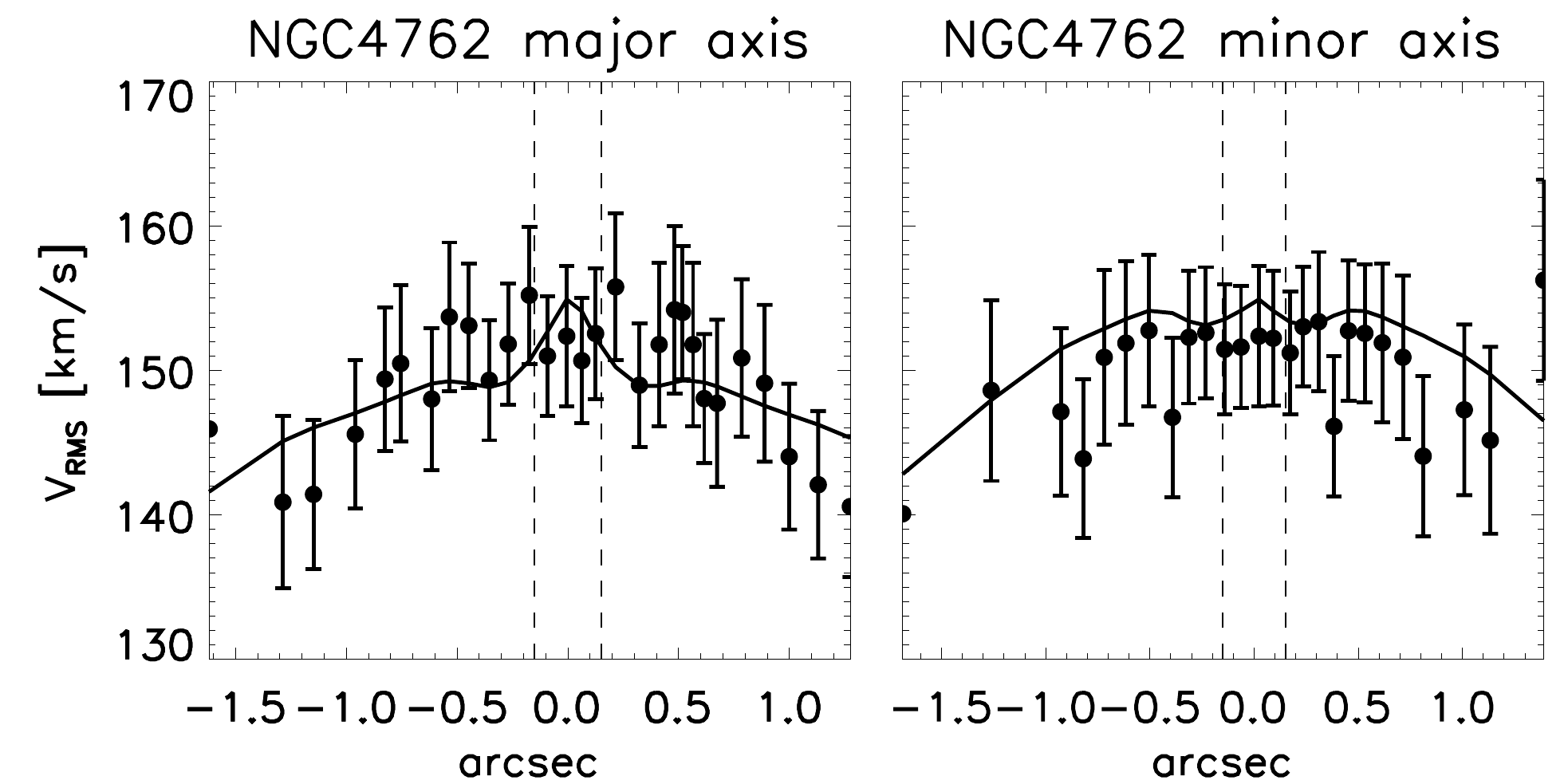}
\caption{Comparison between the V$_{\rm rms}$ extracted from the (symmetrized) V$_{\rm rms}$ maps along the major and minor axes and the JAM models, having the same parameters as in Fig.~\ref{f:jam_comp}. }
\label{f:jam_comp_axes}
\end{figure}
%%%%%%%%%%%%%%%%%%%%%%%%%%%%%%%%%%%%%%%%%%%%%%%%%%%%%%%%%%%

Changing the inclination of the models does not change the results substantially, and the largest difference is seen at the lowest inclination. NGC\,4339 is nominally at only 30\degree\, and putting the models at 89\degree\,changes the shape of the $\chi^2$ contours significantly. At 89\degree\,the best-fitting model is almost isotropic, but the M$_{\rm BH}^{\rm JAM}$  does not change. A similar effect is seen for NGC\,4578 ($i=45$\degree). The change of inclination has a minor effect on the estimated $M/L$. 

As our galaxies are at different distances the NIFS data probe different physical radii in their nuclei. Based on the JAM models, the SoI sizes of the best-fitting black holes span a range from 0.1 to 0.4\arcsec. This means that the dynamical models are constrained by kinematics which covers from about 4 to about 14 times the radius of the SoI. In order to verify that we are indeed probing a sufficient areas to constrain the anisotropy, M$_{BH}$ and $M/L$, we performed the following test. The NIFS data with the largest coverage are for NGC\,4578 and NGC\,4762 (8 and 14 times the obtained radius of the SoI), while the smallest coverage is found for NGC\,4339 and NGC\,4434 (4 times the radius of the SoI). Therefore, we run two grids of JAM models for NGC\,4578 and NGC\,4762 restricting the NIFS field to 4 times their respective radii of SoIs. The resulting grids resemble those in Fig.~\ref{f:jams} in terms of the shape of the $\chi^2$ contours, and also recover similar best-fitting results. In case of NGC\,4578, the best fit is given by the same model as in the case of the full field coverage, except that its vertical ansiotropy $\beta_{\rm JAM}$ is less tangentially biased (by 15 per cent). The NGC\,4762 best-fitting model has 8 per cent smaller M$_{BH}$ and about 30 per cent smaller anisotropy values, which fall within the $3\sigma$ uncertainty level of the grid shown in Fig.~\ref{f:jams}. From these tests we conclude that our NIFS data are adequate to constrain JAM models of all galaxies in the sample. 

\citet{2017MNRAS.464.4789M} constrain their JAM models by propagating uncertainties different from those obtained by extraction of kinematics. Specifically, they assume a constant error on the velocity (5 km/s) and as the uncertainty on the velocity dispersion take 5 per cent its value for each bin. The reason for this is to prevent a biasing of the solution towards the high S/N central pixels. We also run JAM models in such a configuration and recovered the same results. The reason is a relatively small variation of both velocity and velocity dispersion uncertainties across the FoV, as well as the fact that the average uncertainties are actually similar to those proposed by \citet{2017MNRAS.464.4789M}, as can be seen from Table~\ref{t:errors}.

In Fig.~\ref{f:jam_comp}, we compare the second velocity moment maps (parameterized as V$_{\rm rms}$) and the JAM model predictions for the best-fitting models. We compare the JAM models with bi-symmetrized data (see Section~\ref{ss:schw} for details on how we symmetrize maps), as models have such symmetry by construction. A similar comparison is given also in Fig.~\ref{f:jam_comp_axes}, this time along the major and minor axes, showing also the error on the V$_{\rm rms}$. For NGC\,4474 and NGC\,4551 we present models with black hole masses corresponding to derived upper limits. All models reproduce the features on the observed V$_{rms}$ maps reasonably well, while the model maps for NGC\,4551 and NGC\,4762 are somewhat different from data, systematically under- or overpredicting second velocity moments. These difference, however, are typically within the uncertainties, as seen of Fig.~\ref{f:jam_comp_axes}. Models for NGC\,4474 and NGC\,4551 have some difficulties reproducing the extent and the shape of the central decreases in the V$_{\rm rms}$, but they are, except when reproducing the minor axis of NGC\,4551 generally consistent with the uncertainties. We will further discuss the implications of JAM fits to these two galaxies in Section~\ref{ss:nondetect}.

%%%%%%%%%%%%%%%%%%%%%%%%%%%%%%%%%%%%%%%%%%%%%%%%%%%%%%%%%%%
\begin{figure*}
%Fig made by plot_grid_nfis.pro
\includegraphics[width=\textwidth]{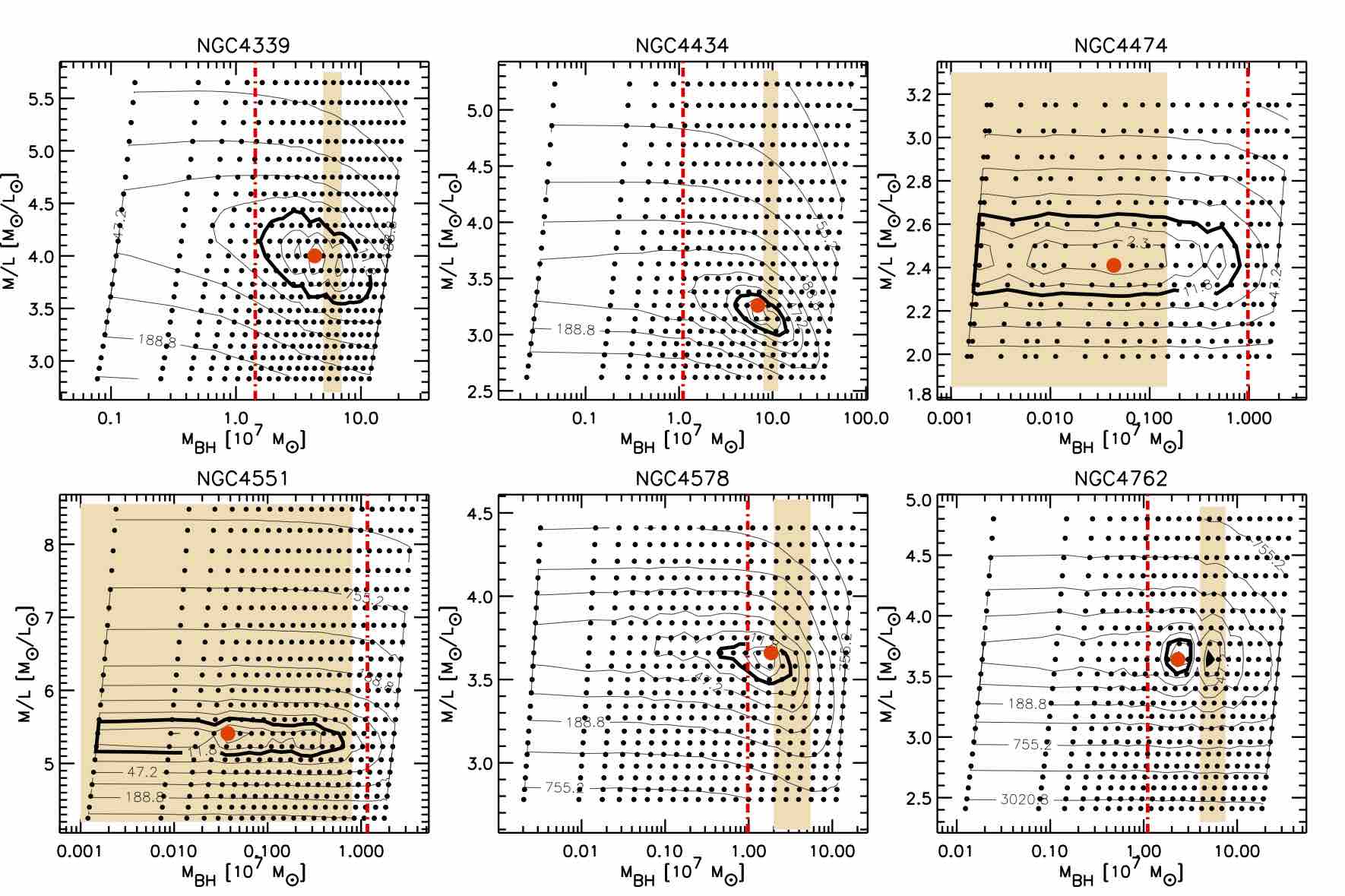}
\caption{Grids of Schwarzschild models (shown by small round symbols) with different $M/L$ and black hole mass. The best-fitting models are shown by a large (red) circles in the formal $\Delta \chi^2$ minimum, while shaded rectangles trace the black hole mass estimates of the JAM modelling. Note that we do not compare the $M/L$ ratios as they are calculated within different regions (see Section~\ref{ss:jam} for details why black hole masses can be compared). Contours are the $\Delta \chi^2 = \chi^2 - \chi^2_{min}$ levels and the thickest contours show the $3\sigma$ level for a two-dimensional distribution. The dash--dotted (red) vertical line indicates the mass of the black hole which has the radius of the sphere of influence 2.5 times smaller than the FWHM of the narrow component of the AO PSF (Table~\ref{t:psf}). This is a very approximate estimate of the lower mass limit to a black hole we expected to be able to detect.  }
\label{f:SCHW_grids}
\end{figure*}
%%%%%%%%%%%%%%%%%%%%%%%%%%%%%%%%%%%%%%%%%%%%%%%%%%%%%%%%%%%

\subsection{Schwarzschild models}
\label{ss:schw}

We used the `Leiden' version of the axisymmetric Schwarzschild code, as it is described in \citet{2006MNRAS.366.1126C}. Briefly, the code takes the MGE parametrization of light 
and, assuming an inclination, an $M/L$ ratio, and axisymmetry, deprojects the surface brightness distribution into a mass volume density, which, after addition of a black hole of a certain mass (M$_{\rm BH}$) specifies the potential. The next step is generation of a representative orbit library for a given set of $M/L$ and M$_{\rm BH}$. Each orbit is specified by three integrals of motion ($E$, $L_z$, $I_3$). We sample $E$ with 21 logarithmically spaced points giving the representative radius of the orbit, and at each energy we use eight radial and seven angular points for sampling $L_z$ and $I_3$ respectively. As orbits can have a prograde and retrograde sense of rotation, the initial set of orbits is doubled and there are 2058 orbital bundles. Each bundle is composed of $6^3$ individual orbits started from adjacent initial conditions. In total, there are $21\times7\times7\times2\times6^3 = 444 528$ orbits which make the basis for the construction of the galaxy. The phase space coordinates are computed at equal time steps and projected on the sky plane as triplets of coordinates ($x$, $y$, $v_z$), while the sky coordinates ($x$, $y$) are randomly perturbed with probability described by the PSF of the data. The model galaxy is Voronoi binned in the same way as the observed kinematics and at each bin position the code fits the resulting LOSVDs to provide $V$, $\sigma$ and four Gauss-Hermite parameters (up to $h_6$). As the dithering scheme improves on the smoothness of the distribution function, we use only a modest amount of regularization $\Delta=10$ \cite[as defined in][]{1998ApJ...493..613V}, except for NGC\,4762 for which we increased it to $\Delta=4$ in order to make the resulting orbital weights distribution smoother, which also reflected in somewhat smoother $\chi^2$ contours on the model grids (see below)\footnote{We have nevertheless run models at both $\Delta=4$ and 10 regularizations and in all cases, including the NGC\,4762, the results were fully consistent.}.

%%%%%%%%%%%%%%%%%%%%%%%%%%%%%%%%%%%%%%%%%%%%%%%%%%%%%%%%%%%
\begin{figure*}
%Fig made by plot_sch_model_NIFS_sample_comparison.pro
\includegraphics[width=\textwidth]{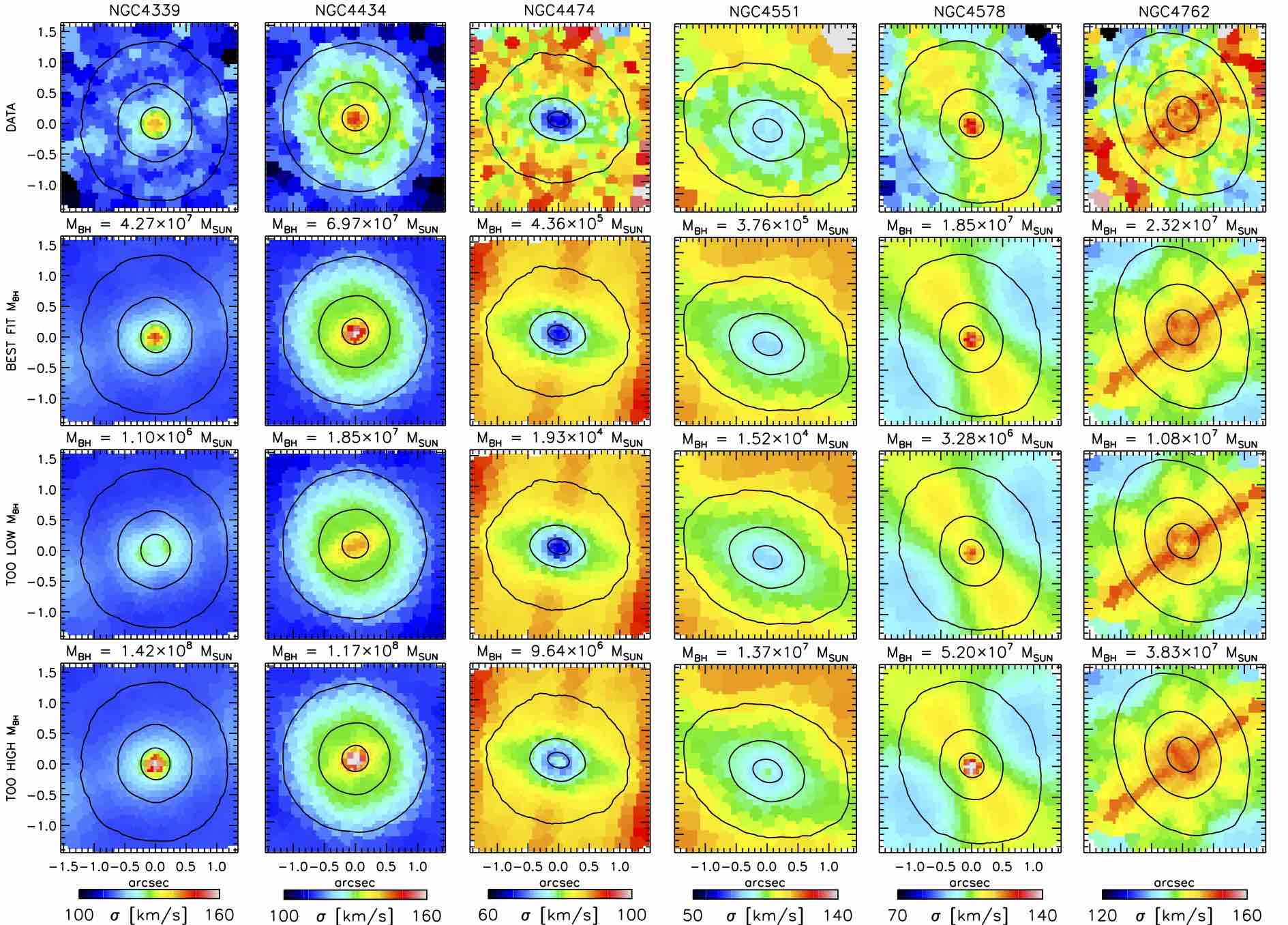}
\caption{Comparison of the velocity dispersion maps between the data and the Schwarzschild models. Each galaxy is in a column (from left to right) as indicated by its name on the top panel. Rows show (from top to bottom) the observed symmetrized velocity dispersion (NIFS data), and the velocity dispersion maps predicted by models with the best fitting, a too low and too high M$_{\rm BH}$, as written next to the maps. The low and high M$_{\rm BH}$ models are selected to be just outside of the $3\sigma$ $\Delta \chi^2$ contours in Fig~\ref{f:SCHW_grids}. For NGC\,4474 and NGC\,4551 the low-mass models are selected to be close to the lowest mass models investigated and outside the $1\sigma$ confidence level. All models are at the best $M/L$ for each case, respectively. The high- and low-mass models are clearly ruled out for all galaxies except for NGC\,4474 and NGC\,4551, for which only the high-mass models show differences compared to the data. }
\label{f:sigma_comp}
\end{figure*}
%%%%%%%%%%%%%%%%%%%%%%%%%%%%%%%%%%%%%%%%%%%%%%%%%%%%%%%%%%%

For each galaxy we run a grid of models specified by $M/L$ and M$_{\rm BH}$. Given the inclination degeneracy \citep{2005MNRAS.357.1113K}, we do not try to fit for the galaxy inclination, but adopt the values given in Table~\ref{t:sample}. A new orbital library is produced for each M$_{\rm BH}$ at the best-fit $M/L$ inferred from the JAM models (Section~\ref{ss:jam}), and scaled to a range of $M/L$ ratios. The initial choice for M$_{\rm BH}$ was the one predicted by the M$_{\rm BH} - \sigma$ relation. The models are constrained with NIFS and SAURON kinematics fitted simultaneously. As the models are by construction bi-(anti-)symmetric, with the minor axis of the galaxy serving as the symmetry axis, we symmetrize the kinematics in the same way. This was done by averaging the kinematics of the positions (($x$,$y$), ($x$, $-y$), ($-x$,$y$), ($-x$,$-y$)). Given that the data are Voronoi binned and that the bins at the four positions are not necessarily of the same shape and size, when there are no exact symmetric points, we interpolated the values on those positions and then averaged them. When averaging we take into account that the odd moments of the LOSVD ($V$ and $h_3$) are bi-anti-symmetric, while the even moments ($\sigma$, $h_4$) are bi-symmetric. We retain the original kinematic errors at each point, however, so as not to underestimate the LOSVD parameter errors and over-constrain the model with artificially low formal errors. During the fitting process we model the full extent of the data and used them to derive the overall $\chi^2$ sum (note that even though we symmetrize the data, we do not fold it or fit only one quadrant). Finally, we exclude the central 1.3\arcsec\,of the SAURON data since these overlap with the high resolution NIFS observations, and we wish to avoid our results being affected by inconsistencies in the relative calibration or assumed PSFs of the two data sets. In practice, however, fitting both data sets in the overlap region does not make a significant difference, as also reported earlier \citep{2009MNRAS.399.1839K,2017A&A...597A..18T}.

%%%%% Table 1. %%%%%%%%%%%%%%%%%%%%%%%%%%%%%%%%%%%%%%%%%%%%%%%%%%%%%
\begin{table*}
   \caption{Summary of dynamical modelling results.}
   \label{t:results}
$$
  \begin{array}{c|ccrc|ccc}
    \hline
    \hline
    \noalign{\smallskip}
    
	&$JAM$ & & & &$Schwarzschild$ &&\\
        $galaxy$ & $M$_{\rm BH} $[M$_\odot$]$& $M/L$ & \beta$  $ & \chi^2_{min}/DOF & $M$_{\rm BH} $[M$_\odot$]$ &  $M/L$ & \chi^2_{min}/DOF \\
                       & 10^7$[M$_\odot$]$                &            &          &                              &10^7 $[M$_\odot$]$                &             &                 \\
             (1)     & (2)                                           &     (3)  &  (4)    &    (5)                      &  (6)            &(7)                                            &     (8)\\
    \noalign{\smallskip} \hline \hline \noalign{\smallskip}
    $NGC$\,4339 & 6.5^{+0.7}_{-1.3}&   3.9 & -0.50  &  1.08 & 4.3^{+4.8}_{-2.3}     & 4.00^{+0.3}_{-0.35}      &  0.46 \\ 
    $NGC$\,4434 & 9.0^{+2.5}_{-0.5}&   3.7 & -0.05  &  1.09 &7.0^{+2.0}_{-2.8}     & 3.26^{+0.04}_{-0.19}  &   0.53   \\ 
    $NGC$\,4474 &<0.15                   &   2.3 & -0.00  &  1.04  &<0.7                        & 2.41^{+0.19}_{-0.11}  &   0.44  \\ 
    $NGC$\,4551 & <0.8                    &   5.9 & -0.17  &  1.47 &<0.5                       & 5.41^{+0.19}_{-0.21}  &    0.51   \\    
    $NGC$\,4578 & 3.5^{+0.4}_{-0.3}&   3.7 & -0.94  &  1.59 & 1.9^{+0.6}_{-1.4}   & 3.65^{+0.15}_{-0.10}    &  0.58  \\ 
    $NGC$\,4762 & 5.2^{+0.9}_{-1.1} &   3.4 & 0.19   &  0.72 &2.3^{+0.9}_{-0.6}    & 3.64^{+0.16}_{-0.14}  &   1.00  \\ 

       \noalign{\smallskip}
    \hline
  \end{array}
$$ 
{Notes -- Column 1: galaxy name; Column 2-6: parameters of the JAM models (black hole mass, mass-to-light ratio, velocity anisotropy, $\chi^2$ of the best fit model per degree of freedom (DOF)); Columns 6 - 8: parameters of the Schwarzschild models (black hole mass, mass-to-light ratio, and $\chi^2$ of the best fit model per DOF ). Uncertainties were estimated by marginalising over the other parameter (M$_{\rm BH}$, $\beta$ or M/L) and assuming a $3\sigma$ confidence level for one degree of freedom. Note that JAM models were constrained using original kinematics, while Schwarzschild models were constrained using symmetrised kinematics. M/L are given in Johnson $R$-band  magnitude system for WFC2 galaxies (NGC\,4339, NGC\,4474 and NGC\,4578) and for F450W-band of the AB magnitude systems for galaxies observed with ACS (NGC\,4434, NGC\,4551 and NGC\,4762).}
\end{table*}
%%%%%%%%%%%%%%%%%%%%%%%%%%%%%%%%%%%%%%%%%%%%%%%%%%%%%%%%%%%%%%%%%%

Figure~\ref{f:SCHW_grids} shows the model grids for all galaxies in the sample. We measure M$_{\rm BH}$ at the $3\sigma$ confidence level for four of the six objects, with the remainder giving only upper limits. The results are again outlined in Table~\ref{t:results}, where one can see that the results from the JAM modelling are fully consistent with the Schwarzschild results within $3\sigma$ level or better. The parameter uncertainties from the Schwarzschild modelling were estimated by marginalizing along the best-fitting $M/L$ ratio and M$_{\rm BH}$ and for the $3\sigma$ confidence level (for one degree of freedom). We choose the $3\sigma$ level because $1\sigma$ levels often do not have a single well defined minimum, due to numerical discretization noise. In all cases $M/L$ is well constrained, and the $\Delta \chi^2$ contours are regular and show only a minor degeneracy between M/L and M$_{\rm BH}$ (e.g. NGC\,4339). For NGC\,4762, there were multiple local minima apparent even at the $3\sigma$ confidence level. Increasing the level of regularization from $\Delta=10$ to $\Delta=4$ avoided this issue, but there is still evidence for degeneracy towards higher black hole masses. As shown below, this is not significant and the mass of the black hole is well constrained. For NGC\,4474 and NGC\,4551, the $\Delta \chi^2$ contours do not close over 2-3 order of magnitude in M$_{\rm BH}$ (down to a few times $10^4$ M$_\odot$). In both cases, there are tentative $1\sigma$ level lower limits for best-fitting black holes of $4.4\times10^5$ and $3.8\times10^5$ M$_\odot$, respectively, but these should not be considered reliable, as discussed below. 

Our results can be compared to those of \citep{2013MNRAS.432.1709C} who modelled the full ATLAS$^{\rm 3D}$ sample, in particular with their $M/L$ ratios. They work in SDSS $r$ band and we need to first adjust for the difference in the filter and the photometric systems. Using the web tool based on \citet{1994ApJS...95..107W} models\footnote{\href{http://astro.wsu.edu/dial/dial_a_model.html}{http://astro.wsu.edu/dial/dial\_a\_model.html}}, we estimate that to change from our Johnson R and F450W AB systems to SDSS $r$ AB system we need to apply factors of 1.25 and 0.85, respectively. This means that our $M/L$ based on the Schwarzschild models (and approximately within one effective radius) are 5.0, 2.8, 3.0, 4.6, 4.6 and 3.1 for NGC\,4339, NG\,4434, NGC\,4474, NGC\,4551, NGC\,4578 and NGC\,4762, respectively. A comparison with the values published in \citep{2013MNRAS.432.1709C} shows that the differences between respective $M/L$ are small, typically below 10 per cent, except for NGC4762 where they are about 17 per cent. 

This comparison, as well as girds on Fig.~\ref{f:SCHW_grids}, show that the results of Schwarzschild models compare well with the results of the JAM models within a $3\sigma$ uncertainty level. This is both encouraging and revealing, as our approach is purely empirical. In this paper we use two different methods (one more general, but containing possible numerical issues, and another less general, but numerically accurate) to measure the same quantity. This is motivated by the agreement found for a number of other black hole mass estimates, as outlined in Section~\ref{ss:methods}, including the most precisely known black holes in the Milky Way and the megamaser galaxy NGC\,4258, as well as in the somewhat controversial case of NGC\,1277 (which we show in Appendix~\ref{app:n1277}). The fact that such different modelling techniques, with their inherent assumptions and sources of systematics, as well as constrained by different data sets (NIFS only for JAM and SAURON and NIFS for Schwarzschild), give consistent results, allows us to trust the derived M$_{\rm BH}$. On the other hand, a disagreement between JAM and Schwarzschild results would have to be taken with a caution. The lack of generality of JAM models could be the source of such a disagreement \citep[for potential issues with Jeans models in spherically symmetric case see][]{2016MNRAS.462.2847M}, but numerical inaccuracies of the Schwarzschild model could also bias the results \citep[e.g. appendix A in][]{2018MNRAS.473.3000Z}. The fact that they are, in our case, consistent typically at $1-2\sigma$ level (except for one galaxy where the difference is just above $3\sigma$ level), and are consistent with previous comparisons between black hole mass estimates \citep[e.g.][]{2009ApJ...693..946S, 2010MNRAS.401.1770V, 2013ApJ...770...86W, 2015MNRAS.450..128D, 2017MNRAS.466.4040F, 2017MNRAS.468.4675D}, discloses how much one can trust black hole mass measurements in general, which should also be considered when discussing the black hole mass scaling relations.

The (symmetrized) data -- model comparison for all galaxies are shown in Figs.~\ref{fapp:nifs} and~\ref{fapp:sau}, indicating that our best-fitting models can reproduce all features rather well, both on the high-resolution NIFS data and the large-scale SAURON data. The quality of fits are best seen in the residual maps, which show the difference between the model and the data relative to the uncertainty for a given kinematic map. Typical disagreement between the best-fitting models and the data are within the uncertainties ($1\sigma$ level), with some $2\sigma$ level discrepancies for individual bins (often towards the edge of the field). Even NGC\,4474 and NGC\,4551 models (at the formally best-fitting M$_{\rm BH}$) reproduce well the full extent of the NIFS and SAURON kinematics. 

In Fig.~\ref{f:sigma_comp} we compare the velocity dispersion maps between the data and the models for the best fit, a lower and a higher mass M$_{\rm BH}$ (just outside the $3\sigma$ contours). The differences between the models are obvious, clearly indicating that the data provide robust upper and lower limits on the masses of black holes. As expected, this is not the case for NGC\,4474 and NGC\,4551, for which the models with the lowest probed M$_{\rm BH}$ show essentially the same kinematics as the observed data. To emphasize this we selected models which are outside of the $1\sigma$ level $\Delta \chi^2$ contours on the lower side of the formally best-fitting model. The differences between these maps and the data are essentially negligible. On the other hand, comparing the data with a model outside of the $3\sigma$ level contour (the upper limit), shows the difference in the central velocity dispersion. This means that one should take the $1\sigma$ results with caution, and we consider only the $3\sigma$  level as a robust estimate on the uncertainty of the Schwarzschild models. In case of NGC\,4474 and NGC\,4551, this means we can trust only the upper limits. Similarly, the robustness of the estimated uncertainties for NGC\,4762 can be seen in comparison of the data with a model with M$_{\rm BH}$ just above the $3\sigma$ limit. This model is located within the $\chi^2$ plateau in Fig.~\ref{f:SCHW_grids}, its M$_{\rm BH}$ creates a marginally different velocity dispersion map from the best fitting model, indicating the level at which the upper limit of M$_{\rm BH}$  can be trusted.

%%%%%%%%%%%%%%%%%%%%%%%%%%%%%%%%%%%%%%%%%%%%%%%%%%%%%%%%%%%
%
% SECTION 5 SECTION 5 SECTION 5 SECTION 5 SECTION 5 SECTION 5
%
%%%%%%%%%%%%%%%%%%%%%%%%%%%%%%%%%%%%%%%%%%%%%%%%%%%%%%%%%%%

\section{Discussion}
\label{s:disc}

\subsection{Possible systematics and their influence on the results of dynamical models }
\label{ss:sys}

{\it The velocity dispersions discrepancy for NGC\,4339.} As shown in Fig.~\ref{f:SAU_NIFS}, there is a minor difference in the measured velocity dispersions using the NIFS and SAURON data of NGC\,4339. In the overlap region between the two data sets, the NIFS velocity dispersion are lower for about 8 per cent. We run the Schwarzschild models using these data. As for all galaxies we excluded the SAURON data in the overlap regions. In order to test the robustness of these result we also run additional models which had the following modifications: (i) both NIFS and the central SAURON data were included, and (ii) we corrected the NIFS data by increasing the velocity dispersions by 8 per cent. Both of these models gave fully consistent results with the base models. The $M/L$ ratio did not change. This was expected as $M/L$ is mostly constrained by the large-scale SAURON data. The mass of the black hole did show a minor change. In the case (i) it increased for 7 per cent to $4.6\times10^7 M_\odot$and in the case (ii) for about 80 per cent to $7.9\times10^7 M_\odot$. Both of these values are within the $2\sigma$ confidence contours and indicate the possible range of M$_{\rm BH}$  due to systematics in the data.  

%%%%%%%%%%%%%%%%%%%%%%%%%%%%%%%%%%%%%%%%%%%%%%%%%%%%%%%%%%%
\begin{figure*}
%Fig made by mbh_sigma_relation_nifs_sample.pro
\includegraphics[width=0.47\textwidth]{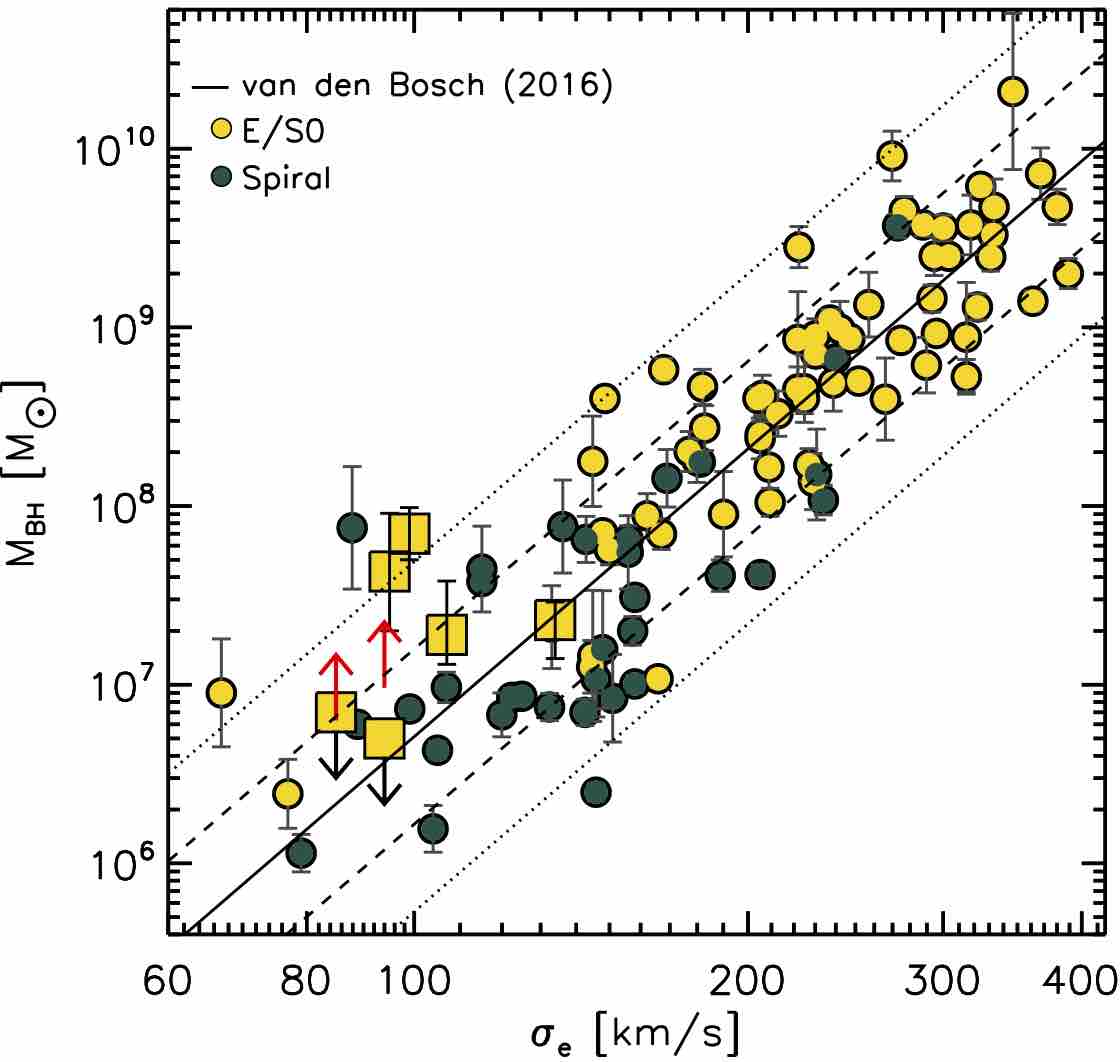}
\includegraphics[width=0.47\textwidth]{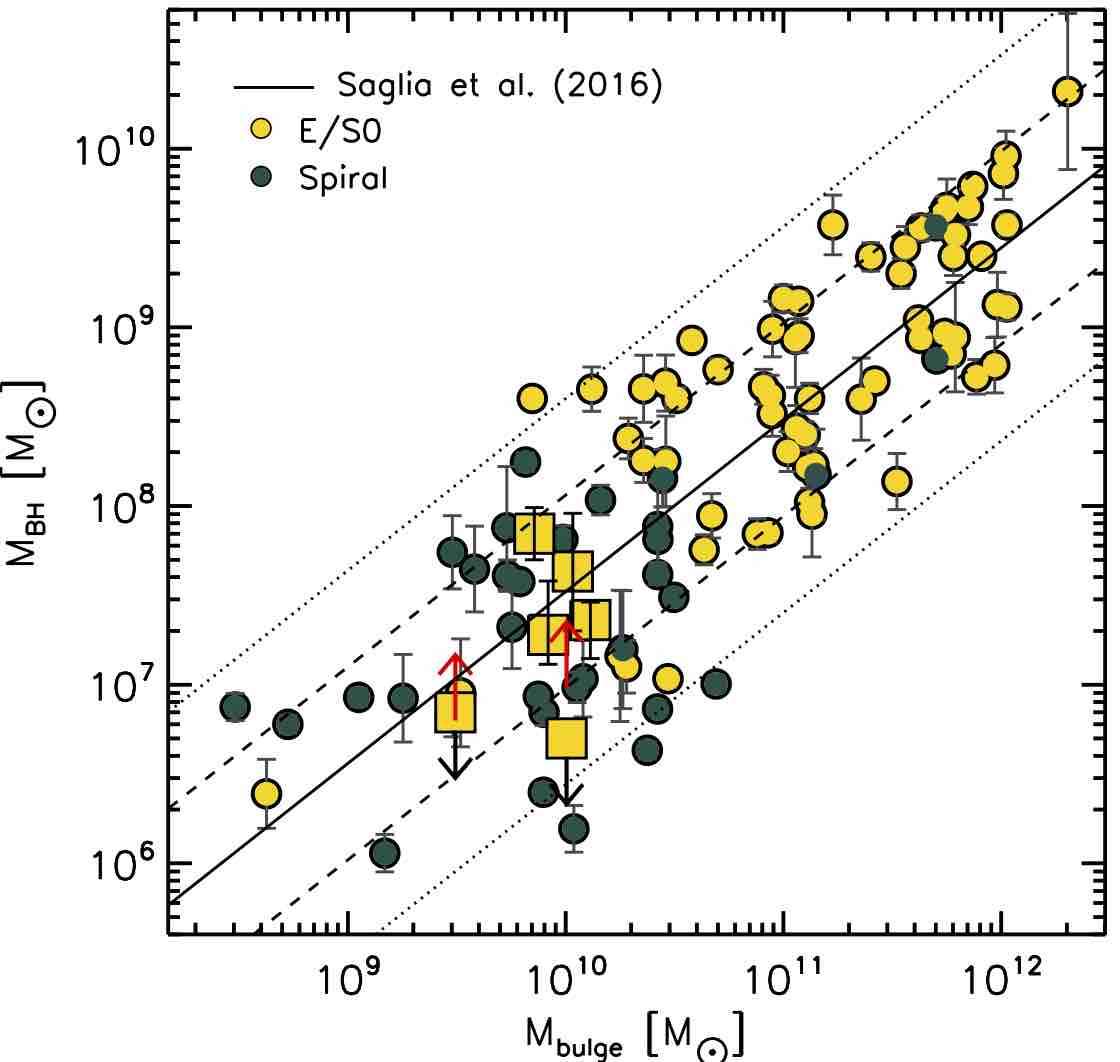}
\caption{{\bf Left:} black hole mass - velocity dispersion relation based on the compilation of galaxies with dynamical M$_{\rm BH}$ measurements from \citet{2016ApJ...818...47S}, divided by their morphological classification as shown on the legend (based on column `Type' from that paper). The solid line is the best-fitting relation from \citet{2016ApJ...831..134V}, and the dashed and dotted lines show the one and two $\sigma$ confidence zone. Galaxies from this study are shown with large squares or as upper limits (square with down pointing black arrows). The red upward pointing arrows indicate the lower limit on M$_{\rm BH}$ that could be expected to be measured at the resolution of our data assuming the SoI argument. Note how our secure mass estimates are all above the relation and in two cases just outside of the $2\sigma$ confidence zone, making them amongst the largest outliers from the relation. {\bf Right:} black hole mass - bulge mass relation based on \citet{2016ApJ...818...47S} compilation. The solid line is the best-fitting relation from that work, while the dashed and dotted lines show the 1$\sigma$ and 2 $\sigma$ confidence zone. Galaxies are divided by their morphology as shown on legend. In this case, our galaxies (filled squares) are within $1\sigma$ confidence zone and consistent with the relation, while the upper limits are below the relation. The likelihood that the galaxies with the upper limit mass estimates do not harbour black holes is discussed in Section~\ref{ss:nondetect}.}
\label{f:msigma}
\end{figure*}
%%%%%%%%%%%%%%%%%%%%%%%%%%%%%%%%%%%%%%%%%%%%%%%%%%%%%%%%%%%

\noindent{\it Possible influence of the dark matter.} \citet{2013MNRAS.432.1862C} and \citet{2017MNRAS.467.1397P} evaluated the dark matter content of our galaxies based on the SAURON kinematics. This was done through JAM dynamical models which had fitted only a total mass profile or a combination of the stellar and dark matter profiles. The results are generally consistent, and, specifically, for our galaxies provided the following fractions of dark matter within one half-radii: $f_{\rm DM}=(0.0, 0.0, 0.05, 0.0, 0.38, 0.23)$ \citep{2013MNRAS.432.1862C} and for $f_{\rm DM}=(0.0, 0.0, 0.22, 0.08, 0.48, 0.24)$ \citep[using Model I in][]{2017MNRAS.467.1397P} for NGC\,4339, NGC\,4434, NGC\,4474, NGC\,4551, NGC\,4578 and NGC\,4762, respectively. It is evident that most of galaxies have a negligible contribution of DM in the regions covered with our kinematics. NCG\,4578 and NGC\,4762 are somewhat different with possibly significant contribution of dark matter. In Section~\ref{ss:schw} we showed that our $M/L$ derived via Schwarzschild modelling are comparable with the JAM $M/L$ ratios of paper \citet{2013MNRAS.432.1709C}, therefore we can relate our results directly to the work of the two studies above. Comparing our M/L ratios with those of stellar populations published in \citep{2013MNRAS.432.1862C} and adjusted for the filter band difference, stellar population $M/L$ based on the \citet{1955ApJ...121..161S} initial mass functions (IMF) are larger than dynamical values we estimate. Therefore, a \citet{2002Sci...295...82K} or a \citet{2003PASP..115..763C} IMF based $M/L$ would be more physical, as expected given the trends between the IMF and the stellar velocity dispersion or mass of the galaxies \citep[e.g.][]{2010Natur.468..940V, 2011MNRAS.415..545T, 2012ApJ...760...71C, 2012Natur.484..485C,2014ApJ...792L..37M, 2014MNRAS.438.1483S}.

Our JAM (Section~\ref{ss:jam}) and Schwarzschild models (Section~\ref{ss:schw}) are based on data that probe very different scales. The JAM models, based only on NIFS data, probe only a few per cent of the half-light regions in our galaxies, while Schwarzschild models take the full extent of the SAURON data into account, which except for NGC\,4762 map the full half-light region of the sample galaxies. In all cases, and specifically for galaxies with possible significant fraction of dark matter, the models give very consistent results (Tabel~\ref{t:results}), in terms of the $M/L$ (less than 10 per cent difference), M$_{\rm BH}$ (within a factor of 2), and the orbital anisotropy (the latter point will be discussed in Section~\ref{ss:orb}). Specifically, the similarity of $M/L$ is rather surprising as they are constrained by very different radii. Both types of models, however, are able to reproduce the main features visible in the kinematics data, and by giving consistent solutions for the free parameters they raise the confidence in the results. For this reason, we do not explore further the influence of the dark matter on the dynamical models. 

\noindent {\it Variations in stellar populations.} \citet{2015MNRAS.448.3484M} presented the star formation histories of ATLAS$^{\rm 3D}$ galaxies and from these data one can derive the radial variation of the stellar population parameters and their mass-to-light ratio $M/L_\ast$. For our galaxies there is a mild radial decrease in $M/L$ of up to 30 per cent across the region covered by SAURON data \citep{2017MNRAS.467.1397P}\footnote{Maps of $M/L$ ratios for ATLAS$^{\rm 3D}$ galaxies are available on the project website: \href{http://purl.org/atlas3d}{http://purl.org/atlas3d}}. Our models do not take into account a variable M/L (see Section~\ref{ss:nondetect}), but we do not find a significant difference in derived $M/L$ from small scale JAM and full scale Schwarzschild models. The grids of Schwarzschild models show no or only a weak dependance between the M/L and M$_{\rm BH}$ (see $\chi^2$ contours in Fig.~\ref{f:SCHW_grids}), the strongest of which is for NGC\,4339. We therefore assume that the impact of the variable $M/L_\ast$ is minor to the final results of this work

\subsection{Additions to the scaling relations}
\label{ss:msigma}

We place our galaxies on the M$_{\rm BH} - \sigma$ and M$_{\rm BH} - $M$_{\rm bulge}$ (bulge mass) relations based on a recent compilation of black hole measurements \citep{2016ApJ...818...47S} in Fig.~\ref{f:msigma}. Looking at the M$_{\rm BH} - \sigma$ relation, there are two galaxies (NGC\,4339 and NGC\,4434) that are significantly above the relation (about $2\sigma$ away), while two (NGC\,4578, and NGC\,4762) lie on the relation. The two upper limits place the galaxies above the relation. Assuming that we could measure a black hole mass that has an SoI 3 times smaller than the resolution of the LGS kinematics, we could just detect black holes in those galaxies with masses within the scatter of the M$_{\rm BH} - \sigma$ scaling relation. 

The situation is somewhat different on the  M$_{\rm BH} - $M$_{\rm bulge}$ relation as all detections are now within the scatter of the relation, while the galaxies with the upper limits fall below the relation. We note that the position of our galaxies on this diagram is critically dependent on the estimates of the bulge mass. We use values from Table~\ref{t:sample}, derived from the total galaxy mass \citep{2013MNRAS.432.1862C} and the disc--bulge decomposition of \citet{2013MNRAS.432.1768K}. 

Kinematically our galaxies are all fast rotators, with regular, disc-like rotations. Morphologically, they seem to consist of a spheroid and an outer disc, which are easily recognizable only in the two galaxies seen at an inclination near 90\degree (NGC\,4474 and NGC\,4762). \citet{2006ApJS..164..334F} fit the HST/ACS surface brightness profiles of five of our galaxies (all except NGC\, 4339) with Sersic \citep{1968adga.book.....S} and core-S\'ersic \citep{2003AJ....125.2951G} profiles. As expected, none of these galaxies require the core-S\'ersic profile and the S\'ersic indices (n) of the single component S\'ersic fit range between $n=2.4$ for NGC\,4434 to $n=4.5$ for NGC\,4578. \citet{2013MNRAS.432.1768K} fit the surface brightness profiles of SDSS $r$-band images with a combination of S\'ersic and exponential profiles. Excluding the central 2.5\arcsec\,of the light profiles they are able to decompose all our galaxies with the indices of the Sersic component ranging between $n=1$ and 2, with the exception of NGC\,4551 ($n=4.2$). This range of indices suggest M$_{\rm BH} = 3\times10^6 - 3\times10^7$ M$_\odot$ \citep[based on the relation from][]{2016ApJ...821...88S}, which compares well with our mass estimates. If we decompose the HST images and include the central regions in the fit, this results in somewhat higher S\'ersic indices for the bulge component, in the range of 2.3 - 4.2. For these S\'ersic indices the predicted black hole masses are between $4\times10^7 - 4\times10^8$ M$_\odot$, approximately consistent, but somewhat higher than our results. 

Two of our galaxies have black holes that are more than $2\sigma$ more massive than predicted by the relation. While these are amongst the largest outliers above the relation, the fact that they exist seems to be consistent with the general behaviour of galaxies around $\sigma_e=100$ km/s. \citet{2016ApJ...826L..32G} find that L$^{*}$ galaxies with mega-masers have a large range of black hole masses and that no single galaxy property correlates with M$_{\rm BH}$ closely. In our case, the two outliers are fast rotator ETGs, with no strong nuclear activity \citep[all galaxies except NGC4339 were observed in the AMUSE X-ray survey][]{2010ApJ...714...25G}, and old stellar populations \citep{2015MNRAS.448.3484M}. These galaxies could simply represent the cases of galaxies which had more efficient feeding histories of their black holes, compared with other galaxies of the same mass range \citep{2013ARA&A..51..511K}. An interesting possibility is that these galaxies assembled around direct-collapse black holes that preceded the formation of the stellar component \citep{2013MNRAS.432.3438A}, and have started their evolution with a more massive black hole.  All our galaxies have similar and relatively high bulge S\'ersic indices and the difference in black holes masses are, therefore, due to specific evolutionary paths of the black holes themselves. It is likely that the duration of the black hole feeding differed, perhaps due to the availability of the gas. This last point might be related to the environment, as most of our galaxies are Virgo Cluster members. NGC\,4434, the largest outlier is however an isolated object and not a member of Virgo. The two galaxies with upper mass limits are discussed in more detail in Section~\ref{ss:nondetect}. 

As our galaxies populate lower ranges of the M$_{\rm BH}$ scaling relations they are also important additions when considering the two black hole -- galaxy growth channels as characterized in \citet{2017arXiv170704274K}. Their addition improves the statistics among moderately large, but low mass galaxies with velocity dispersion around 100 km/s, confirming the claim that for low galaxy masses M$_{\rm BH}$ follows the same trend with velocity dispersion as seen for galaxy properties related to stellar populations, like age, metallicity, alpha enhancement, $M/L$, and gas content.

%%%%%%%%%%%%%%%%%%%%%%%%%%%%%%%%%%%%%%%%%%%%%%%%%%%%%%%%%%%
\begin{figure*}
%Fig made by start_moments_LGSBH.pro
\includegraphics[width=\textwidth]{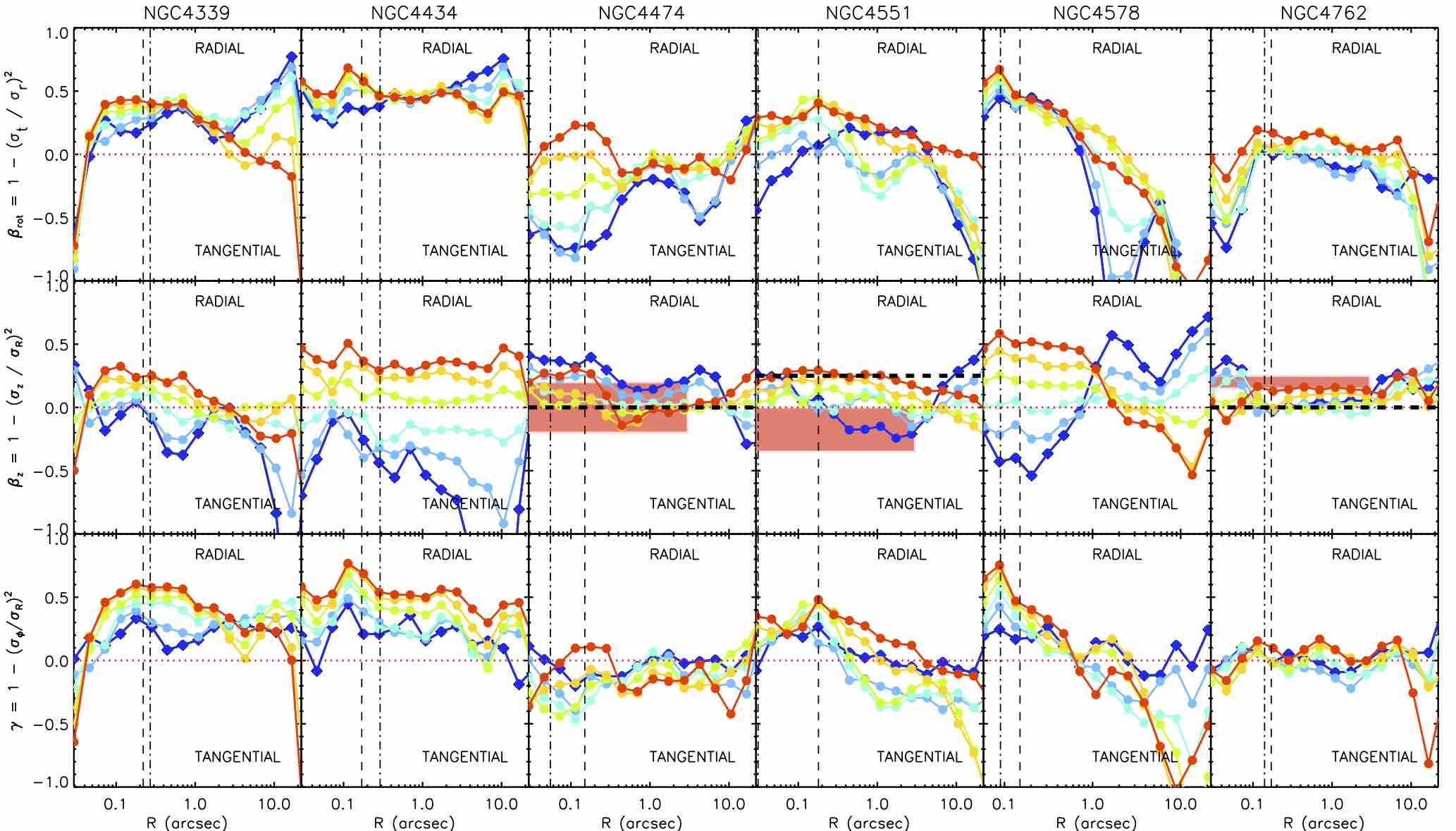}
\caption{Radial profiles of the velocity anisotropy characterizing the best-fittin Schwarzschild dynamical models. Each galaxy is in a column as indicated by names. {\bf Top:} velocity anisotropy $\beta_{rot}$ parameterized in the spherical coordinate system. {\bf Middle:}  velocity anisotropy $\beta_z$ parameterised in the cylindrical co-ordinate system relating the radial and vertical axes (definition is equivalent as in JAM models). {\bf Bottom:} velocity anisotropy $\gamma$ parameterized in the cylindrical coordinate system relating the coordinates of the equatorial plane. In all cases the velocity anisotropy was measured at different polar angles defined in the meridional plane, spanning from 15\degree (close to the equatorial plane, shown with the red line) to 75\degree (close to the symmetry axis, shown with the blue line). The change in colours show the increase of the angle. Dashed vertical lines indicate the achieved resolution with the LGS AO (the FWHM of the narrow component of the PSF) and the dot--dashed vertical lines show the radii of the black hole SoIs using the best fitting M$_{\rm BH}$ (for NGC\,4473 and NGC\,4551 we use the upper limits) from Table~\ref{t:results}. The shaded regions in the middle panels show the range of anisotropies estimated by JAM models for galaxies seen at inclination larger than 60\degr. Thick dashed black line shows JAM $\beta_z$ from \citet{2013MNRAS.432.1709C} constrained by the SAURON data only.
} 
\label{f:anis}
\end{figure*}
%%%%%%%%%%%%%%%%%%%%%%%%%%%%%%%%%%%%%%%%%%%%%%%%%%%%%%%%%%%

\subsection{Orbital structure and the shape of the velocity ellipsoid}
\label{ss:orb}

The internal velocity structure of galaxies is specified by the velocity dispersion tensor, which defines the spread in stellar velocities at every position. Since the velocity dispersion tensor is symmetric, it is possible to choose a set of orthogonal axes in which the tensor is diagonal. These diagonalized coordinate axes specify the velocity ellipsoid. The shape of the velocity ellipsoid, defined by the velocity dispersion along the principle axes, is a useful tool for describing the velocity structure of a system. An `isotropic' system has all velocity dispersions equal, while an `anisotropic' system is a system where the velocity dispersions along the principle axes mutually differ \citep{2008gady.book.....B}. The nature of the Schwarzschild modelling allows for a general investigation of the velocity ellipsoid, as a function of radius and height from the equatorial plane. 

We estimate three anisotropy parameters, one in the spherical ($r$, $\theta$, $\phi$) and two in cylindrical ($R$, $\phi$, $z$) coordinate systems. First, we specify the ratio of radial to tangential velocity dispersion as $\beta_{rot} =1 - (\sigma_t/\sigma_r)^2$, where $\sigma_t = (\sigma_\theta^2 + \sigma_\phi^2)/2$, in spherical coordinates. In this case, the isotropic systems will have $\beta_{rot}=0$, while radially and tangentially anisotropic systems have $\beta_{rot}$ positive or negative, respectively. This is a similar measure to that used in other works \citep[e.g.][]{2003ApJ...583...92G, 2007MNRAS.379..418C,2009ApJ...695.1577G, 2009MNRAS.399.1839K,2014ApJ...782...39T}, but it is best suited for spherical objects. The second anisotropy parameter is $\beta_z = 1- (\sigma_z/\sigma_R)^2$ as in the JAM models. It defines the shape of the velocity ellipsoid in the meridional plane including its two symmetry axes (major and minor), assuming axisymmetry. An additional anisotropy parameter is $\gamma = 1-(\sigma_{\phi}/\sigma_R)^2$, which describes the equatorial plane of the velocity ellipsoid. When $\beta_z=0$ or $\gamma=0$ the cross-sections of the velocity ellipsoids are circles and we speak of an isotropic system. When  $\beta_z>0$ or $\gamma>0$ the orbital distribution is biased towards a radially anisotropy, and when  $\beta_z<0$ or $\gamma<0$ it is biased towards tangential anisotropy.

We show results for our galaxies in Fig~\ref{f:anis}, specifically for models with formally best-fitting black hole masses (for NGC\,4474 and NGC\,4551 we use the models with the upper limit black hole masses). While there are certain differences between the galaxies, a general conclusion is that models have nearly isotropic or mildly radially biased velocity ellipsoid throughout their bodies. In spherical coordinates, NGC\,4339 and NGC\,4434 seem to be somewhat more radially biased, while the velocity ellipsoid of NGC\,4578 has a more complex structure, changing the shape along the major axis from a radially biased within 1\arcsec\, to a tangentially biased shape beyond that radius. In more appropriate cylindrical coordinate system, essentially all galaxies are nearly isotropic, with NGC\,4339, NGC\,4434 and NGC\,4578 showing a larger scatter in anisotropy across different polar angles defined in the meridional plane. Galaxies with the M$_{\rm BH}$ upper limits show a tendency for radial anisotropy in the central regions. The same features can also be seen in the models of these galaxies assuming M$_{\rm BH}$ below the upper limits

The Schwarzschild models of the three galaxies at low inclinations (NGC\,4339, NGC\,4434 and NGC\,4578) show a larger spread in recovered radial anisotropy profiles, while models of galaxies at higher inclinations have less scatter. We do not investigate this in detail, but the effects are likely related to the difficulties in recovering the true properties of systems at low inclinations, as Schwarzschild models also suffer from the inclination--anisotropy degeneracy. We also compare the anisotropy parameter $\beta_z$ obtained from JAM and Schwarzschild models for galaxies with inclination larger than 60\degr. Galaxies with lower inclinations are in the regime of a strong degeneracy between inclination and anisotropy \citep{2012MNRAS.424.1495L}, therefore, we compare only galaxies at higher inclinations. We compare both the results of our JAM models which are restricted to the central $3\arcsec\times3\arcsec$ and JAM models\footnote{Anisotropy values for the full ATLAS$^{\rm 3D}$ sample are available on \href{http://purl.org/atlas3d}{http://purl.org/atlas3d}.} of \citet{2013MNRAS.432.1709C} which sample the full SAURON FoV. For both sets of models there is a rather good comparison with Schwarzschild results, except for NGC4551, for which the JAM models based on NIFS data only favour middy tangentially anisotropic solutions, but still consistent with isotropy. Schawarzschild models favour a more radially anisotropic solution, similar to the JAM models based on SAURON data. We note that this galaxy is seen at a lower inclination ($i=65$\degr), which could also be a source of error in determination of the anisotropy. 

For galaxies with firm black hole mass detections, we do not observe significant changes in the anisotropy parameters as one crosses the black hole SoI (dot--dashed line); the velocity anisotropy profiles remain mostly isotropic or mildly radially anisotropic. A mild change in anisotropy from isotropic to radial anisotropic, or a bias to radial anisotropy, is detectable in NGC\,4474 and NGC\,4551, galaxies with upper limits on black hole mass. We will come back to this issue in Section~\ref{ss:nondetect}. 

The lack of tangentially biased orbits in the region dominated by the gravitational potential of the black holes is consistent with the picture in which the tangential anisotropy is the consequence of the core scouring by interaction of binary black holes \citep{1995ApJ...440..554Q,1997NewA....2..533Q,2001ApJ...563...34M, 2014ApJ...782...39T}. Such behaviour is expected for more massive galaxies and slow rotators in general which grow via relatively frequent dry merging \citep[see review by][]{2016ARA&A..54..597C}. The photometric shapes, the kinematics and the masses can rule out this kind of event in the mass assembly for our galaxies, and now we can also add the shapes of the velocity ellipsoids as evidence that the growth of the galaxies and their central black holes was not dominated by significant dissipation-less merging. 

The evidence for an isotropic to mildly radially anisotropic orbital distribution seems, however, to be contrary to expectations for adiabatic growth of black holes or creation of central stellar disc from nuclear starbursts \citep[e.g.][]{1995ApJ...440..554Q}. The predictions are however done for a spherical system, and if one considers the $\beta_{rot}$ parametrization for spherical coordinates (which might be more applicable for the vicinity of the black hole), our results are compatible with mild tangential anisotropy as predicted and found for other galaxies \citep[e.g. see a compilation of][]{2014ApJ...782...39T}.

\subsection{Are there super-massive black holes in NGC\,4474 and NGC\,4551?}
\label{ss:nondetect}

The two galaxies for which we constrain M$_{\rm BH}$ to be smaller than we can resolve have unusual central dips in the velocity dispersion maps. Their radius is approximately 0.5\arcsec\, for NGC\,4474,  and about 1\arcsec\, for NGC\,4551, or about 35 and 80 pc, respectively. These features have a crucial influence in the construction of the dynamical models. A black hole of $10^7$ M$_\odot$ and a $\sigma_e$ of 100 km/s at the distance of 16 Mpc (like for these two galaxies), would have an SoI of about 0.06\arcsec, which could, in principle, be detected, given our data resolution and the experience with similar data types and quality \citep[both high resolution and large scale IFU data,][]{2009MNRAS.399.1839K, 2010AIPC.1240..211C, 2017A&A...597A..18T}. Fig.~\ref{f:sigma_comp}, however, shows that this is not the case. For both galaxies it can be ruled out that black holes of such a mass are present. Furthermore, the projected angular size of the velocity dispersion dips are a factor of 10-20 larger than radius of the SoI for a $10^7$ M$_\odot$ black hole. What are then the origins of central decreases in the velocity dispersion maps?

%%%%%%%%%%%%%%%%%%%%%%%%%%%%%%%%%%%%%%%%%%%%%%%%%%%%%%%%%%%
\begin{figure}
%Fig made by show_integral_space_polar_ngc4474.pro and show_integral_space_polar_ngc4551.pro
\includegraphics[width=0.95\columnwidth]{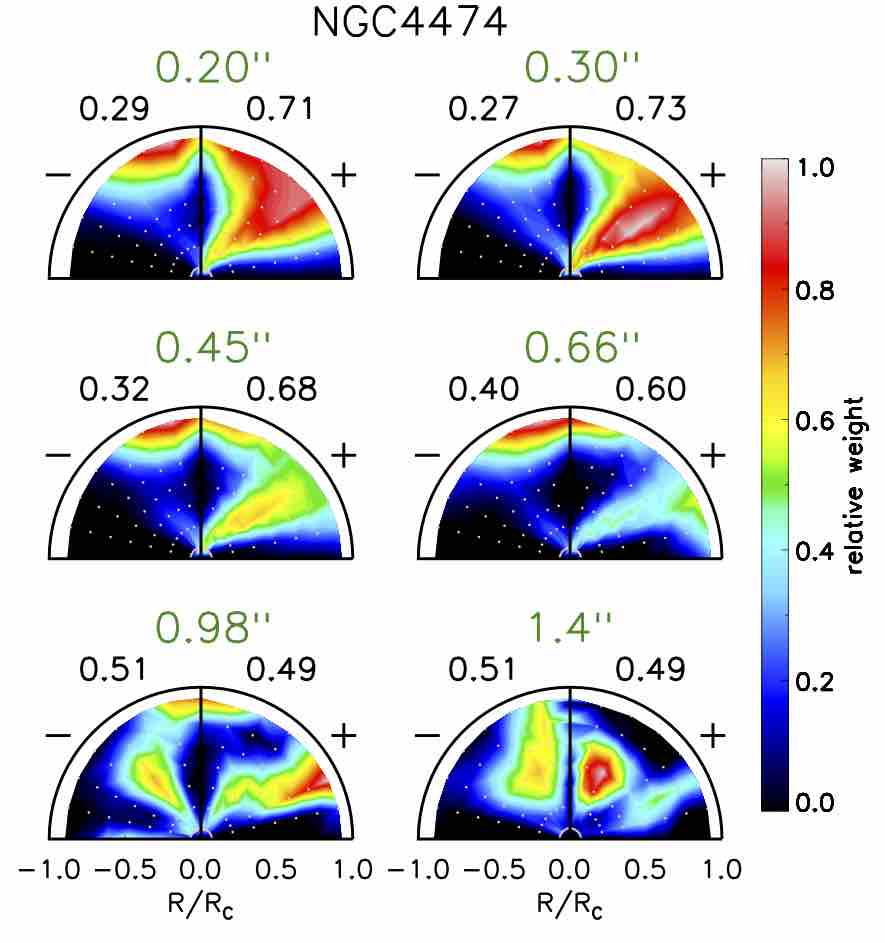}
\includegraphics[width=0.95\columnwidth]{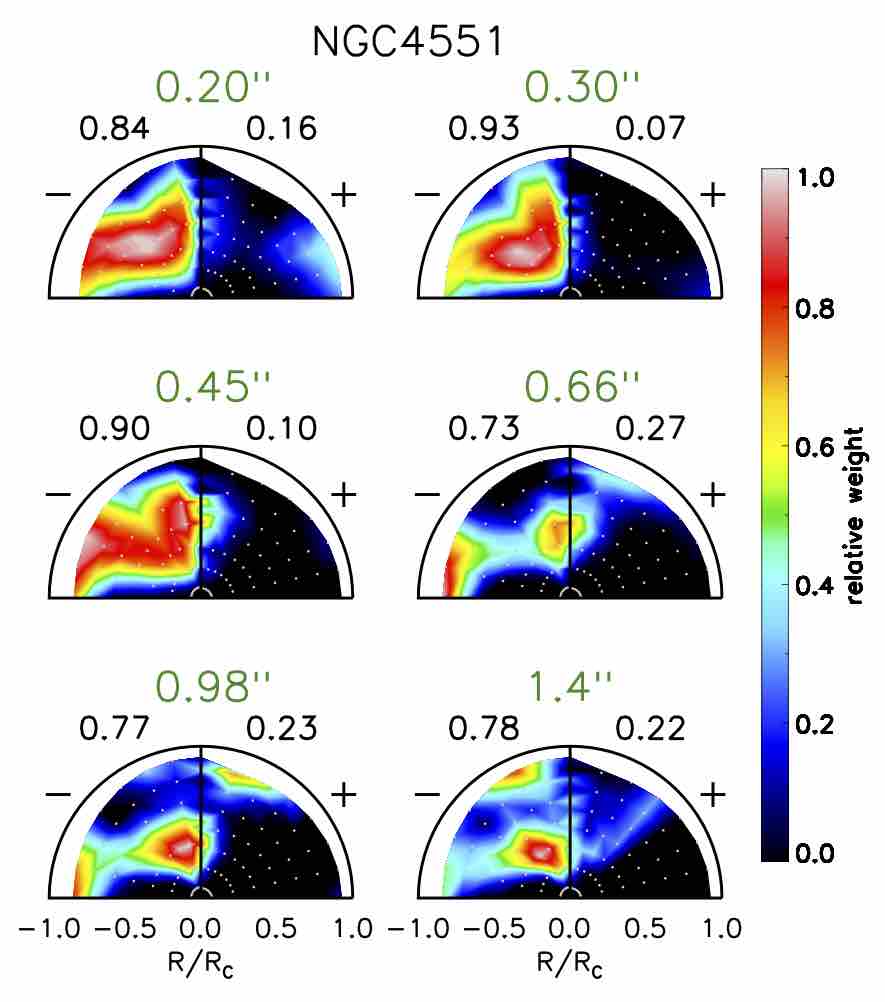}
\caption{Integral space of the best-fitting Schwarzschild dynamical models for NGC\,4474 (top three rows) and NGC\,4551 (bottom three rows). Each semicircle panel plots the meridional plane ($R,z$) for a given energy specified by the radius of the circular orbit in arcseconds (large green numbers above each panel). The starting positions of orbits are shown with white dots, and as orbits can be prograde or retrograde, each panels have a `+' and `--' side showing respective orbits. The horizontal axis specifies the radius of each orbit as a fraction of the radius of the circular orbit at that energy ($R/R_c$). Orbits with the highest angular momenta are found close to $R/R_c=1$, at corners of the semicircles. The coloured contours are the mass fractions assigned to orbits at that given energy. The numbers above each half semi-circler are percentages of the total mass assigned to prograde (above the `+' sign) and retrograde (above the `--' sign) orbits. We show only the subset of orbits covering the central 1.5\arcsec, but excluding the region dominated by the black holes. Note the lack of significant clustering of orbits at the corners of the semicircles.}
\label{f:int_space}
\end{figure}
%%%%%%%%%%%%%%%%%%%%%%%%%%%%%%%%%%%%%%%%%%%%%%%%%%%%%%%%%%%

One possibility is that the dips in the velocity dispersion maps are caused by a specific stellar structure making a nuclear sub-component. The two galaxies are seen at different inclinations, NGC\,4474 is almost edge on, and it has a prominent large-scale stellar disc. NGC\,4551 is, however, at a moderate inclination of 65\degree, which makes it difficult to recognize a large-scale disc morphology. \citet{2006ApJS..164..334F}, based on a sharp change in the `disciness' parameter, $b_4$, identified a nuclear disc in NGC\,4551 seen at scales below 0.1\arcsec, but reported no such a structure in the case for NGC\,4474, even though there is a similar behaviour of its $b_4$. Even if the photometric feature in NGC\,4551 on these scales is really confirmed as a disc, its size is smaller compared to the structure on the NIFS velocity dispersion maps, and it is difficult to make a direct link. Neither of the galaxies have noticeable dust, either filamentary or confined to a plane. Ellipticity of NGC\,4474 is 0.15 at about 1\arcsec\, and increases with a peak of 0.35 at 0.5\arcsec, before dropping to 0.13 in the centre. NGC\,4551 is different in the sense that its ellipticity is relatively high (0.35) between 0.2 and 2\arcsec, but also drops to 0.2 in the centre. The rapid change of ellipticity may indicate the existence of components seen at different inclinations than the rest of the galaxy, but in the two cases the changes in ellipticity are quantitatively different while producing the same kinematic signature. Both galaxies have a central excess of light over a S\'ersic fit to the full light radial profile, which for NGC\,4474 is visible at radii smaller than 0.15\arcsec, while for NGC\,4551 already at about 0.5\arcsec. These excesses are probably the only stellar nuclear structures which could be related to the observed dips in the kinematics. 

Such structures may still correspond to stars that were once arranged in disc, if one considers the impact that minor merger events have on nuclear discs. Indeed, whereas such events easily disrupt the thin structure of discs such that they would no longer be recognized photometrically, the kinematic signature of the disc stars would survive such event, although the remnant disc may not be as dynamically cold as it originally was \citep{2015MNRAS.453.1070S}.

The total gravitational potential is traced by the second velocity moment, adequately approximated by a combination of the velocity dispersion and the mean rotational velocity. Therefore, in the case of a stellar disc,  V$_{rms}$ (or the second velocity moment), should not show a specific central feature, because the (increased) velocity and (decreased) velocity dispersions compensate. V$_{rms}$ maps in Fig.~\ref{f:jam_comp} clearly show that this is not the case: dips in the V$_{rms}$ maps are present and very similar to those seen in the velocity dispersion maps. Therefore, the morphology of the V$_{rms}$ maps does not support the assumption that the kinematic dips are made by dynamically cold stellar structures such as nuclear discs. 

In Fig.~\ref{f:int_space} we show the distribution of orbits in the central 1.5\arcsec\, of both galaxies, covering larger regions than those displaying the velocity dispersion dips. We selected the formally best-fitting Schwarzschild models (with M$_{\rm BH}$ of $4.1\times10^5$ and $3.8\times10^5$ M$_\odot$ for NGC\,4474 and NGC\,4551 respectively), but orbital distributions are quantitatively the same for all models with black hole masses less than the derived upper limits. Each semicircle in Fig.~\ref{f:int_space} represents orbits of a given energy $E$, while the other two integrals of motions defining an orbit are traced within the semicircle: the orbits of the highest $L_z$ are at the radius of the circular orbit ($R_c$), while the orbits with increasing $I_3$ are increasingly more distant from the horizontal axis. Therefore, at a given energy, high angular momentum orbits, which are flat and build discs, are found in the corners of the semicircles, while orbits with high third integral are at the top of the semicircles, and build more spherical structures. 

A striking feature of panels on Fig.~\ref{f:int_space}, for both galaxies, is the lack of high angular momentum orbits confined to the disc plane. Orbits populating a disc should have a low third integral and be found close to the location of the circular orbit. To some extent these orbits are present in NGC\,4551 around 0.5\arcsec\,(seen on third and fourth hemispheres on Fig.~\ref{f:int_space}) from the centre of the galaxy, but they do not extend closer to the assumed location of the black hole. The reason for this is seen on the V$_{rms}$ map (Fig.~\ref{f:jam_comp}), which shows increased values along the major axis at that radius (0.5\arcsec), and is evident on the maps of the mean velocity (Fig.~\ref{f:maps}) as the end points of the outer disc. 

%%%%%%%%%%%%%%%%%%%%%%%%%%%%%%%%%%%%%%%%%%%%%%%%%%%%%%%%%%%
\begin{figure}
%Fig testing_NGC4474_NGC4551_jam_models
\includegraphics[width=\columnwidth]{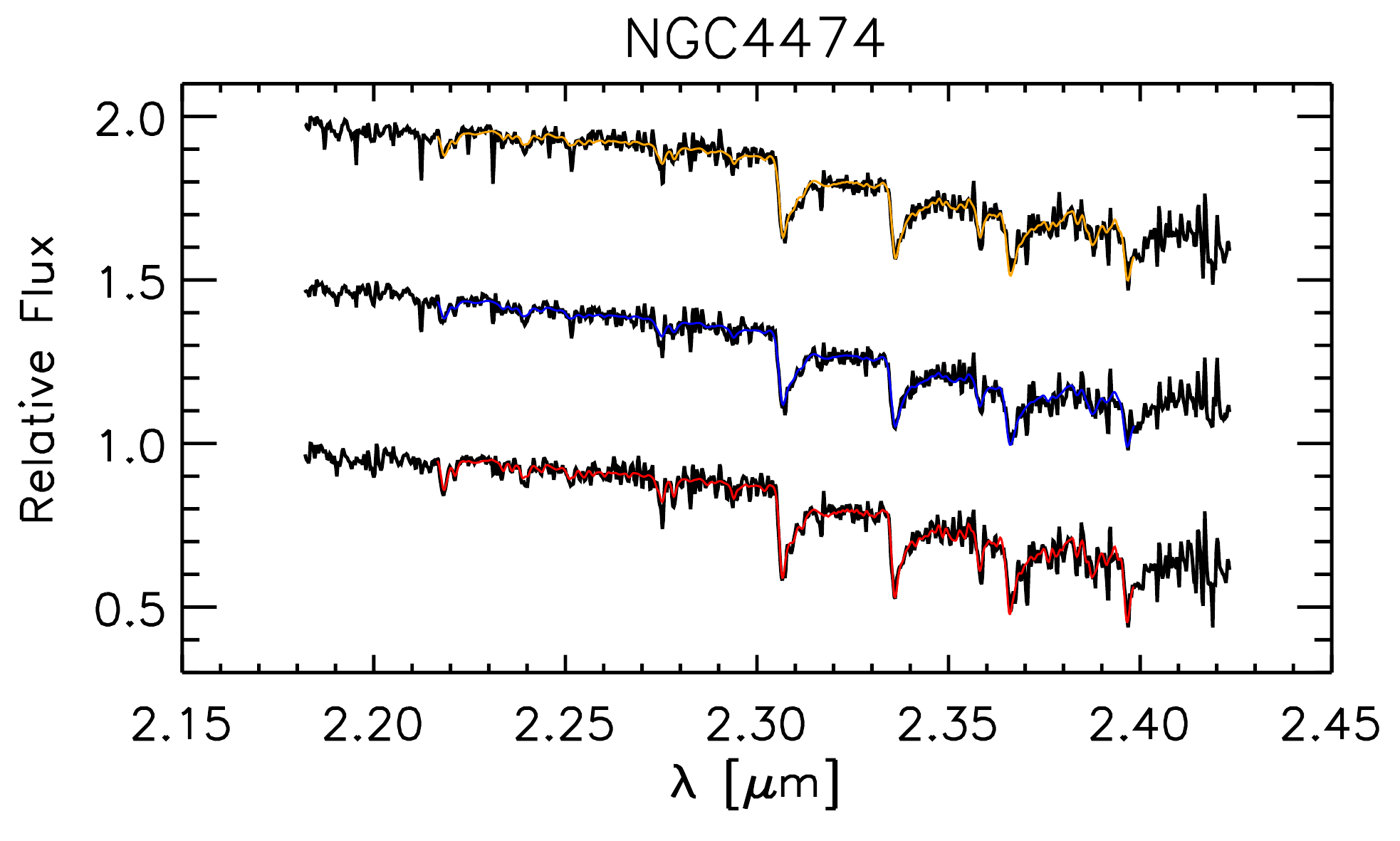}
\includegraphics[width=\columnwidth]{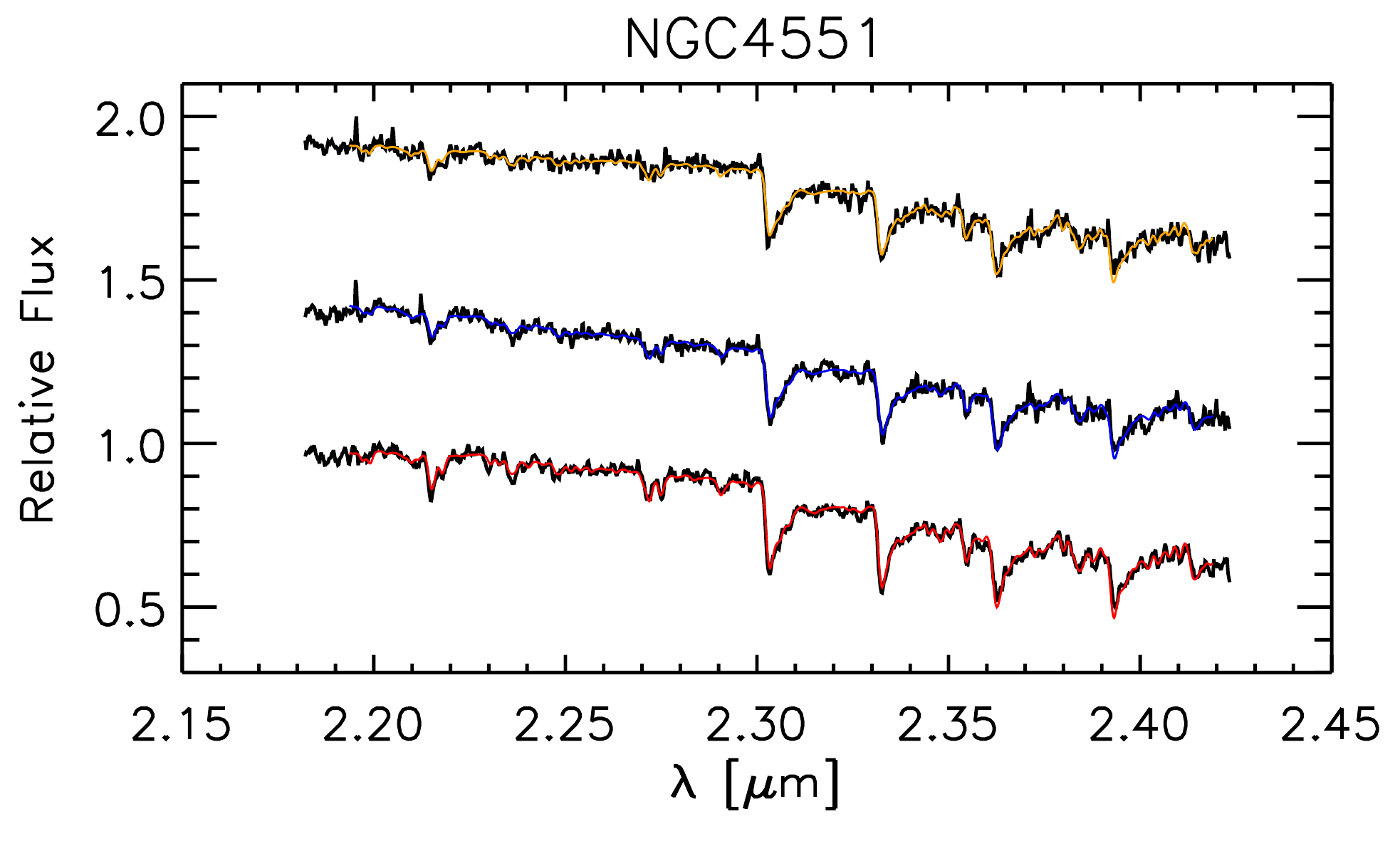}
\caption{Spectra of NGC\,4474 (top) and NGC\,4551 (bottom) at three different locations. In each panel, the bottom spectrum is from the nucleus (within the drop seen in the velocity dispersion maps), the middle spectrum is from location centred on (0.7, 0.7\arcsec), and the top spectrum is centred on (-0.7,-0.7\arcsec). Coloured lines are pPXF fits to the spectra. The central spectrum combines $3\times3$ pixels, while the other spectra are obtained from an aperture of 0.2\arcsec radius. In this way, the spectra have comparable S/rN after the pPXF fit. Note that there are only minor difference in the strengths of NaI and CaI lines.}
\label{f:spec_com}
\end{figure}
%%%%%%%%%%%%%%%%%%%%%%%%%%%%%%%%%%%%%%%%%%%%%%%%%%%%%%%%%%%

Based on stellar populations within the SAURON FoV, NGC\,4474 and NGC\,4551 do not seem to be significantly different from other galaxies in the sample. There is evidence for an intermediate age stellar population  ($>5$ Gyr) within central 2--3\arcsec, but otherwise they have negative gradients in metallicity and $M/L$ ratios similar to those of other galaxies \citep{2015IAUS..311...53K, 2015MNRAS.448.3484M, 2017MNRAS.467.1397P}. The kinematic dips are seen only on the NIFS data and confined to the central 0.5\arcsec, a region fully within a fraction of the SAURON central pixel. Characterizing the stellar populations from NIR spectra is, however, more challenging \citep{2008ApJ...674..194S, 2008A&A...485..425L}. As a test we selected three different regions in both galaxies: a central one encompassing $3\times3$ central pixels, and two diametrically opposite regions centred on (0.7, 0.7\arcsec) and (-0.7, -0.7\arcsec) and approximately 0.2\arcsec\, in radius. In this way all regions have similar S/rN after the pPXF fit.  The spectra are shown on Fig.~\ref{f:spec_com}. We note that both NaI (2.20 $\mu$m) and CaI (2.26 $\mu$m) features seem to be deeper in the central spectra (for both galaxies), but we do not perform a detailed stellar population analysis. Instead, we fit the spectra with pPXF and record the templates \citep[from GNIRS NIR stellar library of][]{2009ApJS..185..186W} needed for the fit. For the central spectrum of NGC\,4474 the fit required stars of the following types: F7III, G7III, K1V and K2III stars. The other two regions required these same stars and in addition G9III and K0IV stars. For the central spectrum of NGC\,4551 the fit required stars of the following types: F7III, G8V, K1V and K2III stars. The other two regions required these same stars and in addition a G7III star. For both galaxies, pPXF fits are of the same quality and are able to reproduce all absorption features. Therefore, we conclude that if there are differences between the stellar population within and outside of the regions with kinematic dips, they are not significant for the observed stellar kinematics.

%%%%%%%%%%%%%%%%%%%%%%%%%%%%%%%%%%%%%%%%%%%%%%%%%%%%%%%%%%%
\begin{figure}
%Fig testing_NGC4474_NGC4551_jam_models
\includegraphics[width=\columnwidth]{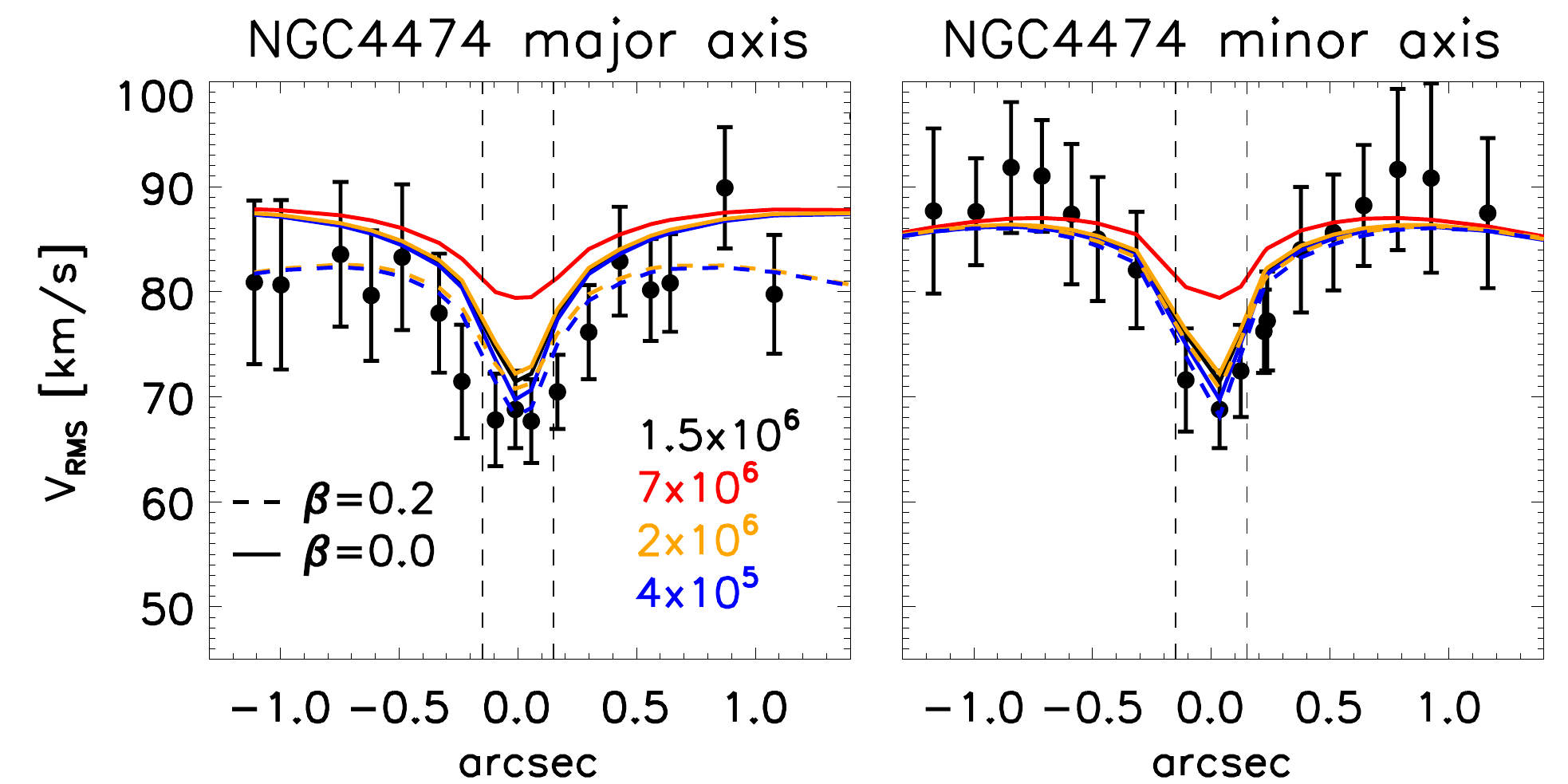}
\includegraphics[width=\columnwidth]{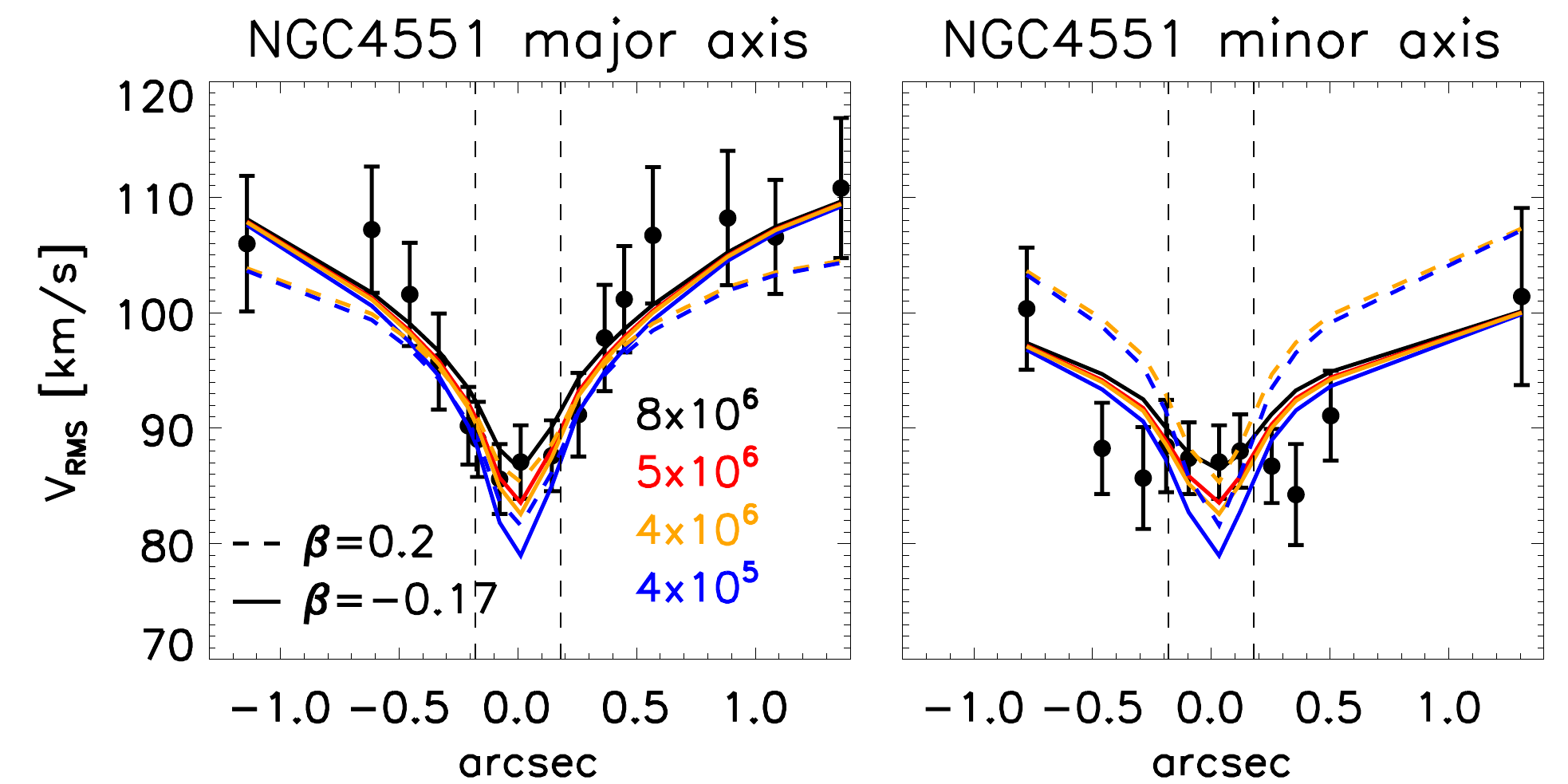}
\caption{Comparison between the observed $V_{\rm rms}$ and JAM models along the major (left) and minor (right) axes for NGC\,4474 and NGC\,4551. JAM models have the same inclination, $M/L$ and anisotropy as the best-fitting JAM models from Section~\ref{ss:jam}, while lines of different colours show models with M$_{\rm BH}$ as specified in the legends, in units of solar masses. These are (from top to bottom of the list: the upper limit from JAM modelling, the upper limit from Schwarzschild modelling, the M$_{\rm BH} - \sigma$  relation prediction, and an order of magnitude smaller value, which approximately corresponds to the formal best fit from the Schwarzschild models. Models with dashed lines correspond to non-optimal anisotropies of $\beta_{\rm JAM}=0.2$ for both galaxies and for models with the predicted and under massive black holes, bracketing the solution space. Vertical dashed lines indicate the FWHM of the PSF.}
\label{f:jam_fake}
\end{figure}
%%%%%%%%%%%%%%%%%%%%%%%%%%%%%%%%%%%%%%%%%%%%%%%%%%%%%%%%%%%

Even though we are not able to use NIR spectra to determine details of the stellar population properties within the velocity dispersion drops, we can consider what would be an effect of variable stellar populations within the NIFS FoV. For example, if IMF changes rapidly from a dwarf-rich (Salpeter) in the centre to a dwarf-poor (e.g. Chabrier) at the edge of the NIFS FoV, the M/L ratio of the stellar populations will drop by 0.25 dex \citep[e.g.][]{2010MNRAS.404.2087B}. Our dynamical models assumed a constant M/L, where such as change in stellar populations M/L would be distributed between the dynamical M/L and M$_{\rm BH}$. Therefore, we run two tests using JAM models with varying M/L$_{pops}$ ratios for both NGC\,4474 and NGC\,4551. In the first test, we assumed that the M/L of the stellar populations ($M/L_{pops}$) decreases from a M/L based on dwarf-rich IMF in the centre to a $M/L$ based on dwarf-poor IMF at the edge of the NIFS FoV. In the second test we assumed an inverse change to $M/L_{pops}$, where the central $M/L$ was based on dwarf-poor IMF and the $M/L$ at the edge was based on dwarf-rich IMF. We introduced the change in the $M/L_{pops}$ by fixing the $M/L_{pops}$ based on Salpeter IMF (dwarf rich) to values from \citep{2013MNRAS.432.1862C}, converted to our bands, and the $M/L_{pops}$ based on Chabrier by subtracting 0.25 dex. The M/L$_{pops}$ ratios changed between these two limiting values logarithmically for each Gaussian of the MGE model which cover the extent of the NIFS FoV. 

We rerun grids of JAM models constrained by the V$_{rms}$ from the NIFS data. For NGC\,4474 there was no change in the estimate of M$_{\rm BH}$, only the upper limits changed. In the case of the dwarf-poor central IMF the M$_{\rm BH}$ upper limit was higher than that obtained for constant $M/L$ in Section~\ref{ss:jam} by a factor of 1.4. Conversely, in the case of dwarf-rich central IMF the upper limit was smaller by a factor of 1.6 than the nominal one. Essentially, the $\chi^2$ grids were the same as in Fig.~\ref{f:jams}, with just a horizontal shift of the contours. We found similar results for NGC\,4551, where the Fig.~\ref{f:jams} grid already indicated a possible $1\sigma$ level detection of M$_{\rm BH}$. For dwarf-poor central IMF  the models recovered an M$_{\rm BH}=1.2\times10^7 \pm 0.5$ M$_\odot$. Models based on the dwarf-rich central IMF, however, constrained only an upper limit of M$_{\rm BH}<0.2\times10^7$ M$_\odot$. In this case again the difference from the nominal models were from a factor of 1.5 to a factor of 4.

These results should take into account that recent studies \citep[e.g.][]{2013MNRAS.428.3183D,2016MNRAS.457.1468L,2017ApJ...841...68V} are finding evidence for a change in IMF which corresponds to dwarf-rich in the centre and dwarf-poor in outskirts, if any changes can be detected \citep[e.g.][]{2016ApJ...821...39M}. The consequence is that  black holes in NGC\,4474 and NGC\,4551, if they are present, would be even less massive than what we constrain with our nominal models using a constant $M/L$, while the M$_{\rm BH}$ estimate based on the imposed lower central $M/L$ (due to Chabrier IMF) is not likely. Furthermore, these varying $M/L$ ratio tests indicate a level of the systematics on the estimate of M$_{\rm BH}$ which is typically not taken into account; it seems every M$_{\rm BH}$ should be considered uncertain to a factor of 1.5-2.

The absence of clear structural features (e.g. nuclear discs), which could support the central decreases in the velocity dispersion and V$_{rms}$ maps, suggest that the origin of the dips is at least partially linked to the non-detection of the black holes. As shown by other galaxies in the sample, black holes exert influence on larger radii than the radii of their SoIs. Using the M$_{\rm BH} - \sigma$ scaling relation \citep[e.g.][]{2016ApJ...818...47S}, the predicted masses of black holes for NGC\,4474 and NGC\,4551 are $\sim2\times10^6$ and $\sim4\times10^6$ M$_\odot$, respectively. Such black holes would have R$_{\rm sph} \approx 0.017$\arcsec\, and 0.025\arcsec, respectively, a factor 7-8 times smaller than the achieved resolution. In Fig.~\ref{f:jam_fake} we compare JAM models of various M$_{\rm BH}$ with the data extracted along the major and minor axes. The models are not able to reproduce all points along both major and minor axes at the same time, but essentially all models are able to fit the dips in $V_{\rm rms}$. By changing the anisotropy parameter, one can achieve locally better fits, but they are still not able to discriminate between various black hole models. NGC\,4474 and NGC\,4551 data have the lowest S/N and there are evidences of possible template mismatch. Given the uncertainties on the kinematics and the limited resolution achieved with the AO observations, only an upper limit can be given to black hole masses in these galaxies. 

NGC\,4474 and NGC\,4551 have X-ray upper limits of $10^{38.5}$ and $10^{38.3}$ erg s$^{-1}$, respectively \citep{2010ApJ...714...25G}. These galaxies lack warm ionized gas in their nuclei and NGC\,4551 is also not detected in the radio continuum with an upper limit of $10^{18.61}$ W Hz$^{-1}$ at 5 GHz \citep{2016MNRAS.458.2221N}. These black holes are dormant, and it is highly speculative to estimate their black hole masses from X-ray and radio observations of the nuclei. Using \citet{2009ApJ...706..404G} relation between radio and X-ray flux and black hole mass, for NGC\,4551 upper limits, we obtain M$_{\rm BH}<2-3\times10^{7}$ M$_\odot$, about an order of magnitude larger than allowed by the models. The uncertainly on the accretion level prohibits a more robust estimate.

It is nevertheless interesting to consider a possibility that NGC\,4474 and NGC\,4551 are candidates for galaxies with significantly undermassive or no central black holes, joining a few more objects with stringent M$_{\rm BH}$ upper limits below what is predicted by scaling relations, based on either unresolved or resolved nuclear ionised-gas, or stellar kinematics \citep[e.g.][]{2006MNRAS.366.1050C, 2009ApJ...692..856B,2011ApJ...741...38G}.

If NGC\,4474 and NGC\,4551 have undermassive black holes, they are representative of galaxies where black holes did not co-evolve with hosts, perhaps due to low level and sporadic feeding \citep{2013ARA&A..51..511K}. Among other scenarios, NGC\,4474 and NGC\,4551 might have never had black holes, or they lost them during one of the past merging events, when the black holes were kicked out of the galaxies. The first scenario is difficult to reconcile with current ideas about the quenching of star formation \citep[e.g.][]{2014MNRAS.444.1518V, 2015MNRAS.450.1937C,2015ARA&A..53..115K}. There are, however, several mechanism which could be invoked for the removal of the black holes \citep{2010ApJ...717L...6B}, of which the gravitational wave recoil originating in the merger of a binary black hole system seems the most favourable as it can provide recoil velocities in excess of 3000 km/s \citep[e.g.][]{2007PhRvL..98w1102C, 2011PhRvD..83b4003L}, adequate to eject black holes from galaxies \citep[e.g.][]{ 2007ApJ...659L...5C, 2010MNRAS.404.2143V}. The main difficulty with the second scenario is that such kicks should not occur often, requiring a special geometry and black hole spin alignment, mostly resulting in black holes oscillating around the centre of the gravitational potential instead of being completely ejected \citep[e.g.][]{2004ApJ...607L...9M,2008ApJ...678..780G,2012PhRvD..85h4015L}.

If a black hole from NGC\,4474 or NGC\,4551 was kicked out of its host, the progenitor of the host had to go through a merger event which allowed for the creation of a binary black hole. This means that progenitors had to be of a similar mass (so that their black holes are of similar masses to about a factor of two). The light profiles of NGC\,4474 or NGC\,4551 show mild excesses in the centres, and these galaxies are fast rotators with large scale discs. A dry major merger, typically invoked for enabling a creation of a binary black hole \citep[e.g.][]{2009ApJS..182..216K}, is therefore not a likely choice. Another possibility is that these galaxies first experienced a dry merger and lost their black holes followed by a subsequent accretion event that made the outer disc and the central light excess. A possible hybrid scenario is a variant to a scenario proposed by \citet{2013MNRAS.433.2812K} to explain cores in fast rotator galaxies. The black hole binary was created during a wet merger, but the coalescence of black holes in this case was on a shorter time-scale than the duration of starbursts \citep[see also][]{1997AJ....114.1771F}. As the starburst continued after the black holes merged and were ejected by the kick from the gravitational wave emission, it was able to create the nuclear excess of light. Both of these scenarios, like the scenario with no black holes to start with, leave open the question of how the star formation was quenched. The surroundings of these galaxies provide a possible solution. These galaxies are members of Virgo Cluster, and live in the densest environment compared with other galaxies in our sample \citep{2011MNRAS.416.1680C}, implying that the (final) quenching of the star formation could have been purely environmental. 

As outlined earlier (see Section~\ref{ss:orb}), a merger of black holes should bias the orbital distribution towards tangential anisotropy. Fig.~\ref{f:int_space} shows that the Schwarzschild models require a mild increase in the radial anisotropy of the velocity tensor in the central 0.3\arcsec\, for NGC\,4474 (although an isotropic solution is also acceptable in this case) and NGC\,4551. Parameterizing the anisotropy in spherical coordinates, NGC\,4474 shows a tendency to tangential orbits, but the scatter is large and it is not possible to distinguish between trends for adiabatic black hole growth or core scouring. Taking all data together, we conclude that such orbital configurations within the scenario of the ejected black holes is possible only if the nuclear starburst is of longer duration than the merging of the black hole binary, and, therefore can replenish stars on radial orbits that were initially removed by the binary.

The evidence presented cannot prove the non-existence of central black holes in NGC\,4474 and NGC\,4551. The data support that these galaxies could harbour highly unusual black holes, about an order of magnitude smaller than the predictions. While it is difficult to choose between possible scenarios (a black hole that did not co-evolve, a black hole that was kicked out, or never having had a black hole at all), detecting the presence of very low-mass nuclear black holes will be achievable with the next generation of telescope facilities, such as the James Webb Space Telescope or from the ground-based Extremely Large Telescope.

%%%%%%%%%%%%%%%%%%%%%%%%%%%%%%%%%%%%%%%%%%%%%%%%%%%%%%%%%%%
%
% SECTION 6 SECTION 6 SECTION 6 SECTION 6 SECTION 6 SECTION 6
%
%%%%%%%%%%%%%%%%%%%%%%%%%%%%%%%%%%%%%%%%%%%%%%%%%%%%%%%%%%%

\section{Conclusion}
\label{s:con}

We observed six ETGs (NGC\,4339, NGC\,4434, NGC\,4474, NGC\,4551, NGC\,4578 and NGC\,4762) selected from the ATLAS$^{\rm 3D}$ sample with NIFS LGS AO in order to determine the masses of the central black holes. Galaxies were selected to be fast rotators of relatively low mass ($\approx10^{10}$ M$_\odot$, except NGC\,4762 which is $\approx10^{11}$ M$_\odot$) and velocity dispersion within the effective radius of $\sigma_e\approx 100$ km/s ($\approx135$ km/s for NGC\,4762). Four galaxies are members of the Virgo Cluster (at the distance of about 16 Mpc), while two are isolated objects at larger distance (at about 22 Mpc).

We extract the kinematics in the NIR and show that they are consistent with the optical observations with SAURON from the ATLAS$^{\rm 3D}$ survey. The spatial resolution achieved with the LGS AO is below 0.2\arcsec, allowing us to probe the stellar kinematics in the vicinity of the central black hole. We parameterize the light distributions of our galaxies using space- and ground-based imaging, and then construct the three-dimensional mass distribution assuming axisymmetry and inclinations derived in previous studies. We build two types of dynamical models: JAM, based on the Jeans equations allowing for a velocity anisotropy, and Schwarzschild models based on the superposition of orbits. The JAM models are only constrained by the high spatial resolution NIFS data, while Schwarzschild models use both large-scale SAURON data and NIFS data presented in this work. The two approaches give remarkably consistent results, in all but one galaxy the differences between JAM and Schwarzschild models are between $1-2 \sigma$, while for one galaxy the difference is just above $3\sigma$ confidence level. Both methods are able to constrain black hole masses for four galaxies, while for two galaxies the models favour black hole masses that are smaller than we can resolve. 

Two of our galaxies (NGC\,4339 and NGC\,4434) have black hole masses that are more massive than the expectation from the latest M$_{\rm BH} - \sigma$ relation by about twice the reported scatter of the relation ($4.3^{+4.8}_{-2.3}\times10^7$ and $7.0^{+2.0}_{-2.8}\times10^7$ M$_\odot$, respectively, at $3\sigma$ confidence level). Two galaxies (NGC\,4578 and NGC\,4762) have black hole masses as predicted by the  M$_{\rm BH} - \sigma$ relation ($1.9^{+0.6}_{-1.9}\times10^7$ and $2.3^{+0.9}_{-0.6}\times10^7$ M$_\odot$, respectively, at $3\sigma$ confidence level). The scatter between these measurements could be related to the differences in the feeding of the black holes, but there are no correlations with the type of internal structure (all have relatively high bulge S\'ersic index and exponential discs) or the environment (objects above and on the relation from our sample comprise both isolated and cluster objects). 

The data for two galaxies (NGC\,4474 and NGC\,4551) are consistent with not having central black holes (with upper limits of $<7\times10^6$ and $<5\times10^6$ M$_\odot$, respectively, at $3\sigma$ confidence level). Their stellar velocity dispersion maps are characterized by large (0.5 and 1\arcsec, respectively) central declines. These cannot be attributed to nuclear discs as those are not visible in the light distribution, nor in the kinematics. The Schwarzschild models also do not require fast rotating discs to reproduce the data. Specifically, second velocity moments ($V_{rms} = \sqrt{V^2 + \sigma^2}$), which trace the gravitational potential, also show distinct decreases in the central values. JAM models with M$_{\rm BH}$ as predicted by the scaling relations ($2\times10^6$ and $4\times10^6$ M$_\odot$, respectively), which are nominally below our resolution, still produces too high central $V_{rms}$ values, not compatible with the data. If there are central black holes in these galaxies they have masses less than a few$\times10^5$ M$_\odot$. Our conjecture is that these are prime candidates for galaxies that lost their black holes, possibly due to gravitational recoils resulting from a coalescence of a black hole binary. 

The nuclear orbital distribution for all our galaxies is consistent with being isotropic or mildly radially anisotropic. Given that our galaxies have S\'ersic profiles or show excess of light in the nuclei, this is consistent with findings that tangentially anisotropic velocity ellipsoids in the vicinity of black holes are preferentially found in galaxies with cores (deficits of light). The mechanism for the excavation of the core via a black hole binary is linked with the removal of stars on radial orbits and modification of the orbital anisotropy. This means that our galaxies either did not experience a major dry merger (required for core excavation) or they went through a subsequent event that replenished the core. As our galaxies have old stellar populations this must have happened at large redshift (z$\sim2$). The galaxies which we consider candidates for not having black holes show a mild increase towards radial anisotropy within their nuclei. As their black holes were possibly kicked out by a recoil generated in the emission of gravitational waves, the coalescence of the black hole binary had to happen on a shorter time-scale than the nuclear starburst that refilled the excavated cores and changed the orbital anisotropy.

\section*{Acknowledgements}

MC acknowledges support from a Royal Society University Research Fellowship. RMcD is the recipient of an Australian Research Council Future Fellowship (project number FT150100333). JF-B acknowledges support from grant AYA2016-77237-C3-1-P from the Spanish Ministry of Economy and Competitiveness (MINECO). Based on observations obtained at the Gemini Observatory, which is operated by the Association of Universities for Research in Astronomy, Inc., under a cooperative agreement with the NSF on behalf of the Gemini partnership: the National Science Foundation (United States), the National Research Council (Canada), CONICYT (Chile), Ministerio de Ciencia, Tecnologia e Innovaci\'on Productiva (Argentina), and Minist\'erio da Ci\^encia, Tecnologia e Inova\c{c}\~ao (Brazil). Based on observations made with the NASA/ESA Hubble Space Telescope, obtained from the data archive at the Space Telescope Science Institute. STScI is operated by the Association of Universities for Research in Astronomy, Inc. under NASA contract NAS 5-26555.

%%%%%%%%%%%%%%%%%%%%%%%%%%%%%%%%%%%%%%%%%%%%%%%%%%



\bibliographystyle{mnras}
%\bibliography{../refs.bib}

%%%%%%%%%%%%%%%%%%%%%%%%%%%%%%%%%%%%%%%%%%%%%%%%%%

%%%%%%%%%%%%%%%%% APPENDICES %%%%%%%%%%%%%%%%%%%%%

\appendix

\section{An illustrative case of NGC\,1277}
\label{app:n1277}

NGC\,1277 is a compact galaxy for which the Schwarzschild model of \citet{2012Natur.491..729V}, based on seeing-limited (long-slit) stellar kinematics, initially indicated a significantly larger black hole (M$_{\rm BH} = (17\pm3)\times10^9$ M$_\odot$) than predicted by the M$_{\rm BH} - \sigma$ relation, highlighting the galaxy as one of the most distant outliers from the black hole scaling relations. For the purpose of this section, the Schwarzschild models of van den Bosch et al. predicted a black hole in NGC\,1277 much larger than the value of M$_{\rm BH}\approx5\times10^9$ M$_\odot$ \citep{2013MNRAS.433.1862E}. The later models are based on an $N$-body realization having the first and second moments identical to a JAM model \citep[when computed via equations 19 -- 21 in][]{2008MNRAS.390...71C}, and were constrained with the same data as the Schwarzschild models. \citet{2013MNRAS.433.1862E} did not explore the full parameter space, but assumed instead a fixed inclination \citep[75\degr, as in][]{2012Natur.491..729V}, and an ad hoc value $M/L$ =10. This discrepancy of results made NGC\,1277 the only galaxy for which JAM-like- and Schwarzschild-based M$_{\rm BH}$ estimates did not agree.  NGC\,1277 is a fast rotating galaxy, consistent with axisymmetry \citep{2015MNRAS.452.1792Y} and as such is suitable for JAM modelling. The seeing-limited JAM-equivalent black hole estimate was also established by the Schwarzschild models of \citet{2016ApJ...817....2W} finding M$_{\rm BH} = (4.9\pm0.2)\times10^9$ M$_\odot$ ($1\sigma$ formal errors ignoring systematics) when using high-resolution NIFS IFU data. 

%%%%%%%%%%%%%%%%%%%%%%%%%%%%%%%%%%%%%%%%%%%%%%%%%%%%%%%%%%%
\begin{figure}
%Fig made by Michele
\includegraphics[width=\columnwidth]{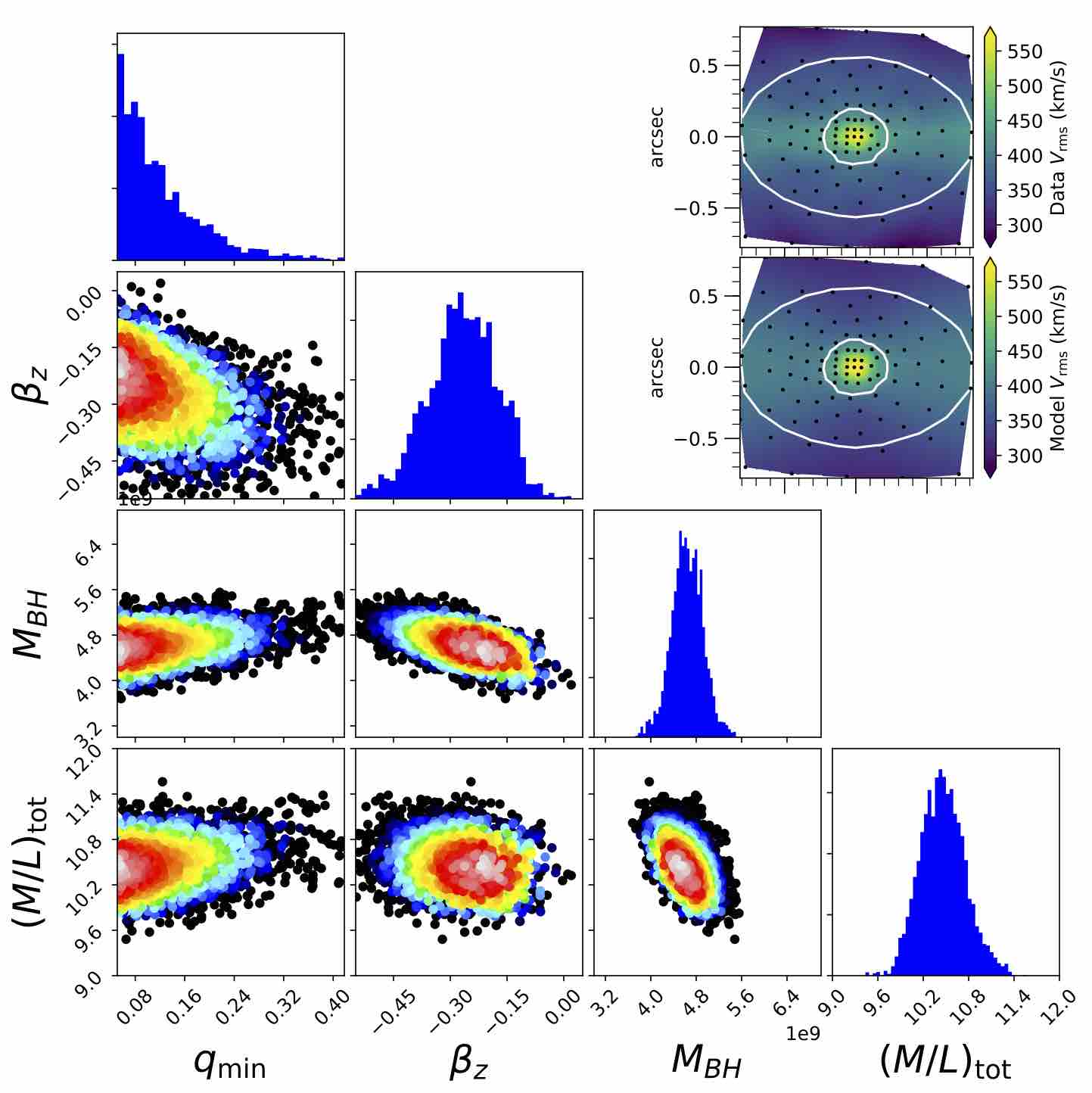}
\caption{ A set of plots showing the posterior probability distribution for the non-linear model parameters of JAM modelling ($q_{\rm min}$, $\beta_z$, $M/L_{tot}$ and M$_{\rm BH}$), marginalized over two dimensions (colour contours) and one dimension (blue histograms). The symbols are coloured according to their likelihood: white corresponds to the maximum value and dark blue to a $3\sigma$ confidence level. The inset in the upper right corner shows the data $V_{rms}$ (bottom) and $V_{rms}$ of the best-fitting model (top). }
\label{fapp:corner}
\end{figure}
%%%%%%%%%%%%%%%%%%%%%%%%%%%%%%%%%%%%%%%%%%%%%%%%%%%%%%%%%%%

As an additional test for this paper, we used the JAM method to fit a mass-follows-light model to NGC\,1277 using the {\it same} published IFU kinematics and errors of \citet{2016ApJ...817....2W}, where we defined $V_{rms}=\sqrt{V^2+\sigma^2}$, ignoring $h_3$ and $h_4$. For maximum consistency, we adopted their published PSF parametrization, distance, and the same MGE model from \citet{2012Natur.491..729V}. We fit for the inclination (parameterized as axial ratio $q_{\rm min}$, which is the intrinsic axial ratio of the flattest Gaussian in the MGE), anisotropy $\beta_z$, total $M/L_{\rm tot}$ and M$_{\rm BH}$ in a Bayesian framework. The calculation of the posterior probability distribution is done using the adaptive \citet{1953JChPh..21.1087M} algorithm of \citet{2001Bernoulli.7.223}. This algorithm, while very efficient, is strictly speaking non-Markovian, but it has the correct ergodic properties and can be used to estimate the posterior distribution as in standard Markov chain Monte Carlo method. We assumed ignorant (constant) priors on all model parameters. The corner plots of the probability distribution for the non-linear model parameters are shown in Fig~\ref{fapp:corner}. The data model comparison is shown in  the upper-right corner of Fig.~\ref{fapp:corner}. Our best-fitting model has M$_{\rm BH}= (4.61 \pm 0.37)\times10^9$ M$_\odot$,  $\beta_z = -0.25 \pm 0.18$, $M/L_{tot}=10.3 \pm0.3$ ($1\sigma$ formal errors), and $q_{\rm min}=0.05\pm 0.19$. This quantity gives the best fitting inclination \citep[using equation 14 from][]{2008MNRAS.390...71C} of $i=66\degr$. Within the $3\sigma$ level this value is consistent with 75\degr, the assumed inclination in \citet[][based on a dust ring]{2012Natur.491..729V}, and subsequently used in all other above mentioned papers. The black hole mass is consistent with \citet{2016ApJ...817....2W}, while their stellar $M/L_\ast$=$9.3 \pm1.6$ and total $M/L$ = 9.5 (Walsh, private communication), are somewhat smaller, but still within the uncertainties with our estimates. The velocity anisotropy in the spherical coordinate system of NGC\,1277 is published in \citet{2015MNRAS.452.1792Y} based on Schwarzschild models of large-scale data. While their anisotropy cannot be directly compared with JAM's $\beta_z$ in the cylindrical coordinates, we note that in both cases the models favour tangentially anisotropic velocities. This adds NGC\,1277 to those galaxies that show consistent results between the Schwarzschild and JAM modelling, in this case, for the first time when using the same data. 

%%%%%%%%%%%%%%%%%%%%%%%%%%%%%%%%%%%%%%%%%%%%%%%%%%%%%%%%%%
\begin{figure}
%Fig made by Michele
\includegraphics[width=\columnwidth]{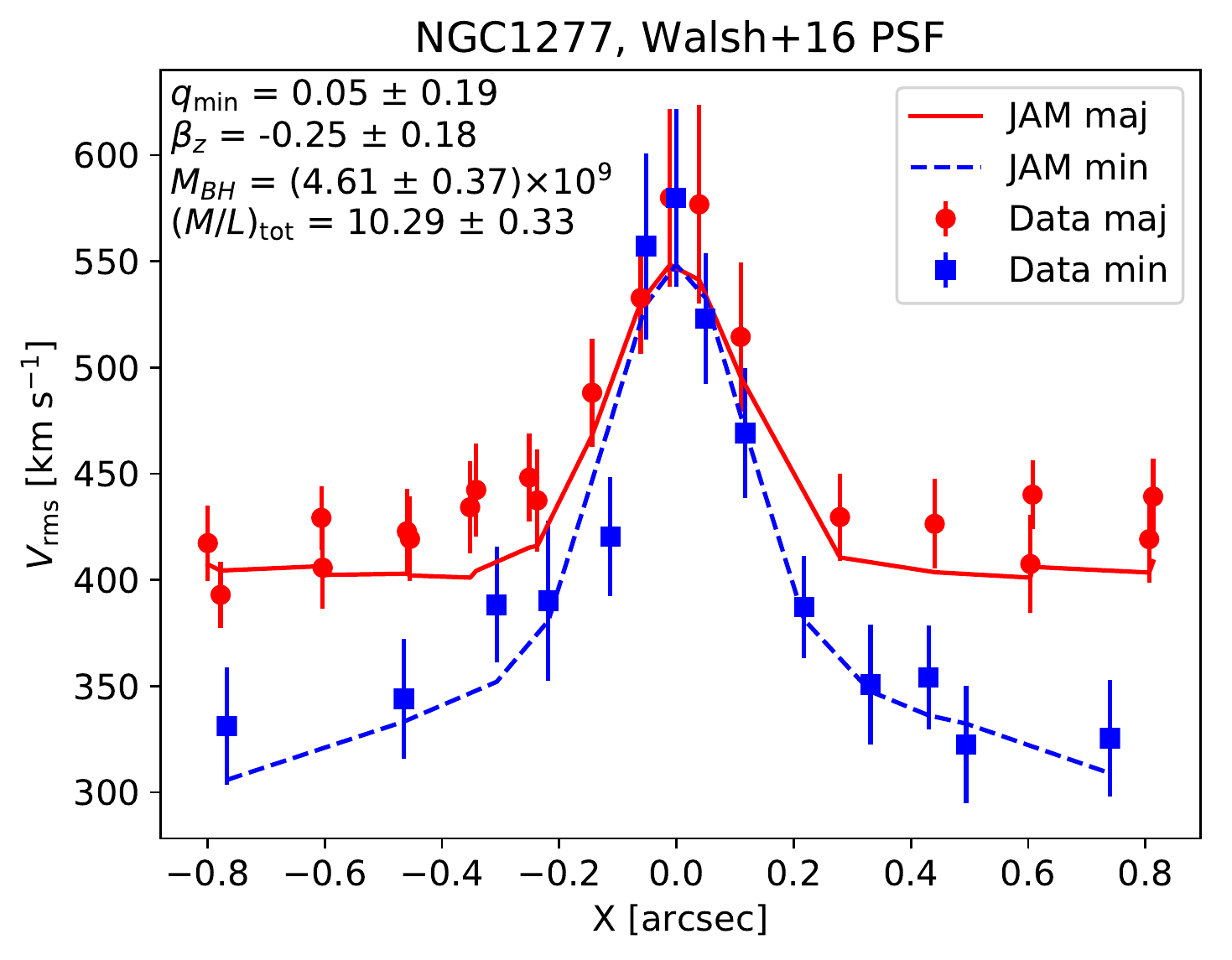}
\includegraphics[width=\columnwidth]{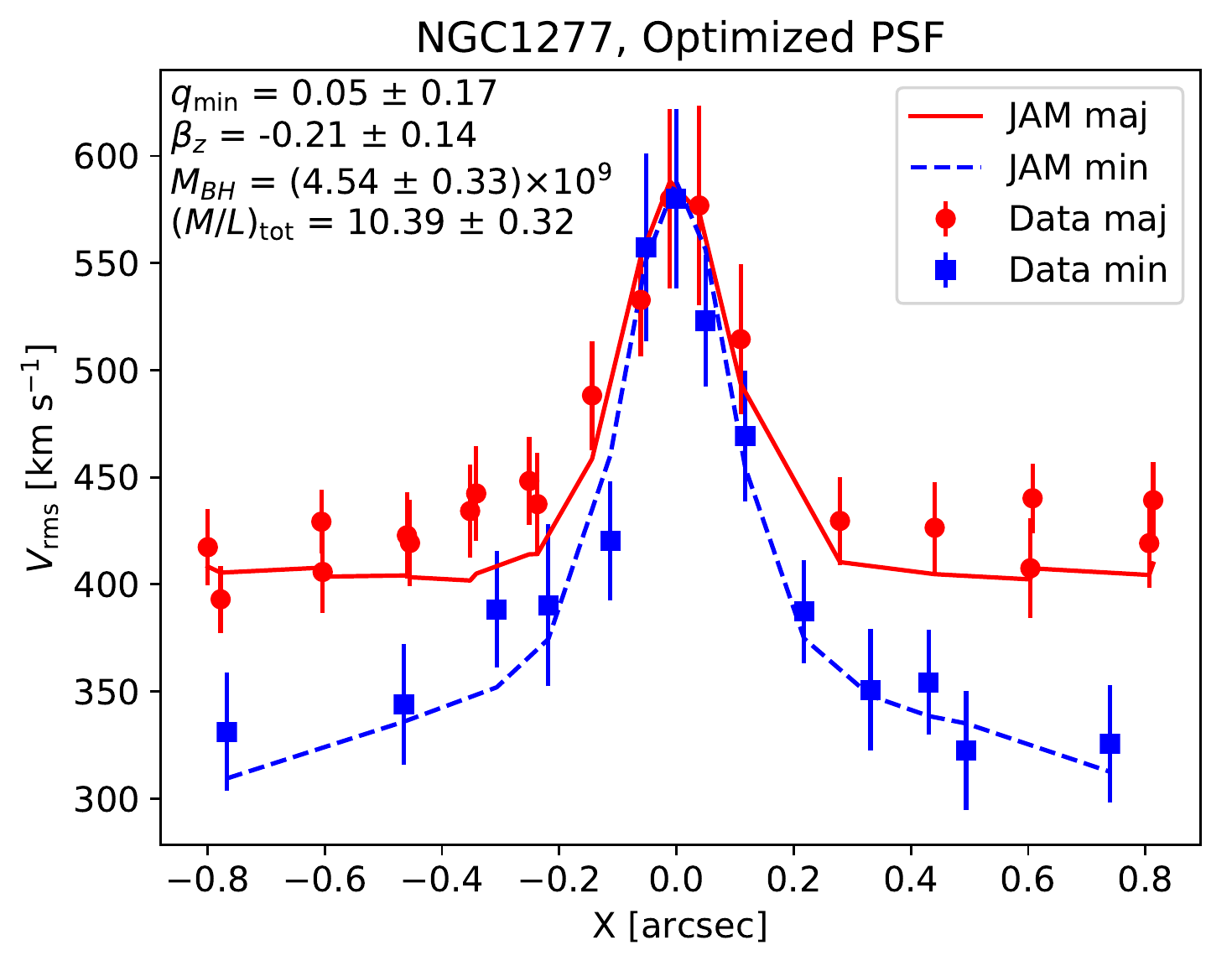}
\caption{A comparison between JAM models and NIFS data for NGC\,1277 from \citet{2016ApJ...817....2W} along the major (blue) and minor (red) axes. The JAM model in the top panel was done using the published PSF, while in the bottom panel we optimized the PSF until a better fit was achieved, as described in the text. Both the changes in the PSF parameters and the best-fitting black hole mass are within the quoted or expected errors. }
\label{fapp:1277}
\end{figure}
%%%%%%%%%%%%%%%%%%%%%%%%%%%%%%%%%%%%%%%%%%%%%%%%%%%%%%%%%%%

Looking for a more illustrative comparison between the data and the model we show cuts along the major and the minor axes on the top panel of Fig.~\ref{fapp:1277}. The model seems not to be able to reproduce the central peak in the $V_{rms}$. Similar disagreement can also be seen on fig.~4 of \citet{2016ApJ...817....2W}. A better fit can be achieved for a small change of the assumed PSF parameters. We changed the FWHM of the inner Gaussian component (of two Gaussian parametrization) from the 0.07\arcsec, preferred by \citet{2016ApJ...817....2W}, to 0.05\arcsec. This change is within the errors one can expect on the determination of the PSF (as also shown in this paper). The new PSF allows JAM model to fit essentially all data points (extracted along the major and minor axes on the figure) as is shown on the bottom panel of Fig.~\ref{fapp:1277}, while the change in the black hole mass and the anisotropy is essentially negligible (M$_{\rm} = (4.54 \pm 0.33)\times 10^9$ M$_\odot$ and $\beta_z = -0.21 \pm 0.14$), and still fully consistent with the $1\sigma$ uncertainties quoted by \citet{2016ApJ...817....2W}. %A similar decrease in black hole mass is expected from the Schwarzschild modelling using the modified PSF value. 

While NGC\,1277 is a more massive galaxy that those presented in this paper, its properties are in many ways similar to those of our galaxies (e.g. disc-dominated fast rotators). Therefore, it is reassuring that even for this galaxy (with initial contradictory claims) consistent black hole massed are recovered when the same, high-quality data are used, with two vastly different modelling techniques.

\section{Parametrisation of the PSFs }
\label{app:PSF}

In Table~\ref{tapp:psf} we show MGE parameters describing the PSF of the HST images, which were used for comparison with the NIFS reconstructed images and the determination of the NIFS PSF. In Fig.~\ref{tapp:psf} we show the comparison along two axis between the NIFS reconstructed images and the HST convolved by the best-fitting PSF listed in Table~\ref{t:psf}. 

%%%%% Table 1. %%%%%%%%%%%%%%%%%%%%%%%%%%%%%%%%%%%%%%%%%%%%%%%%%%%%%
\begin{table*}
   \caption{MGE parametrization of the HST PSF for images in Table~\ref{t:obs}}
   \label{tapp:psf}
$$
%  \begin{array}{c|ccccc|cccc}
  \begin{array}{cc|cc|cc|cc|cc|cc}
    \hline
    \hline
    \noalign{\smallskip}

        $NGC\,4339$   &$(WFPC2)$ & $NGC\,4434$ & $(ACS)$ &  $NGC\,4474 $ &$(WFPC2)$&  $NGC\,4551$  &$(ACS)$&  $NGC\,4578$ & $(WFPC2)$& $NGC\,4762$&$(ACS)$ \\
                                 &$F606W$    &                        & $F475W$    &                      &  $F702W$ &                          &$F475W$&                      & $F606W$    &                       &$F475W$\\
        $Norm$ & \sigma&  $Norm$ & \sigma&  $Norm$ & \sigma&  $Norm$ & \sigma &  $Norm$ & \sigma &  $Norm$ & \sigma\\
    \noalign{\smallskip} \hline \hline \noalign{\smallskip}
       0.2374 & 0.3800  &  0.6770  & 1.1456  & 0.2596 & 0.4208   & 0.0685 & 0.438   &  0.2382   & 0.3800  & 0.0954  & 0.4270 \\
       0.5500 & 1.1182  &  0.1584  & 1.9440  & 0.5593 & 1.2320   & 0.6084 & 0.962   &  0.5502   & 0.1142   & 0.5791 & 0.9235 \\ 
       0.0982 & 3.1355  &  0.0967  & 5.1000  & 0.0839 & 3.5586   & 0.1886 &1.821    & 0.0968    & 3.1171   & 0.2021 & 1.7319 \\
       0.0761 & 7.3108  &  0.0524  &15.9560 & 0.0708 & 8.4162   & 0.0786 & 5.137   & 0.0735    & 7.0799   & 0.0727  & 4.6360 \\
       0.0382 & 21.2612&  0.0098  & 27.642  & 0.0333 & 22.5037 & 0.0426 & 15.848  & 0.0412   & 20.3981 & 0.0508  & 12.1414\\
       --          &  --         &  0.0050  & 54.9590&  --         & --           & 0.0087 & 26.321  &  --           & --            &  --          & --           \\
       --          & --          &  --          & --          &  --          & --           & 0.0046 & 55.171   & --           & --            &  --          & --           \\

       \noalign{\smallskip}
    \hline
  \end{array}
$$ 
{Notes -- For each galaxy the PSF is parameterized by 5--7 circular Gaussians. The `{\it Norm}' column has the normalization of the Gaussian (such that the sum it equal to unity) and the `$\sigma$' column has the dispersion of the Gaussian given in pixels. The conversion factor to physical units for WPFC2 is 0.0445 pixels per arcsec and for ACS is 0.05 pixels per arcsec.  }
\end{table*}
%%%%%%%%%%%%%%%%%%%%%%%%%%%%%%%%%%%%%%%%%%%%%%%%%%%%%%%%%%%%%%%%%%

%%%%%% Figure A1%%%%%%%%%%%%%%%%%%%%%%%%%%%%%%%%%%%%%%%%%%%%%%%
\begin{figure*}
\includegraphics[width=0.315\textwidth]{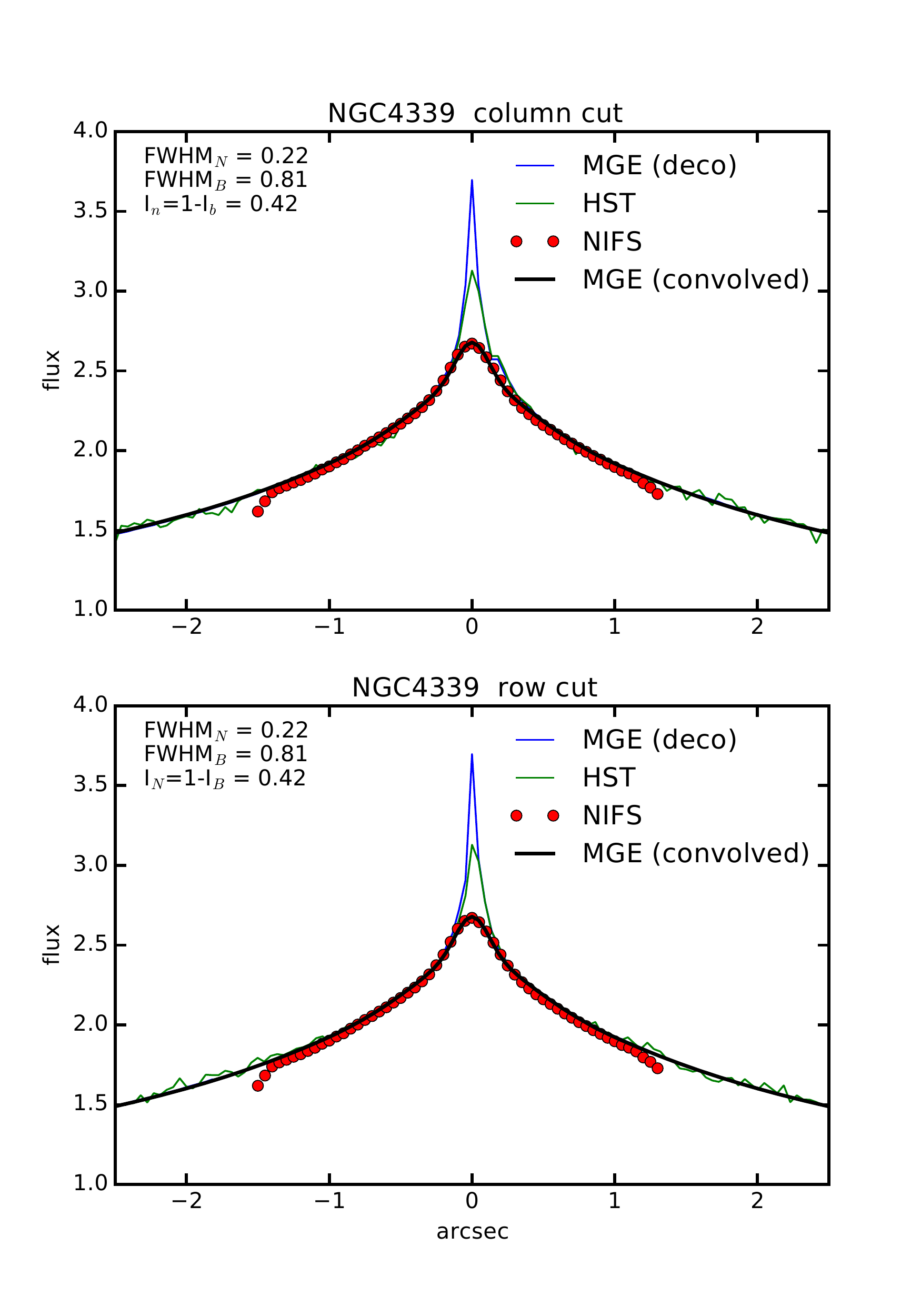}
\includegraphics[width=0.315\textwidth]{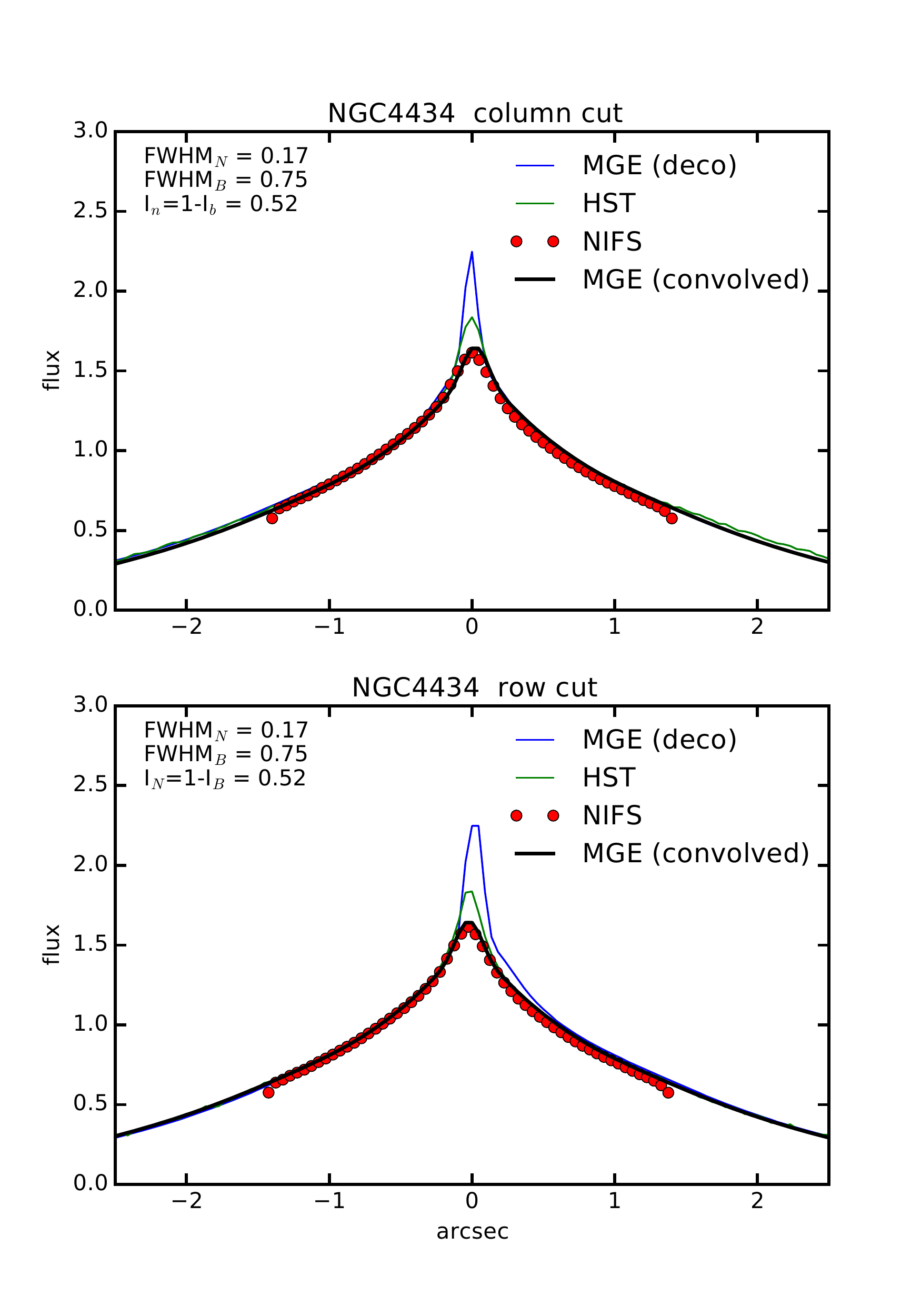}
\includegraphics[width=0.315\textwidth]{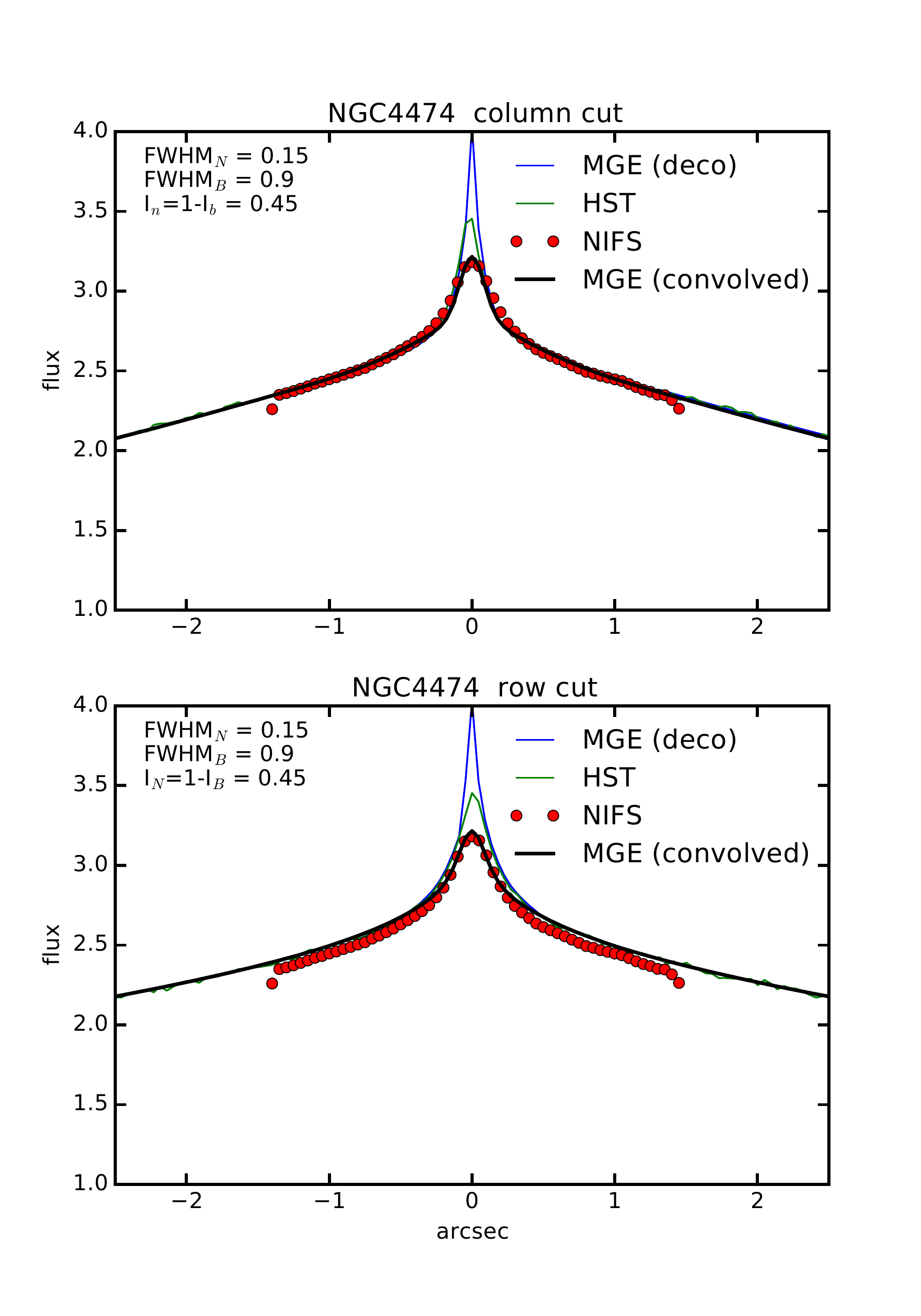}
\includegraphics[width=0.315\textwidth]{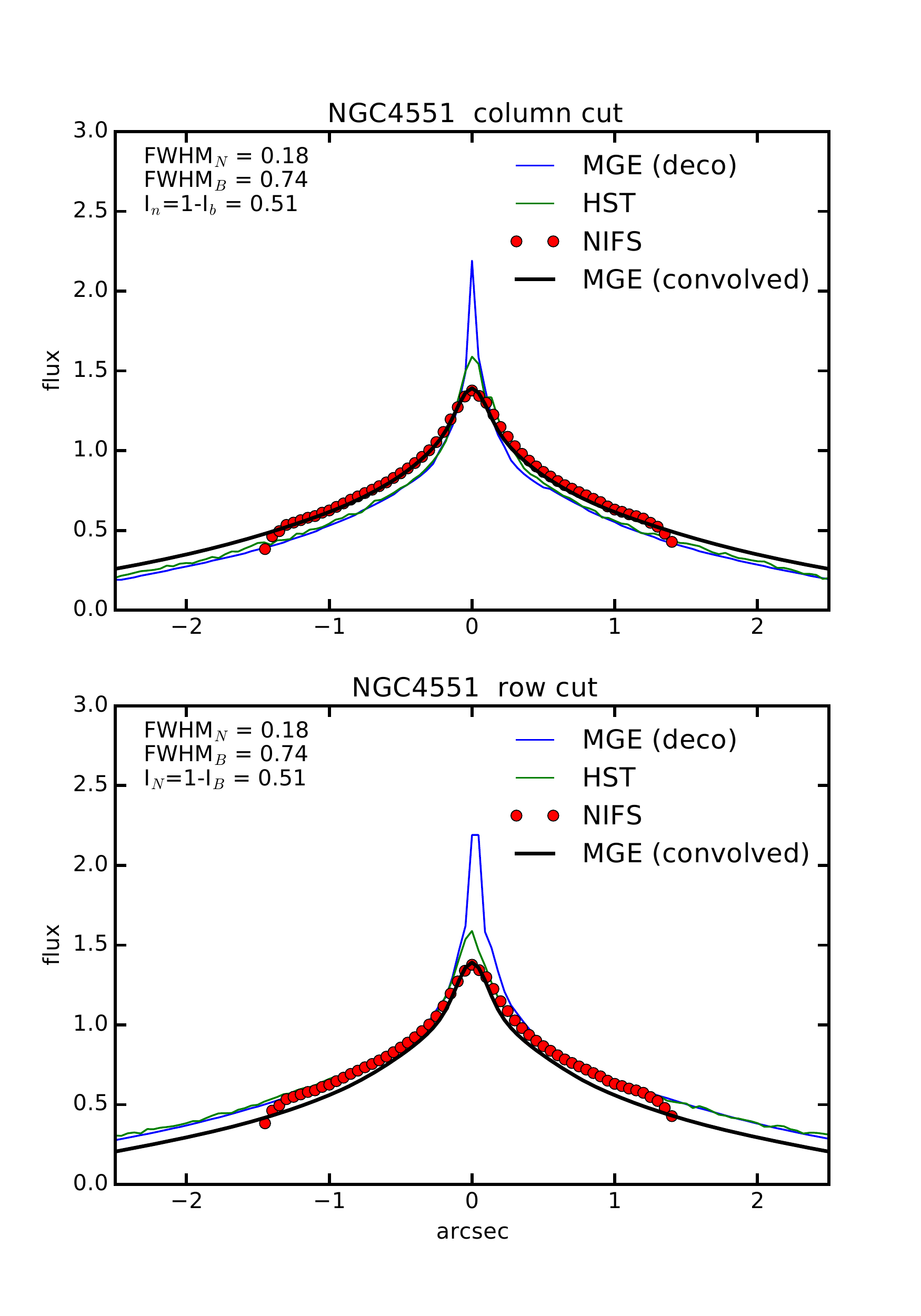}
\includegraphics[width=0.315\textwidth]{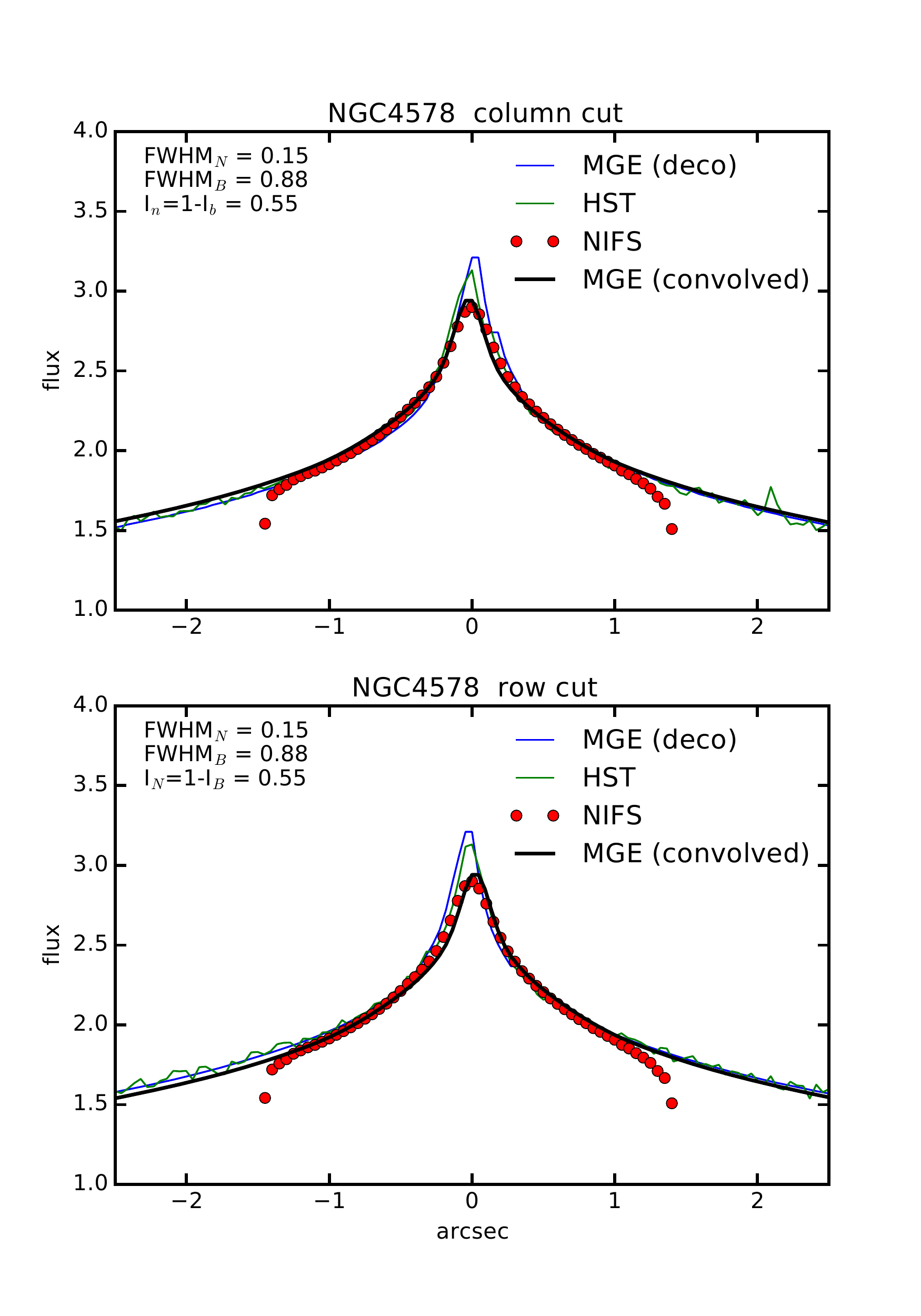}
\includegraphics[width=0.315\textwidth]{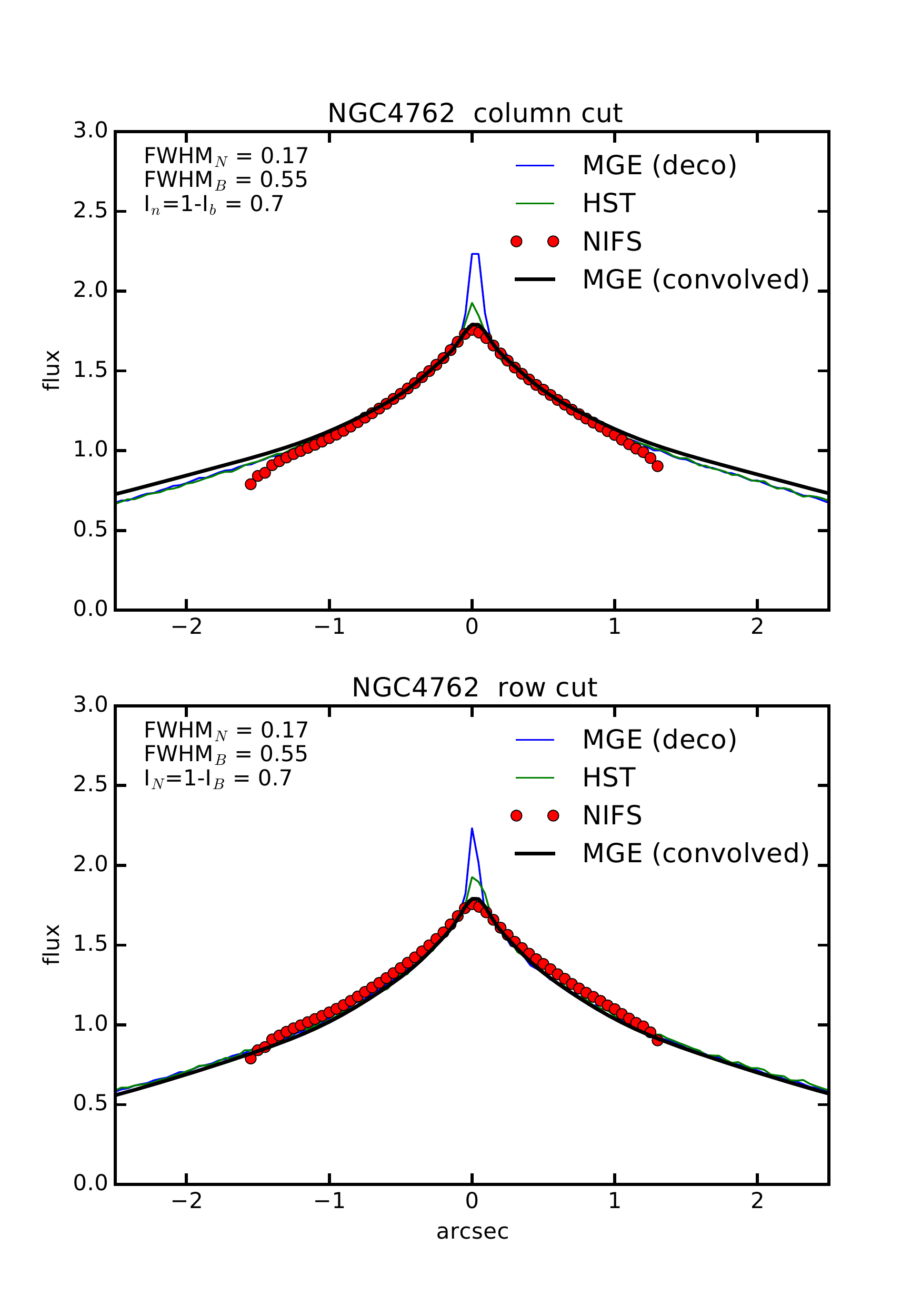}
\caption{Comparison between the light profiles extracted from the NIFS reconstructed image (red circles) and the convolved MGE models (thick black line) for galaxies in our sample. Also shown are the light profiles for the HST image (green line) used for the construction of the MGE model and the deconvolved MGE model (blue line) using the PSF of the HST image. Note that these MGE models are not the same as those shown in Fig.~\ref{f:mge} and used in the dynamical modelling. The purpose of these MGE models (constrained only by the HST data) was to estimate the resolution of the NIFS LGS AO observations only. In the upper left corner of each image are the parameters of the double Gaussian, which describes the PSF of the NIFS data cube. This PSF was used to convolve the deconvolved MGE model and produce the thick black line, which fits the NIFS image well. Images were oriented like the NIFS cubes, with north up and east left and the light profiles were extracted along vertical (top panels) and horizontal (bottom panels) axis passing through the centre of the galaxy. For more details see Section~\ref{ss:psf}.}
\label{fapp:psf}
\end{figure*}
%%%%%%%%%%%%%%%%%%%%%%%%%%%%%%%%%%%%%%%%%%%%%%%%%%%%%%%%%

\section{SAURON kinematics}
\label{app:sauron}

The extraction of SAURON kinematics\footnote{Available from http://purl.org/atlas3d} was presented in \citet{2011MNRAS.413..813C} and the velocity maps of our galaxies were already shown in \citet{2011MNRAS.414.2923K}. Here, for completeness, we show the maps of the mean velocity, the velocity dispersion, and the $h_3$ and $h_4$ Gauss-Hermite moments, characterizing the full LOSVD. Typical error on the SAURON kinematics is noted in Table~\ref{t:errors}. Note that these maps are different from those shown in Appendix~\ref{app:comparison} as the data there were symmetrized for the modelling. 

%%%%%% Figure A2%%%%%%%%%%%%%%%%%%%%%%%%%%%%%%%%%%%%%%%%%%%%%%%
\begin{figure*}
%Fig made by make_plots_all_kin_maps_LGS_SAURON.pro
\includegraphics[width=\textwidth]{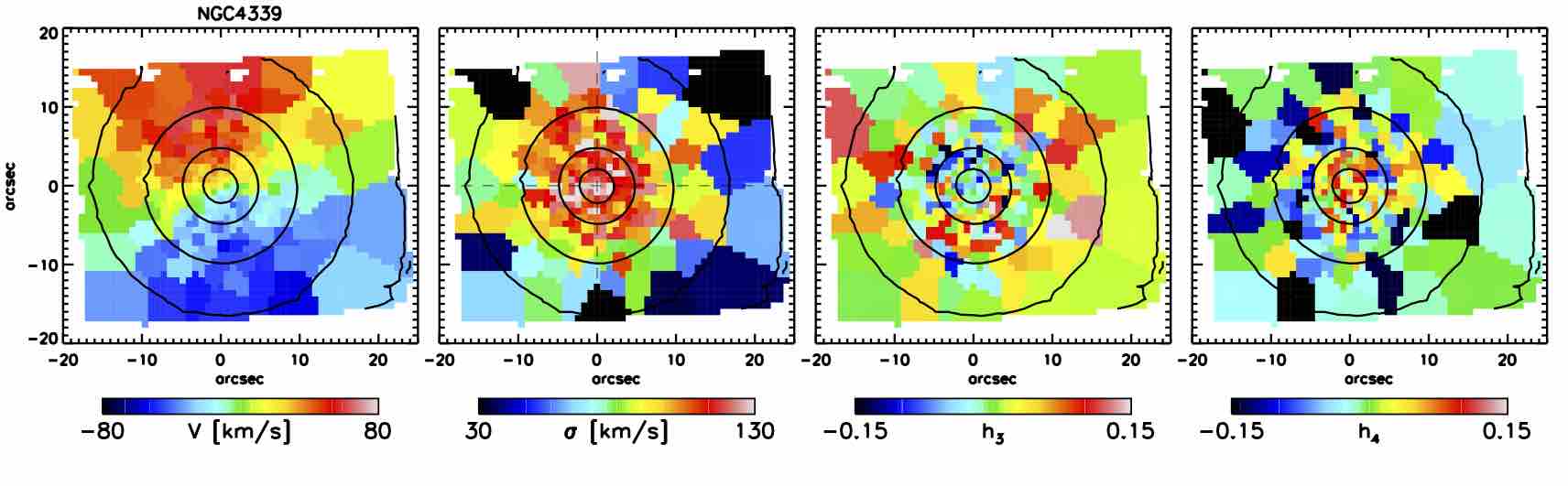}
\includegraphics[width=\textwidth]{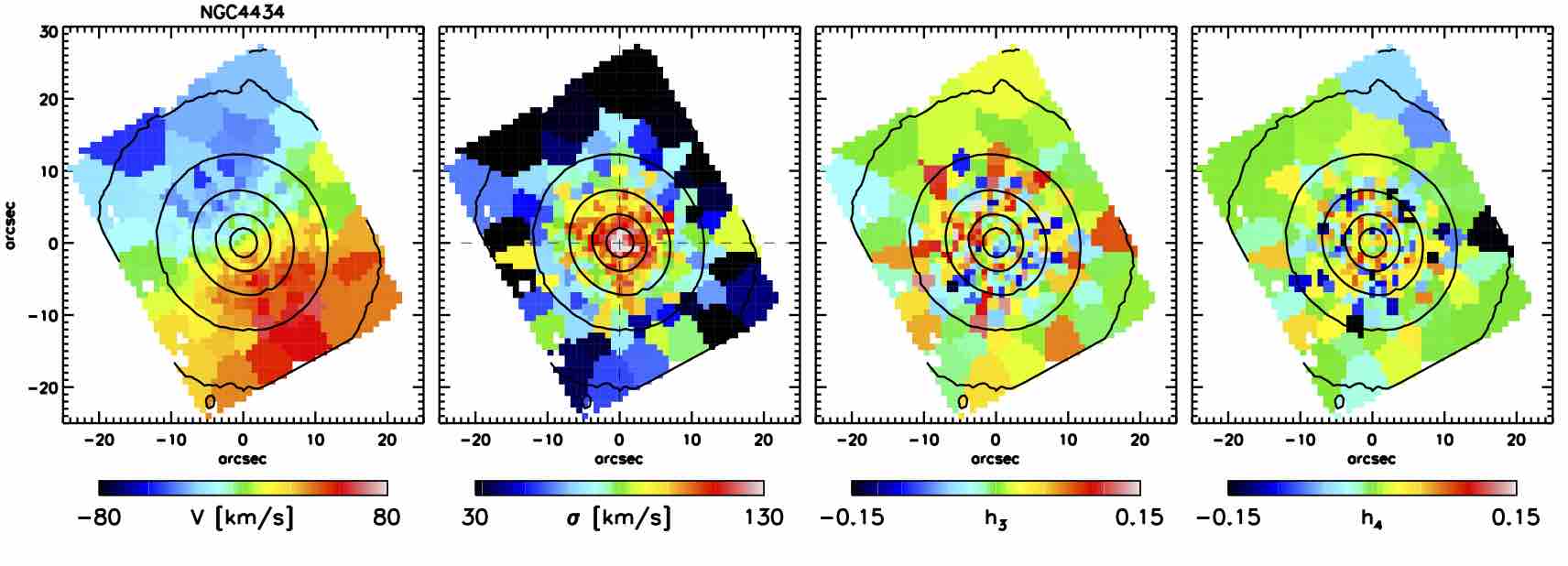}
\includegraphics[width=\textwidth]{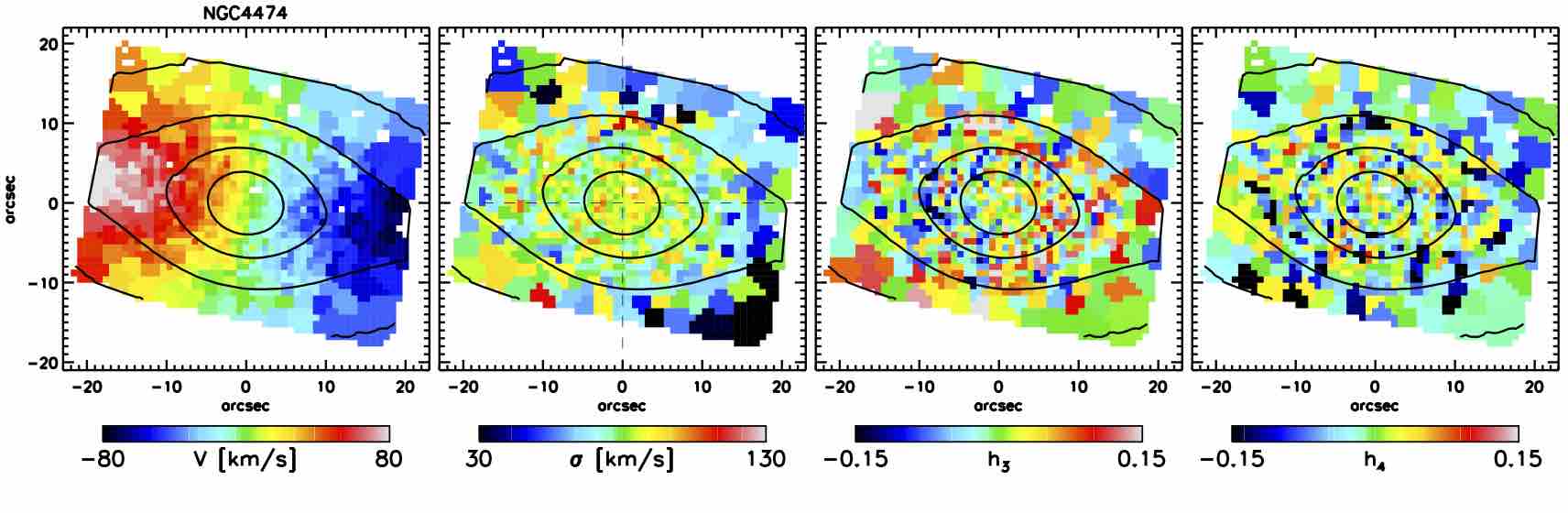}
\caption{SAURON kinematic of the first three target galaxies comprising maps of the mean velocity, the velocity dispersion, and $h_3$ and $h_4$ Gauss-Hermite coefficients. North is up and east to the left. These galaxies were observed as part of the ATLAS$^{\rm3D}$ survey, and the data reduction and the extraction of the kinematics are described in \citet{2011MNRAS.413..813C}}
\label{fapp:sauron1}
\end{figure*}
%%%%%%%%%%%%%%%%%%%%%%%%%%%%%%%%%%%%%%%%%%%%%%%%%%%%%%%%%

%%%%%% Figure A2%%%%%%%%%%%%%%%%%%%%%%%%%%%%%%%%%%%%%%%%%%%%%%%
\begin{figure*}
%Fig made by make_plots_all_kin_maps_LGS_SAURON.pro
\includegraphics[width=\textwidth]{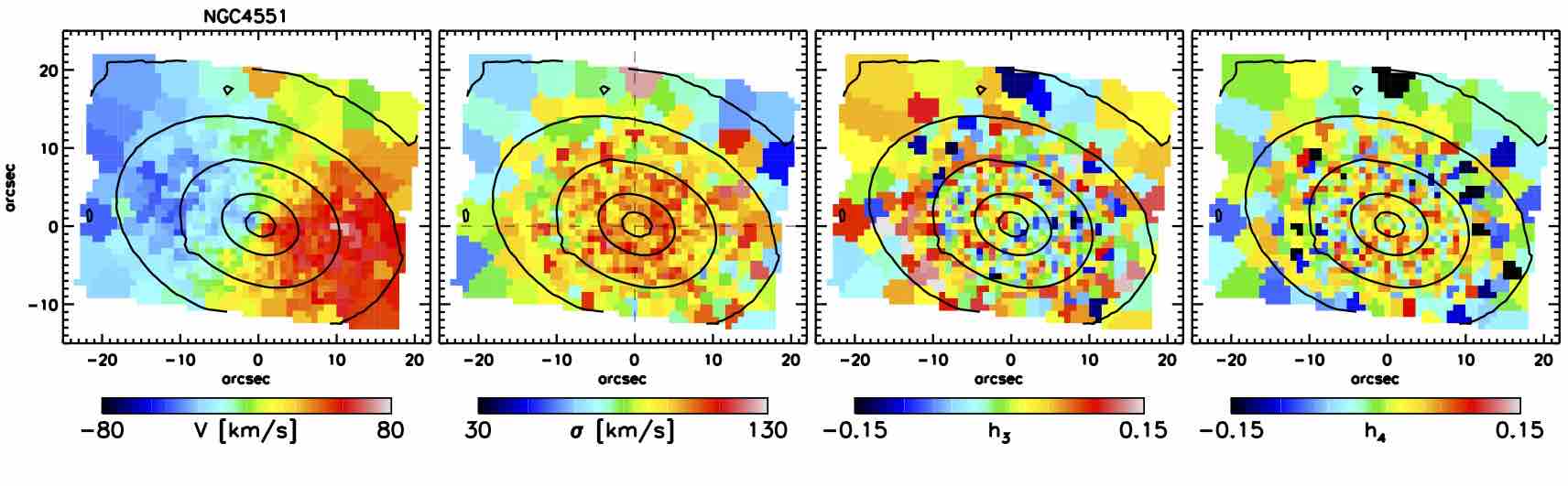}
\includegraphics[width=\textwidth]{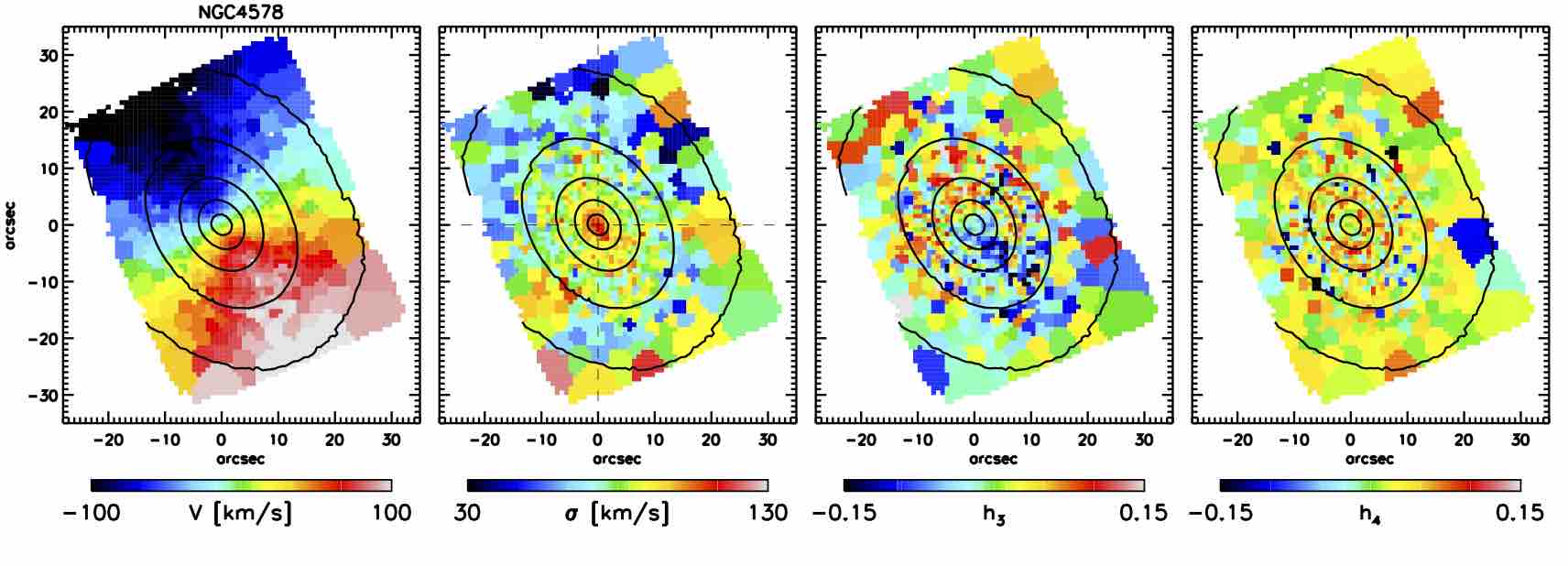}
\includegraphics[width=\textwidth]{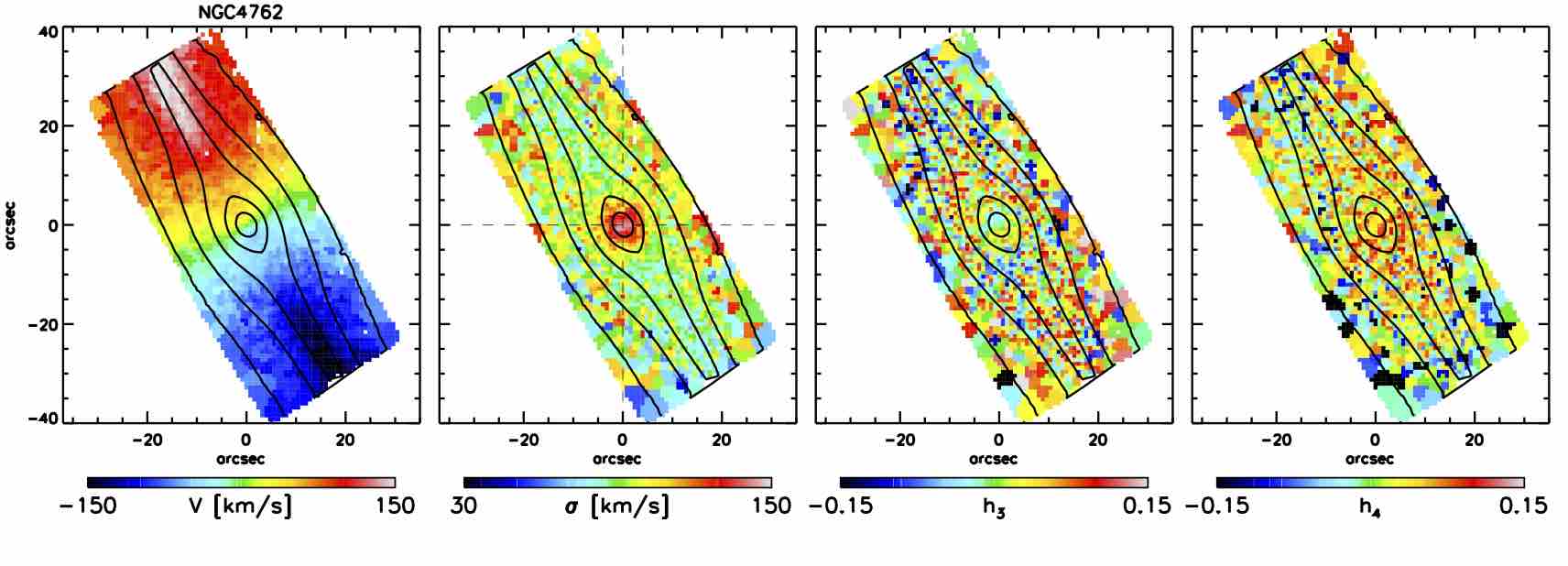}
\caption{Same as Fig~\ref{fapp:sauron1} for the remaining three galaxies}
\label{fapp:sauron2}
\end{figure*}
%%%%%%%%%%%%%%%%%%%%%%%%%%%%%%%%%%%%%%%%%%%%%%%%%%%%%%%%%

\section{MGE parametrization of the light distributions of sample galaxies}
\label{app:mge}

%%%%% Table 1. %%%%%%%%%%%%%%%%%%%%%%%%%%%%%%%%%%%%%%%%%%%%%%%%%%%%%
\begin{table*}
   \caption{MGE parametrization of the light distribution for sample galaxies}
   \label{tapp:mge}
$$
  \begin{array}{l rrr | rrr | rrr}
    \hline
    \hline
    \noalign{\smallskip}

%    \multicolumn{4}{c|}{$NIFS$ } &  \multicolumn{3}{c}{$HST$}\\ 
%   \hline
%    \noalign{\smallskip}
              &$NGC\,4339 $    &($R-band$)    &           & $NGC\,4434 $  &($F450W-band$) &                           &$NGC\,4474 $   &($R-band$)   &\\
  \hline
        $j$ & \log I_j               & \log \sigma_j & $q$_j & \log I_j               &  \log \sigma_j      & $q$_j                 & \log I_j              &  \log \sigma_j & $q$_j \\
              & (L_\odot pc^{-2}) & (arcsec)           &           & (L_\odot pc^{-2}) & (arcsec)                 &                           & (L_\odot pc^{-2}) & (arcsec)           & \\
    \noalign{\smallskip} \hline \hline \noalign{\smallskip}
   1&    5.457    &   -1.770   &  1.00    & 5.0578  &  -1.409  & 0.98    &  5.182    &  -1.770   &  0.66\\
   2&    4.711    &   -1.244   &  1.00    &  4.140   &  -0.780  & 0.97    &  4.626    &  -1.366   &  0.69\\
   3&    4.112    &   -0.879   &  1.00    &  3.862   &  -0.471  &  0.99   &  4.459    &  -1.022   &  0.64\\
   4&    3.939    &   -0.539   &  0.96    &  3.704   &  -0.084  & 0.98    &  3.844    &  -0.642   &  0.45\\ 
   5&    3.693    &   -0.164   &  0.96    &  2.955   &   0.234  & 0.97    &  3.611    &  -0.618   &  0.86\\
   6&    3.452    &   0.109    &  0.97    &  2.923   &   0.580  & 0.91    &  3.412    &  -0.235   &  0.96  \\
   7&    3.002    &   0.407    &  0.93    &  2.308   &   0.852  & 0.97    &  3.169    &  -0.196   &  0.42 \\ 
   8&    2.683    &   0.525    &  1.00    &  2.054   &   1.142  & 0.96    &  3.508    &   0.090   &  0.92  \\
   9 &   2.814    &   0.845    &  0.96    &  1.209   &   1.449  & 0.90    &  3.360    &   0.401   &  0.78\\
  10&   2.133    &   1.141    &  0.90    & -0.509   &   1.783  & 0.80    &  3.019    &   0.662   &  0.77 \\
  11 &  1.682    &   1.386    &  1.00    & --           & --           & --        &  2.306    &   0.985   &  0.74\\
  12&   1.776    &   1.466    &  0.90    &--            &  --          &--         &  2.188    &   1.250   &  0.11 \\
  13&   1.292    &   1.921    &  0.90    & --           &   --         & --        &  2.224    &   1.305   &  0.20 \\
  14&    --          &         --    &  --        & --            &   --         & --        &  2.161    &   1.340   &  0.49\\
  15 &   --          &         --    &  --        & --            &   --         & --        &  1.704    &   1.519   &  0.71\\
  16&   --           &         --    &  --        & --            &   --         & --        &  0.835    &   1.830   &  0.97\\
        \noalign{\smallskip}
    \hline
    \hline
        \noalign{\smallskip}
       \noalign{\smallskip}
       \noalign{\smallskip}
    \hline
    \hline

   &$NGC\,4551$&$(F450W-band)$&  &$NGC\,4578$& $(R-band)$& &$NGC\,4762$&$(F450W-band)$&\\
  \hline
        $j$ & \log I_j &  \log \sigma_j & $q$_j & \log I_j &  \log \sigma_j & $q$_j & \log I_j & \log \sigma_j & $q$_j \\
               & L_\odot pc^{-2} & arcsec  & & L_\odot pc^{-2} & arcsec  & & L_\odot pc^{-2} & arcsec  & \\
    \noalign{\smallskip} \hline \hline \noalign{\smallskip}
1   &    4.522   &  -1.114    &   0.56    &   5.353   &  -1.658 & 1.00     &  4.820   &  -1.367   &  1.00\\
 2  &    3.955   &  -0.695    &   0.63    &   4.829   &  -1.161 & 1.00     &  4.267   &  -0.650   &  0.80\\
3   &    3.435   &  -0.351    &   0.68    &   4.303   &  -0.752 & 0.89     &  4.062   &  -0.257   &  0.77\\
4   &    3.384   &  -0.025    &   0.64    &   3.833   &  -0.406 & 0.78     &  3.837   &   0.157   &  0.68\\
5    &   3.111    &  0.263     &   0.74    &   3.607   & -0.103  & 0.77     &  3.352   &   0.469   &  0.84\\
6    &   2.829   &   0.560    &   0.74    &   3.060   &  0.190  & 0.92      &  2.682   &   0.941   &  0.64\\
7    &   2.696   &   0.842    &   0.72    &   3.175   &  0.199  & 0.65      &  1.890   &   1.382   &  0.54\\
8    &   2.281   &   1.155    &   0.69    &   3.224   &  0.530  & 0.75      &  2.832   &   1.481   &  0.05\\
9    &   1.564   &   1.436    &   0.82    &   2.732   &  0.780  & 0.74      &  2.645   &   1.592   &  0.11\\
10  &  -2.000   &   2.110    &   1.00    &   2.488   & 1.110   &  0.72     &  2.095   &   1.902   &  0.12\\
11  &         --    &       --      &       --     &   1.958   & 1.568   & 0.69      &  1.543   &   1.910   &  0.31\\
12  &          --   &        --     &        --    &   1.287   & 1.962   &  0.74     &  0.766   &   2.062   &  0.62\\
        \noalign{\smallskip}
    \hline
 
  \end{array}
$$ 
{Notes: Galaxies from the ACS data are calibrated to the AB magnitude system (F450W band), while WFPC2 are on Vega system (Johnson R).}
\end{table*}
%%%%%%%%%%%%%%%%%%%%%%%%%%%%%%%%%%%%%%%%%%%%%%%%%%%%%%%%%%%%%%%%%%

\section{Comparison of the Schwarzschild dynamical models with the symmetrised data}
\label{app:comparison}

%%%%%%%%%%%%%%%%%%%%%%%%%%%%%%%%%%%%%%%%%%%%%%%%%%%%%%%%%%%
\begin{figure*}
%Fig made by plot_sch_model_NIFS_sample_comparison.pro
\includegraphics[width=0.98\textwidth]{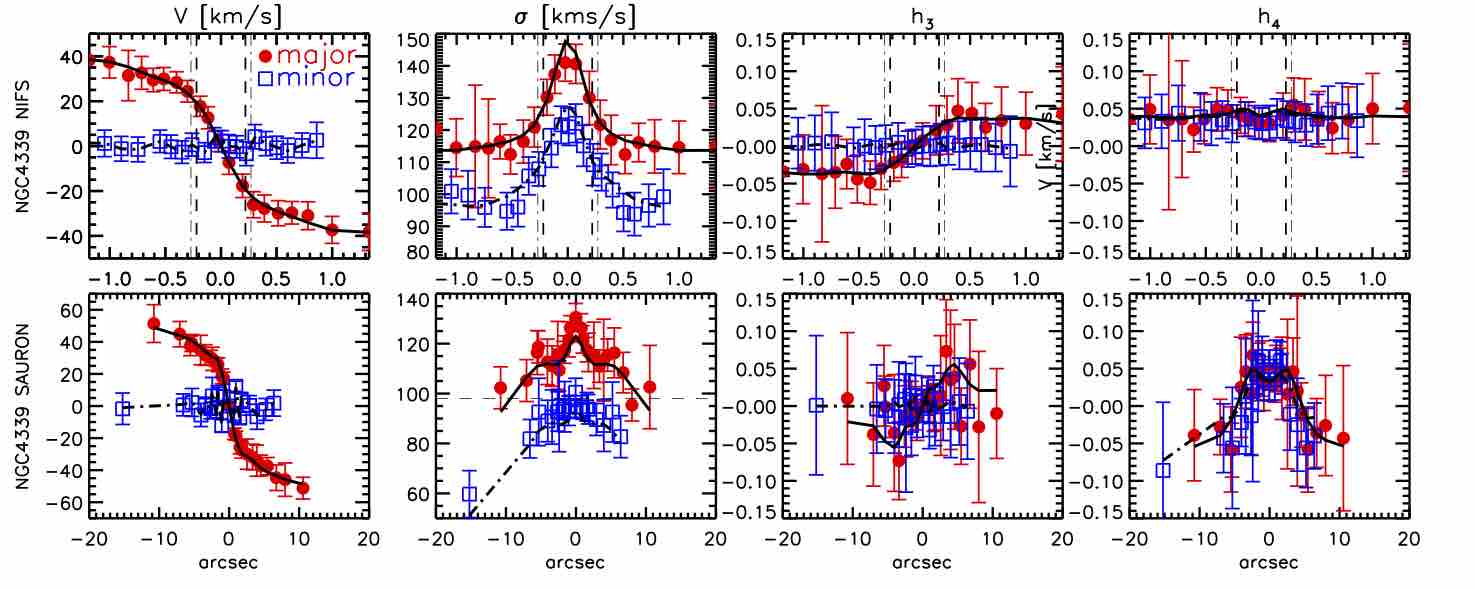}
\includegraphics[width=0.98\textwidth]{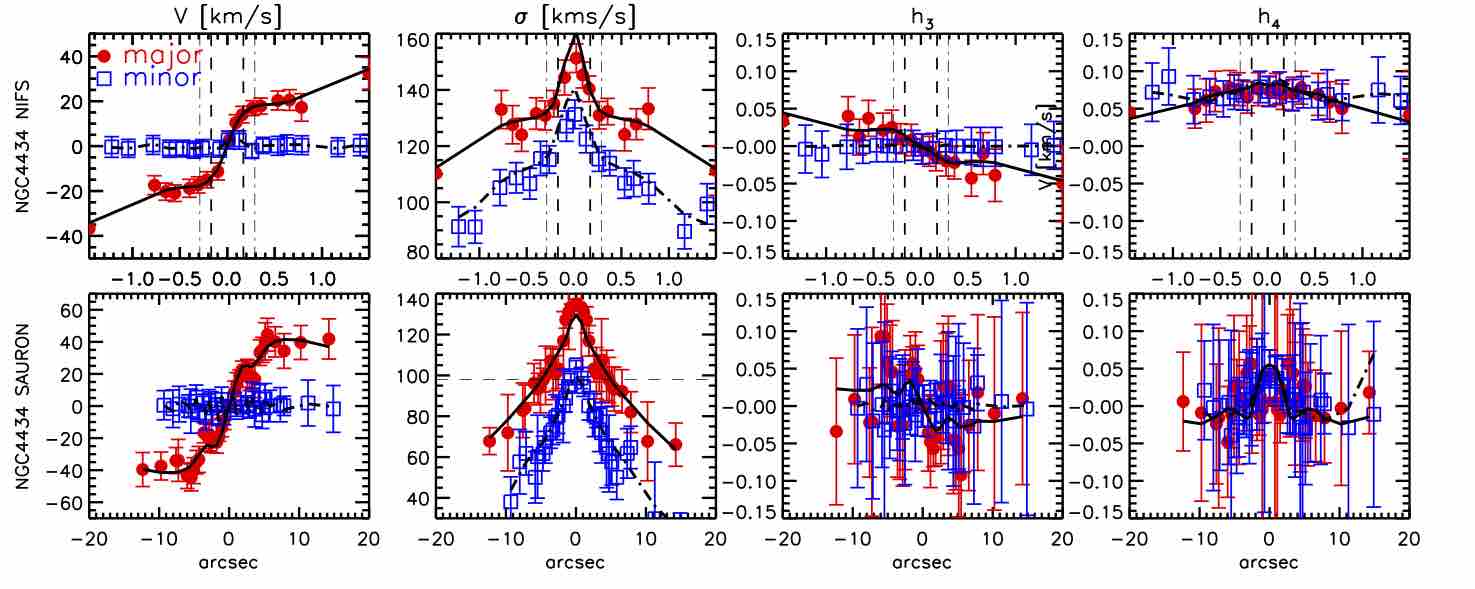}
\includegraphics[width=0.98\textwidth]{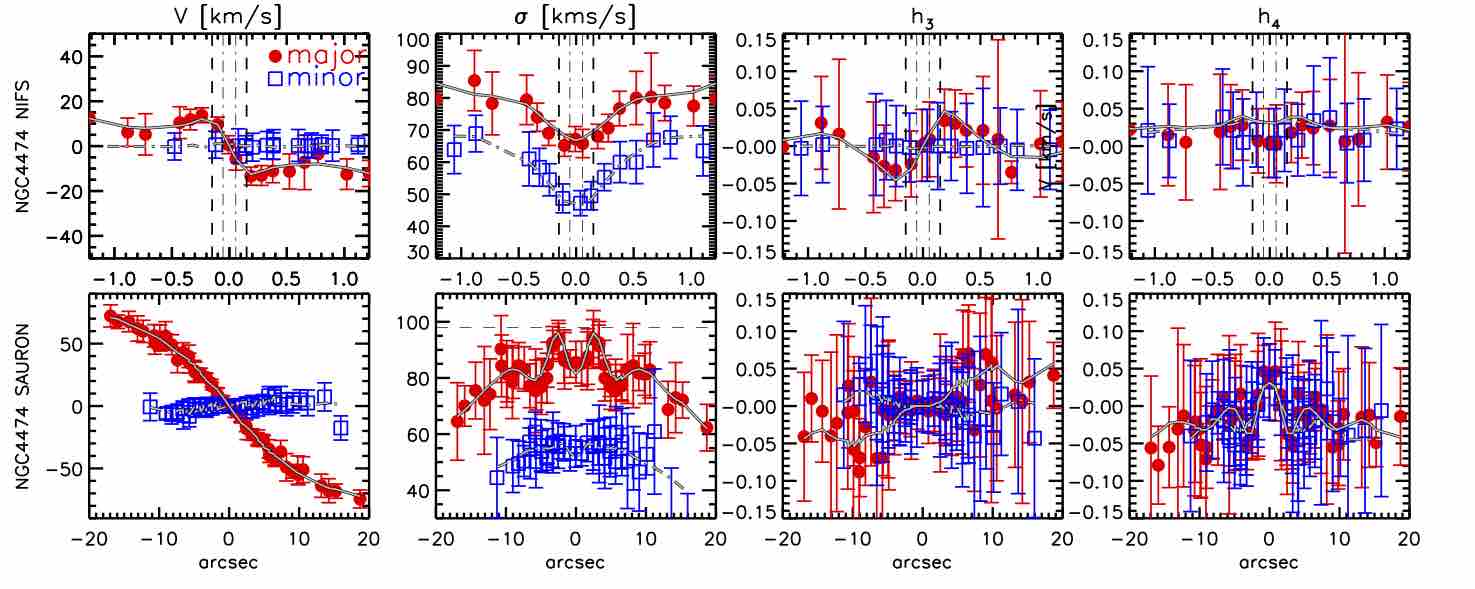}
\caption{Comparison between the best-fitting Schwarzschild models and the symmetrized kinematics extracted along the major and minor axes for both NIFS and SAURON data of NGC\,4339, NGC\,4434 and NGC\,4474 (for NGC\,4474 we used both the formally best fitting model and the model providing the upper limit on M$_{\rm BH}$). Major axis data and models are shown with a solid line and red circles, while the a dash--dotted line and blue open squares represent the positions along the minor axis. Upper limit models are shown with grey lines. The data were extracted along pseudo-slits of width 0.05\arcsec\, and 0.8\arcsec for NIFS and SAURON data respectively. Each galaxy is represented by two rows (top -- NIFS, bottom -- SAURON) of four panels showing the mean velocity, the velocity dispersion, and $h_3$ and $h_4$ Gauss--Hermite moments. The minor axis (both for NIFS and SAURON) velocity dispersion values are offset downwards for better visibility. Vertical dashed lines on the plots pertaining to NIFS data-model comparison indicate the FWHM of the PSF, while vertical dot--dashed lines indicate the estimated size of the SoI of the best-fitting model or the upper limit (in case of NGC\,4474 and NGC\,4551).  Horizontal line on the velocity dispersion plot for SAURON data indicates the SAURON spectral resolution. NIFS spectral resolution ($\sim$30km/s) is below the plotting range.}
\label{f:sch_longslit1}
\end{figure*}
%%%%%%%%%%%%%%%%%%%%%%%%%%%%%%%%%%%%%%%%%%%%%%%%%%%%%%%%%%%

%%%%%%%%%%%%%%%%%%%%%%%%%%%%%%%%%%%%%%%%%%%%%%%%%%%%%%%%%%%
\begin{figure*}
%Fig made by plot_sch_model_NIFS_sample_comparison.pro
\includegraphics[width=\textwidth]{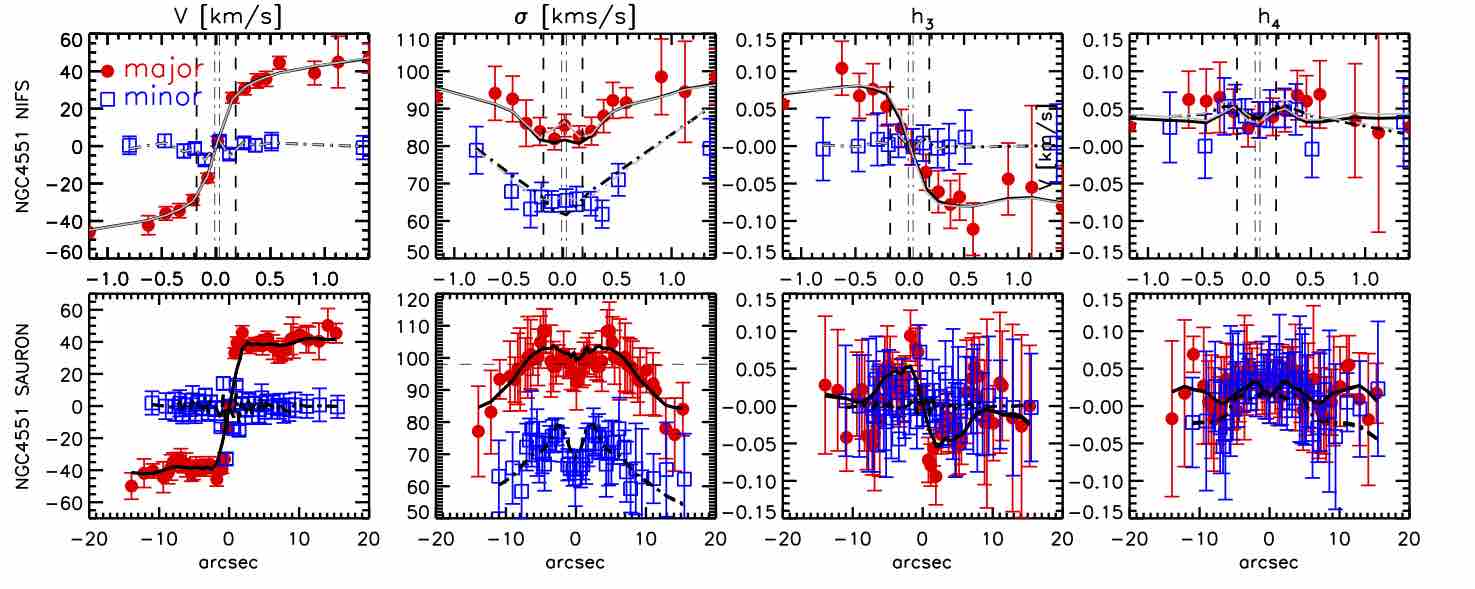}
\includegraphics[width=\textwidth]{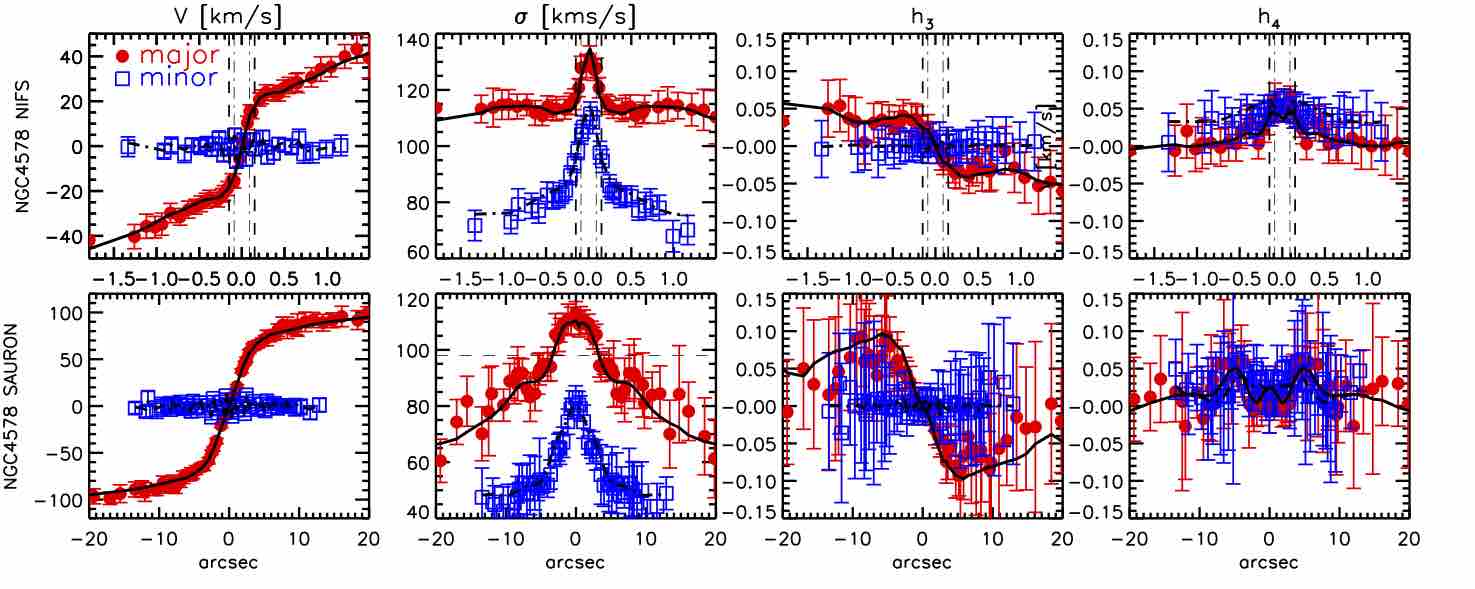}
\includegraphics[width=\textwidth]{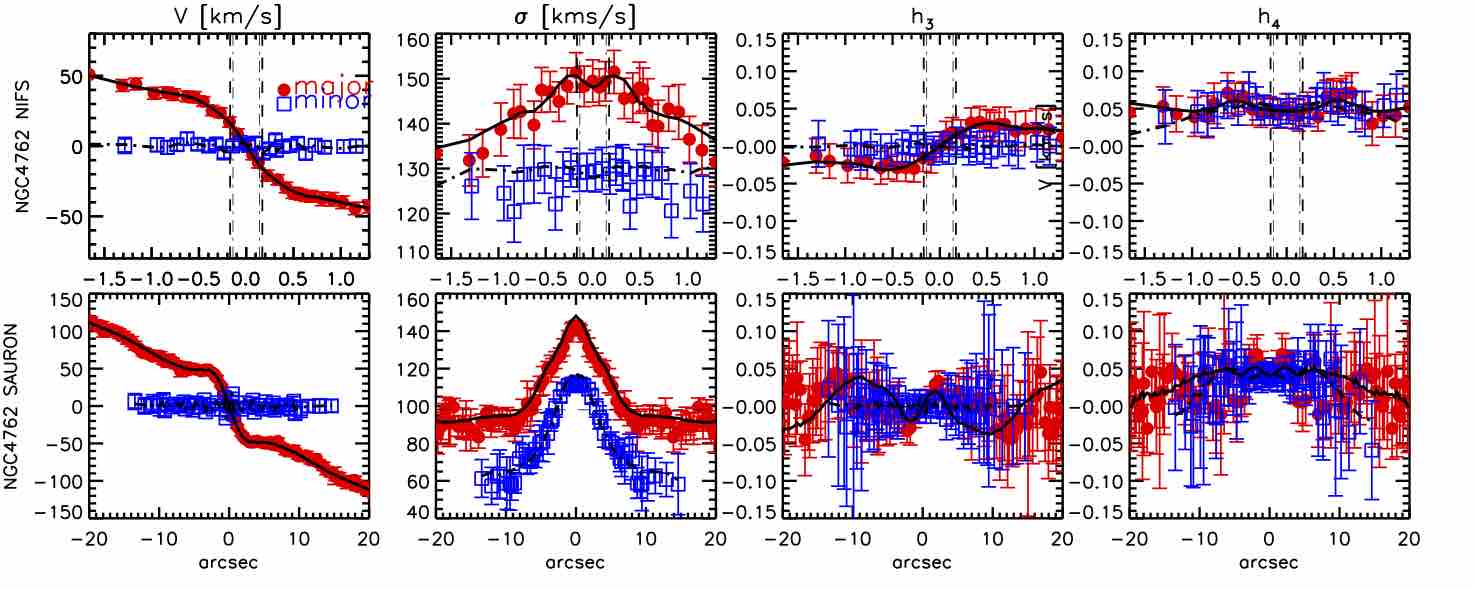}
\caption{Same as Fig.~\ref{f:sch_longslit1} but for NGC\,4551, NGC\,4578 and NGC\,4762. }
\label{f:sch_longslit2}
\end{figure*}
%%%%%%%%%%%%%%%%%%%%%%%%%%%%%%%%%%%%%%%%%%%%%%%%%%%%%%%%%%%

%%%%%% Figure A2%%%%%%%%%%%%%%%%%%%%%%%%%%%%%%%%%%%%%%%%%%%%%%%
\begin{figure*}
%Fig made by ngc4339_plot_schwarz_model_maps_onefig
\includegraphics[width=0.45\textwidth]{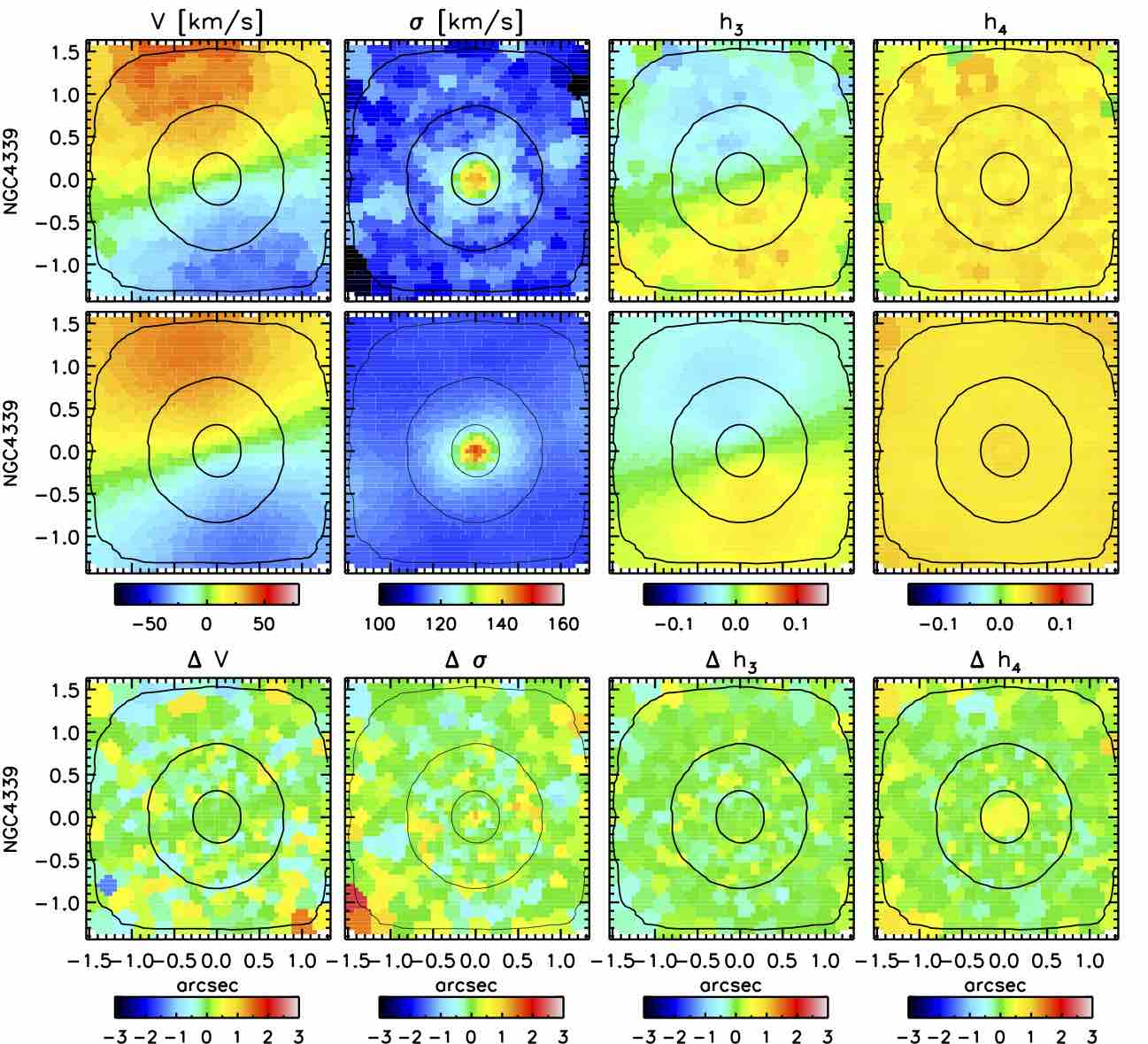}
\includegraphics[width=0.45\textwidth]{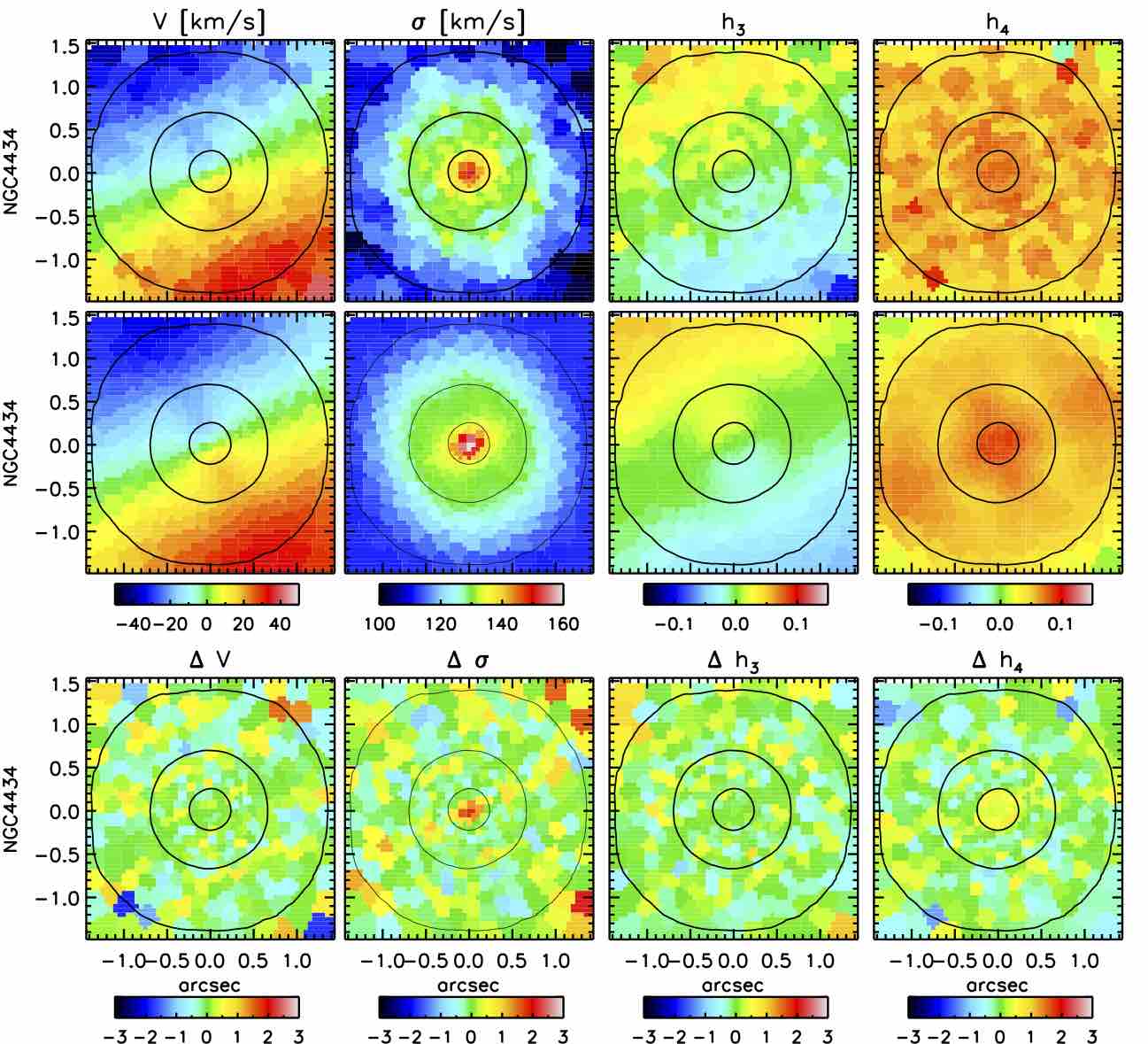}
\includegraphics[width=0.45\textwidth]{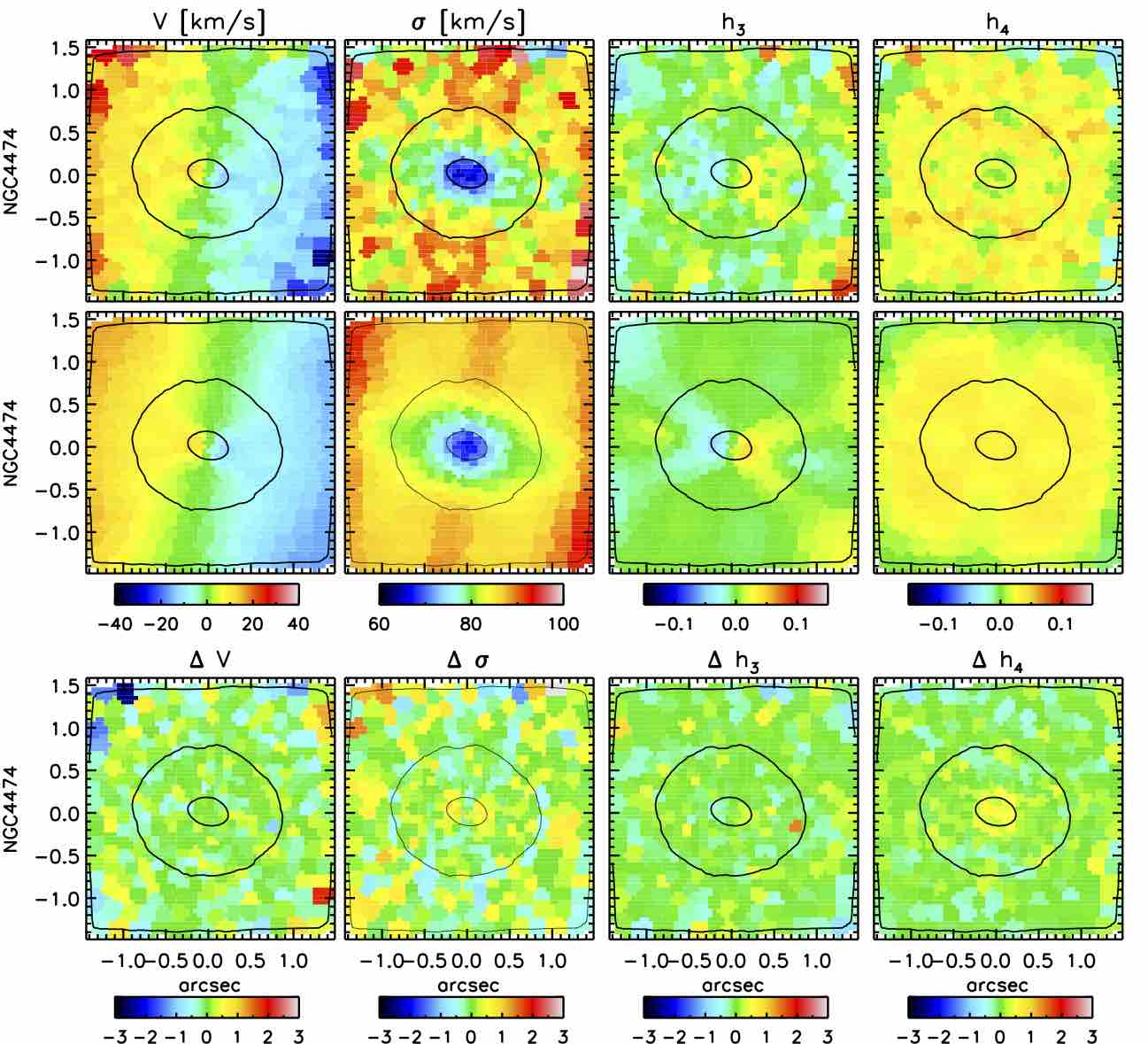}
\includegraphics[width=0.45\textwidth]{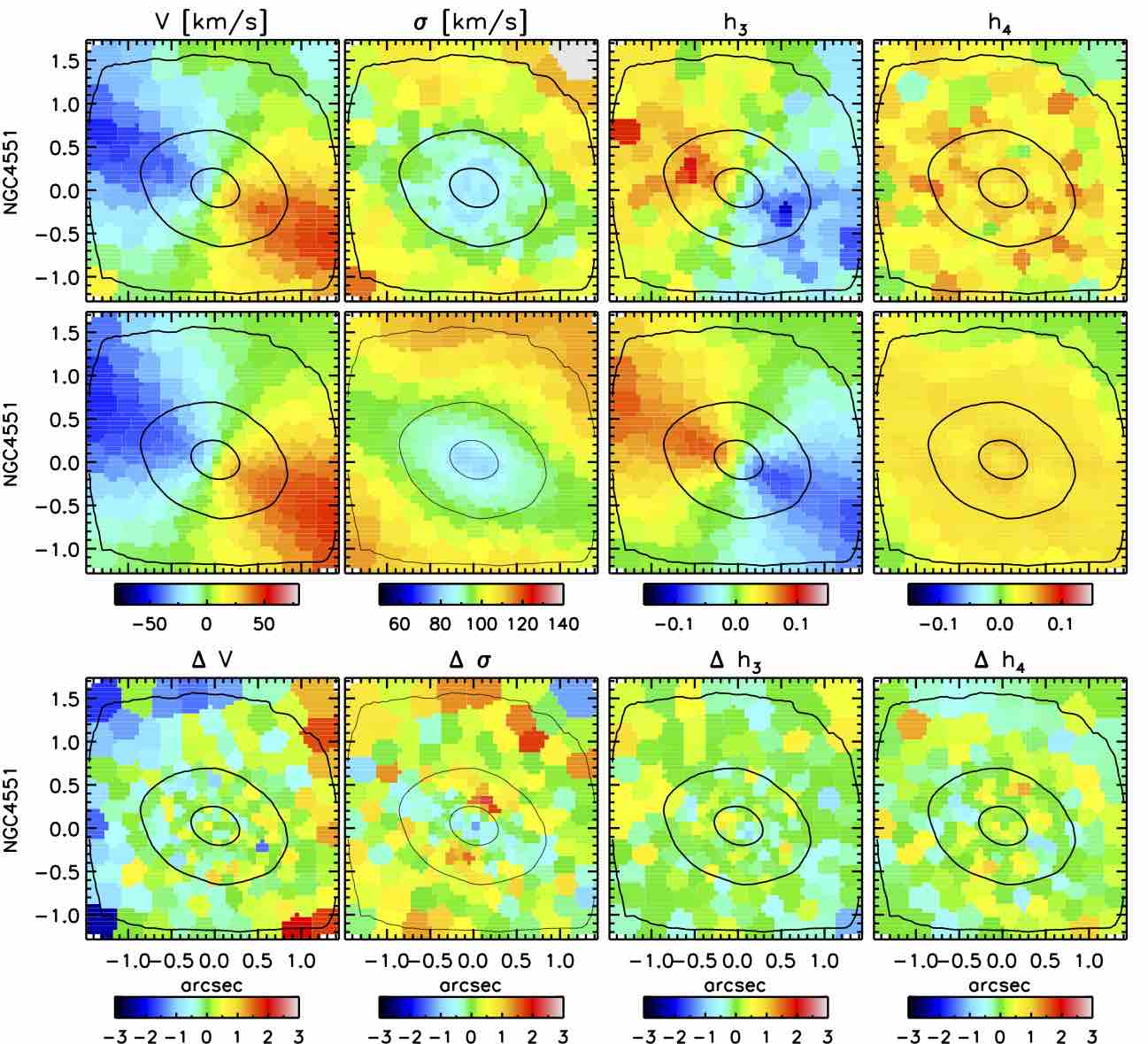}
\includegraphics[width=0.45\textwidth]{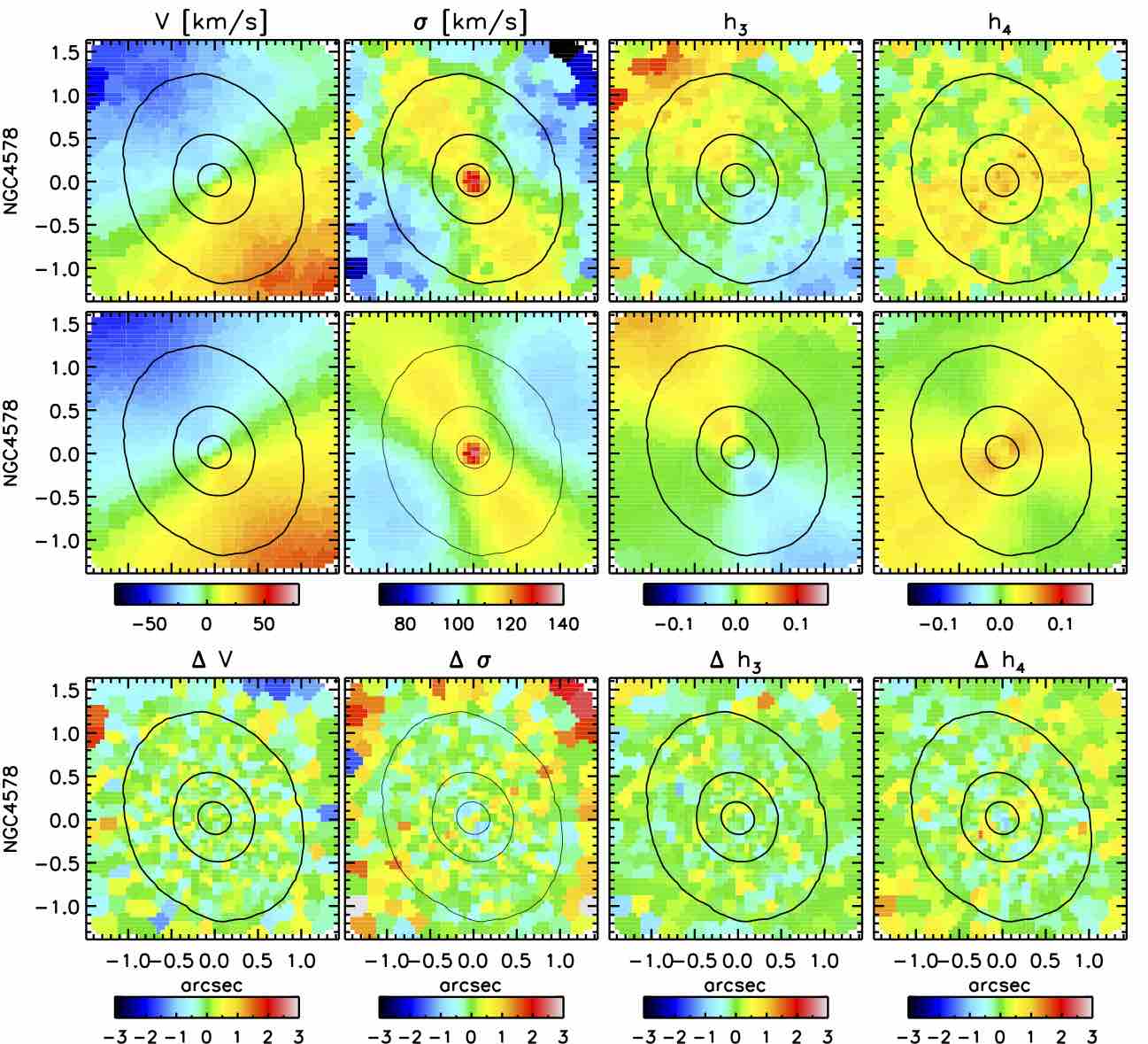}
\includegraphics[width=0.45\textwidth]{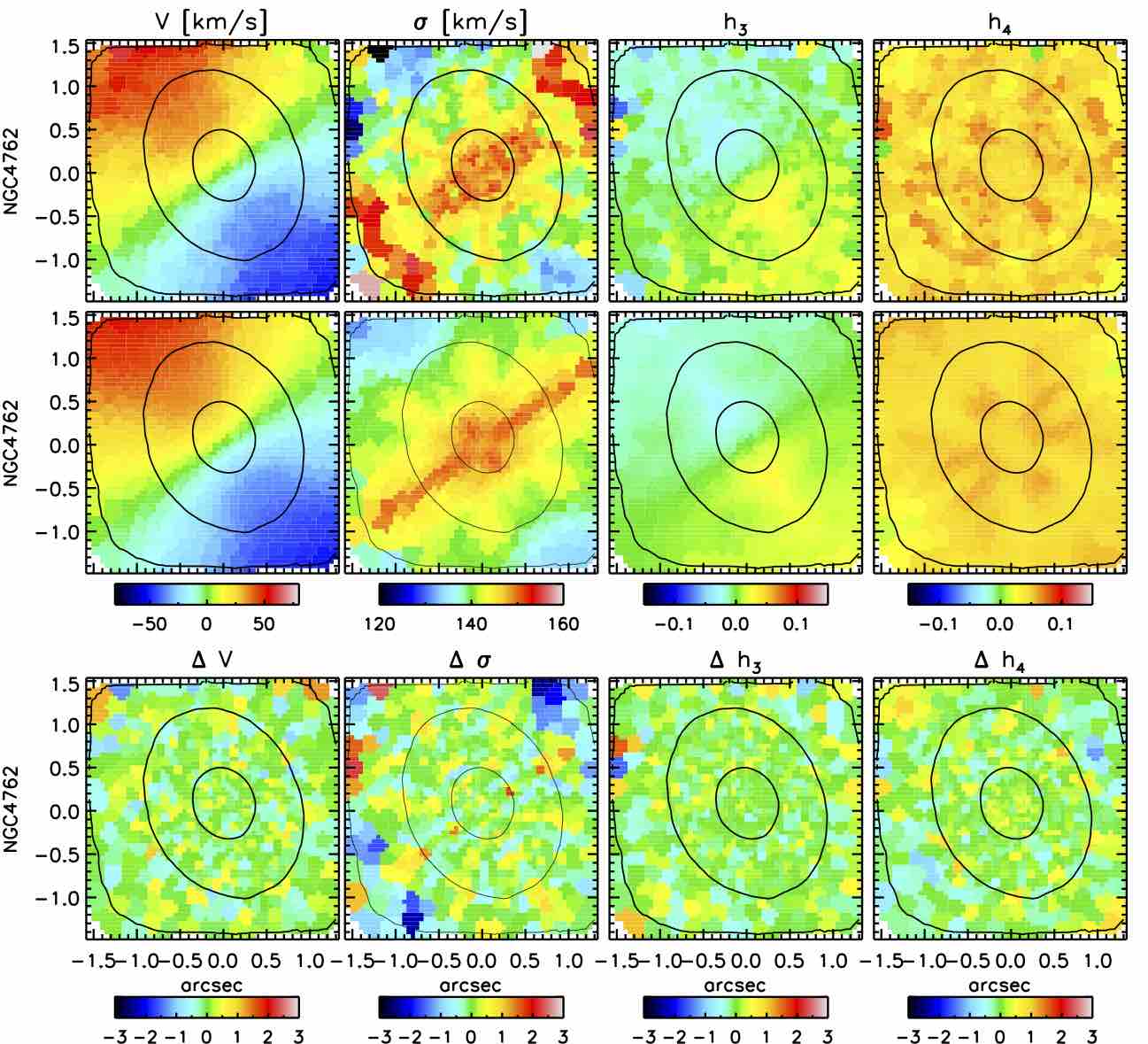}
\caption{Comparison between the maps of symmetrized kinematics observed with NIFS and the best fitting Schwarzschild models. Each galaxy has three rows of maps showing the mean velocity, velocity dispersion, and $h_3$ and $h_4$ Gauss-Hermite moments. Galaxies are (top left to bottom right): NGC\,4339, NGC\,4434, NGC\,4474, NGC\,4551, NGC\,4578 and NGC\,4762. Observations are in the first row, Schwarzschild models in the second row and the third row shows residuals (model data for a given kinematic map) divided by the uncertainties. This means that colours trace the level at which the models differ to the data relative to the errors, and the extremes are set to 3 times the uncertainty. Black contours are light isophotes as observed and objects are oriented such that north is up and east to the left.}
\label{fapp:nifs}
\end{figure*}
%%%%%%%%%%%%%%%%%%%%%%%%%%%%%%%%%%%%%%%%%%%%%%%%%%%%%%%%%

%%%%%% Figure A2%%%%%%%%%%%%%%%%%%%%%%%%%%%%%%%%%%%%%%%%%%%%%%%
\begin{figure*}
%Fig made by ngc4339_plot_schwarz_model_maps_onefig
\includegraphics[width=0.45\textwidth]{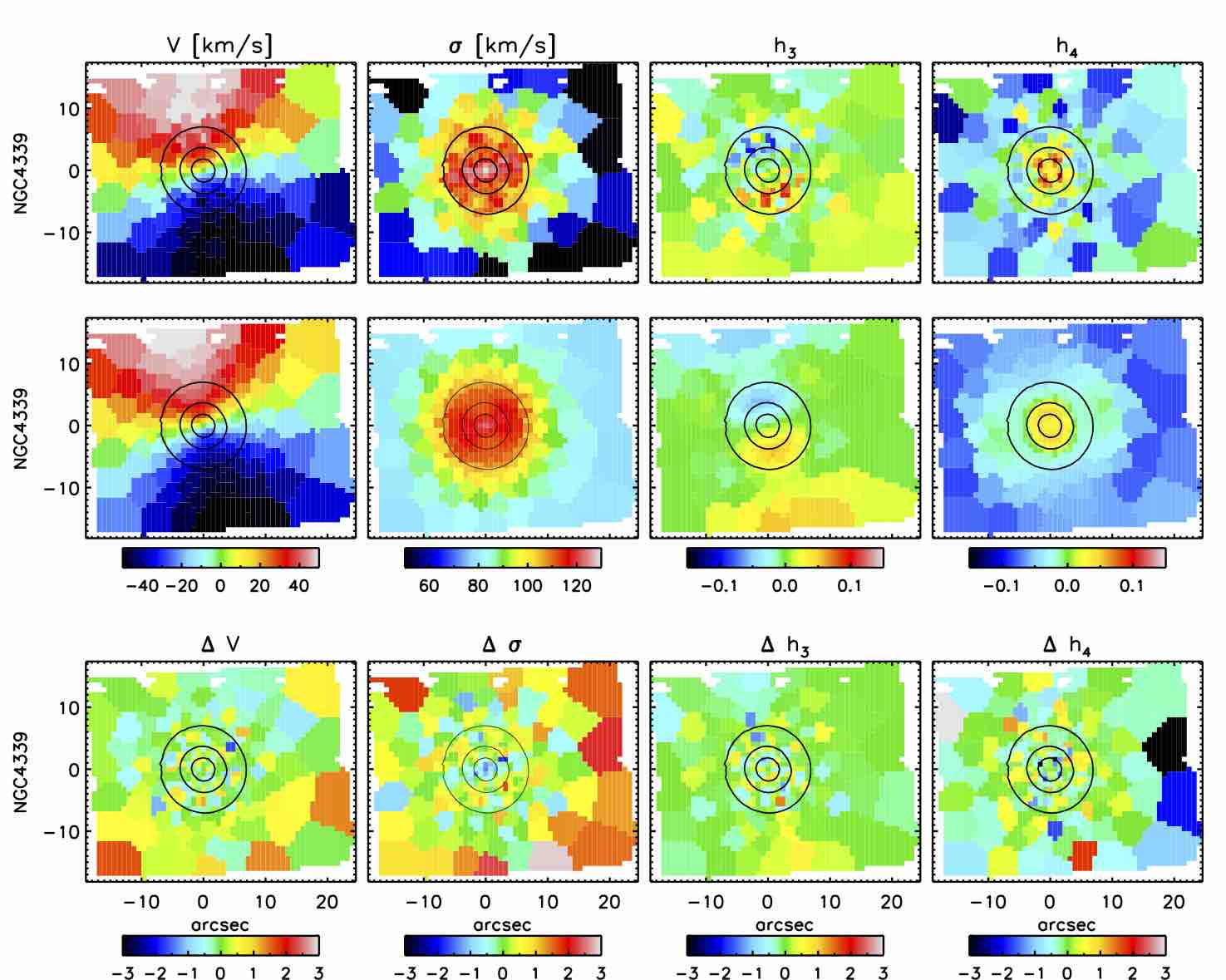}
\includegraphics[width=0.44\textwidth]{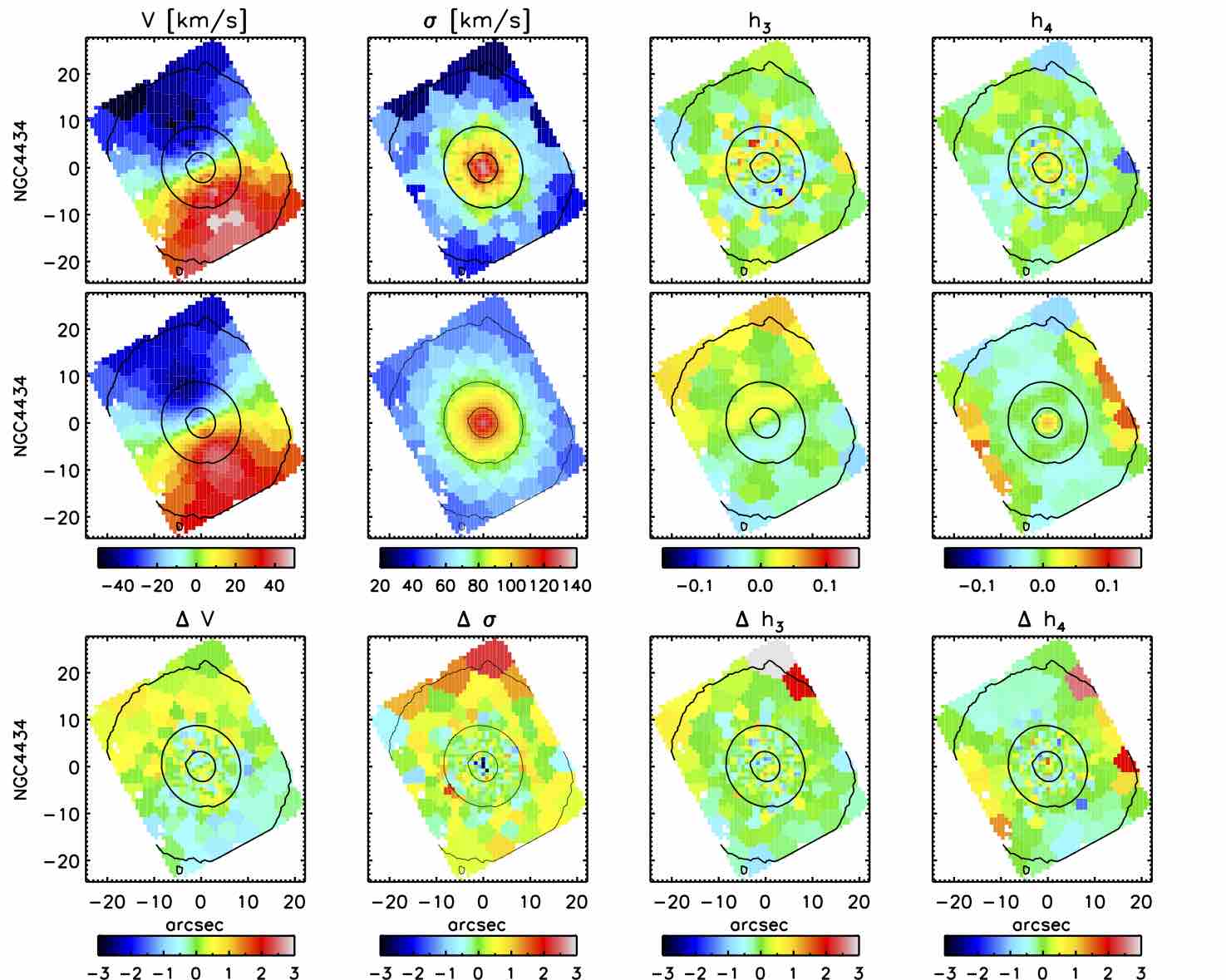}
\includegraphics[width=0.44\textwidth]{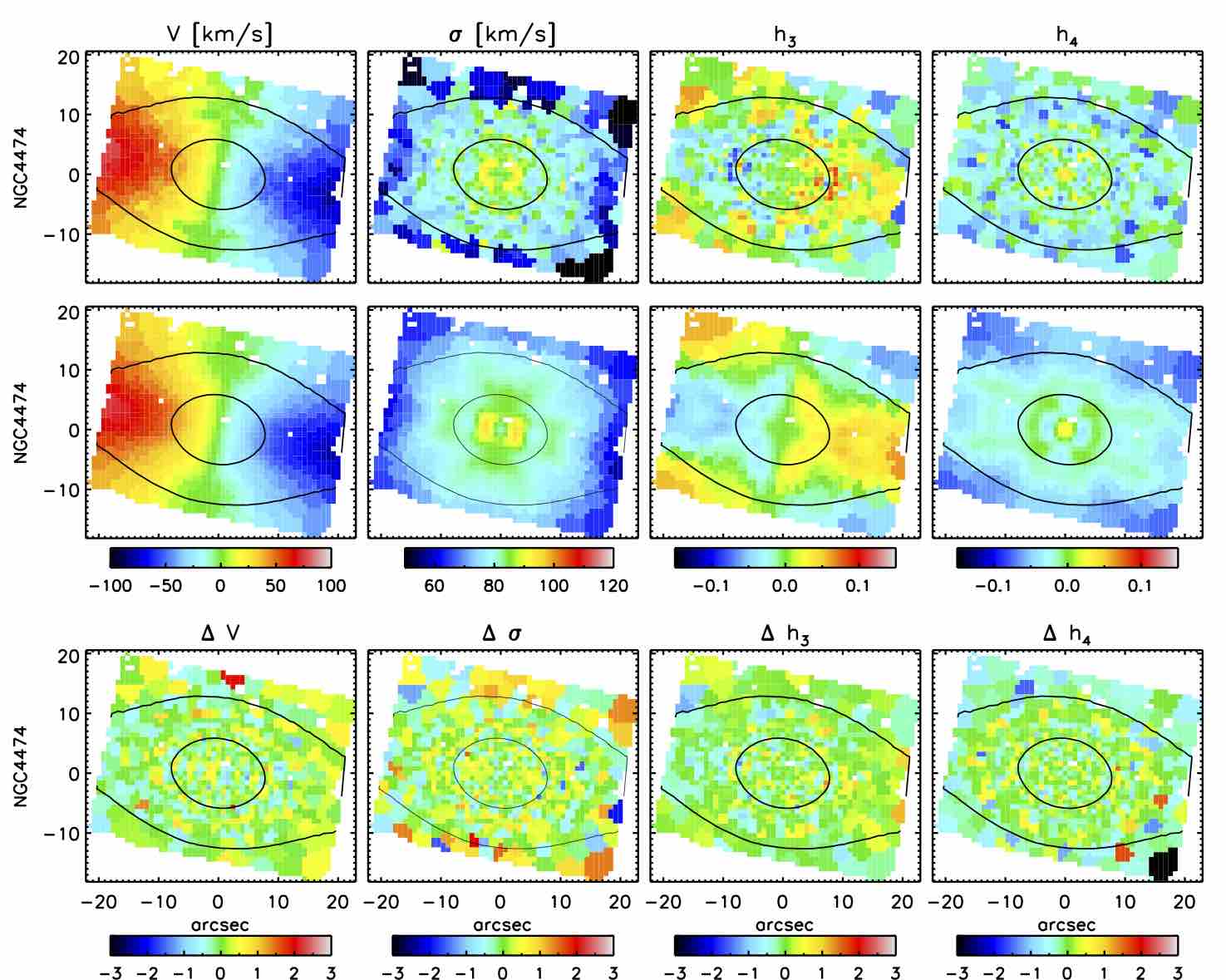}
\includegraphics[width=0.48\textwidth]{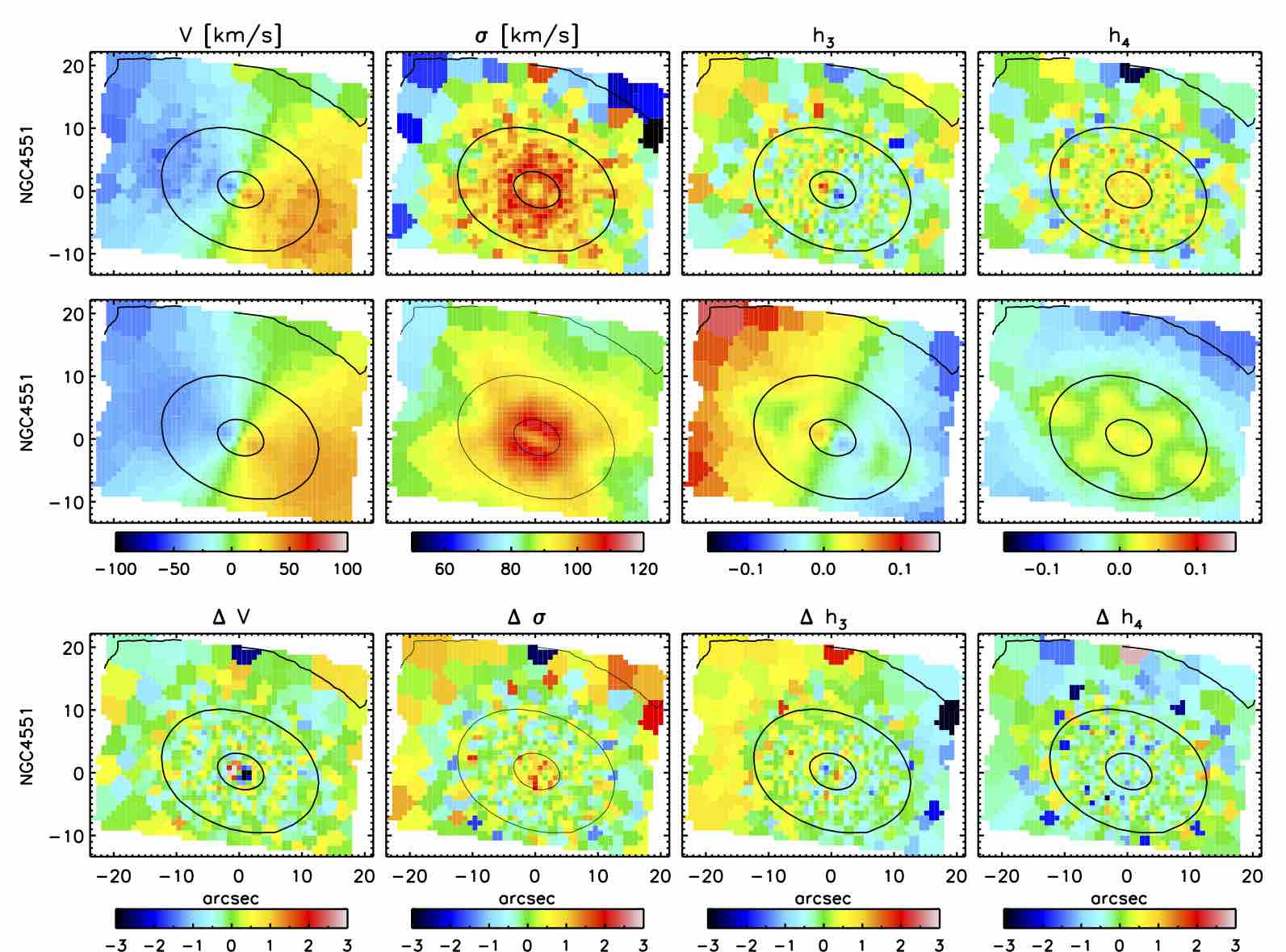}
\includegraphics[width=0.48\textwidth]{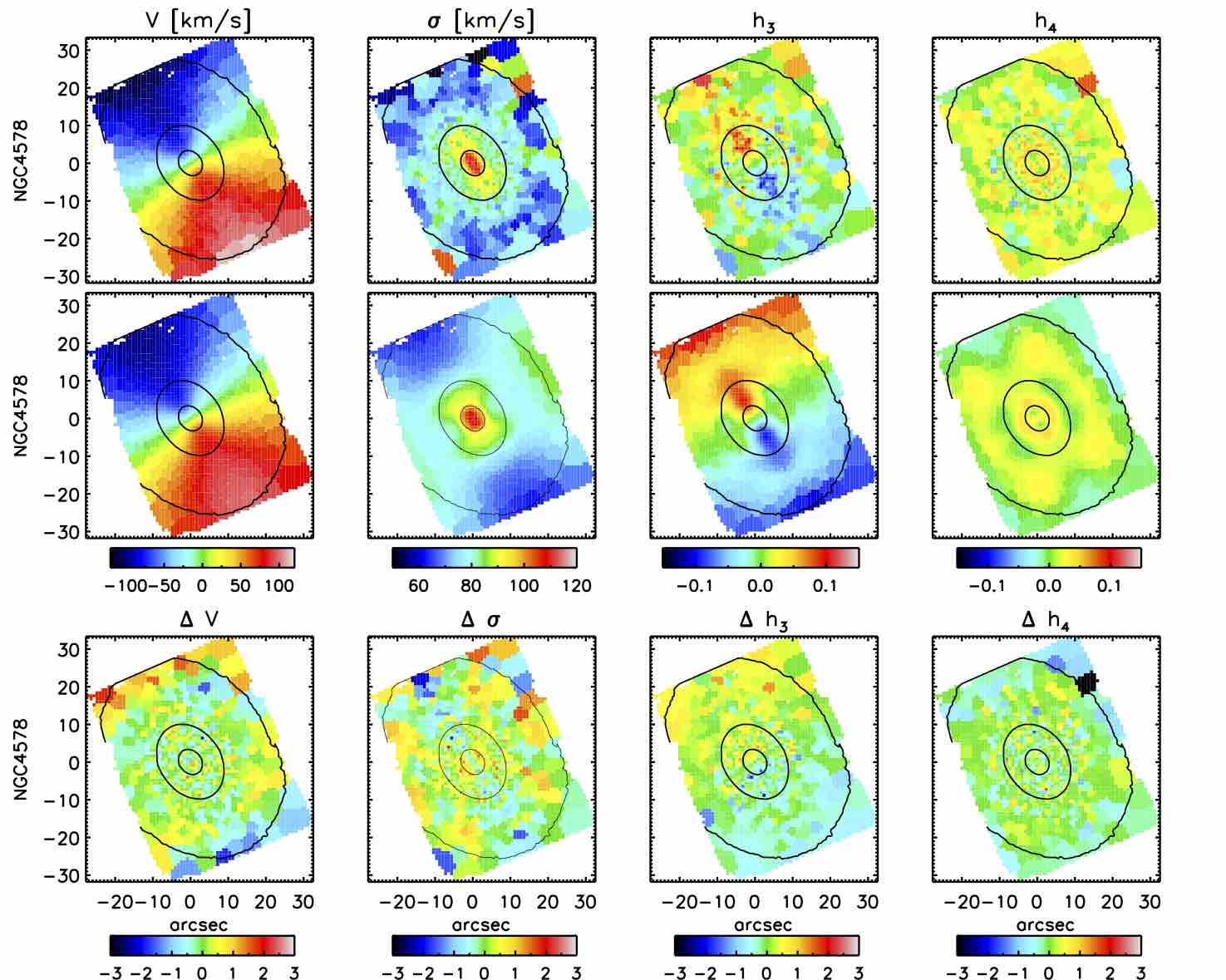}
\includegraphics[width=0.43\textwidth]{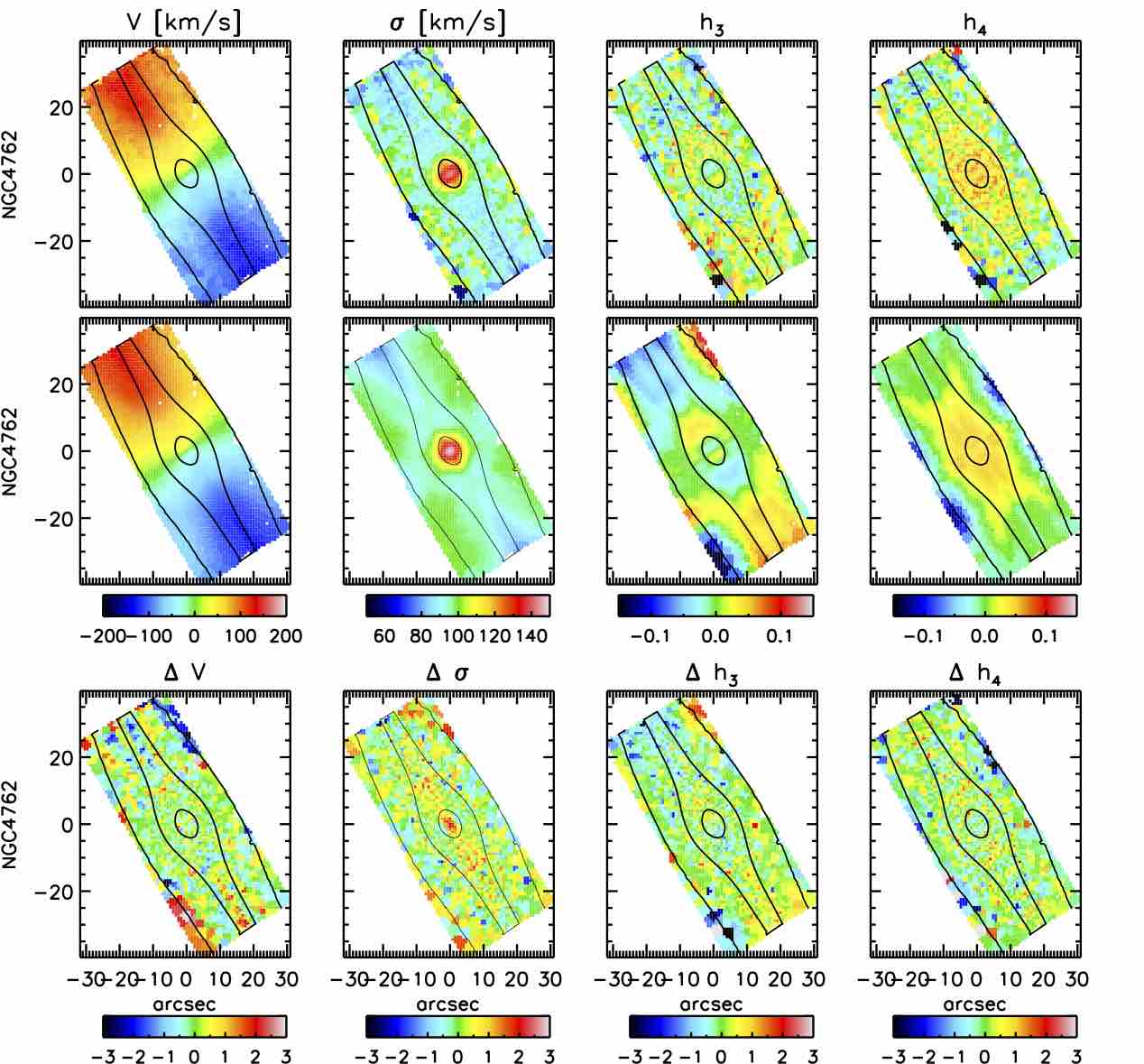}
\caption{Same as Fig~\ref{fapp:sau} but reproducing the SAURON kinematics. }
\label{fapp:sau}
\end{figure*}
%%%%%%%%%%%%%%%%%%%%%%%%%%%%%%%%%%%%%%%%%%%%%%%%%%%%%%%%%

%%%%%%%%%%%%%%%%%%%%%%%%%%%%%%%%%%%%%%%%%%%%%%%%%%

% Don't change these lines
\bsp	% typesetting comment
\label{lastpage}
\end{document}